\documentclass[pdftex,twocolumn,epjc3_preprint,runningheads]{svjour3}

\usepackage{fix-cm}
\usepackage{graphicx}
\usepackage{latexsym}
\usepackage[numbers,sort&compress]{natbib}
\usepackage[colorlinks,citecolor=blue,urlcolor=blue,linkcolor=blue]{hyperref}
\usepackage{xspace}
\usepackage[dvipsnames,x11names,svgnames]{xcolor}
\usepackage[utf8]{inputenc}
\usepackage{listings}
\usepackage{amsmath}
\usepackage{amsfonts}
\usepackage{amsbsy}
\usepackage{subdepth}
\usepackage{afterpage}

\usepackage{etoolbox}
\makeatletter
\def\@mkboth#1#2{}
\newlength\appendixwidth
\preto\appendix{\addtocontents{toc}{\protect\patchl@section}}
\newcommand{\patchl@section}{%
  \settowidth{\appendixwidth}{\textbf{Appendix }}%
  \addtolength{\appendixwidth}{1.5em}%
  \patchcmd{\l@section}{1.5em}{\appendixwidth}{}{\ddt}%
}
\makeatother

\allowdisplaybreaks

\newcommand\ie{{i.e.}\xspace}
\newcommand\eg{{e.g.}\xspace}
\newcommand{\smoking}{\textsf{xsec}\xspace}
\newcommand{\Smoking}{\textsf{Xsec}\xspace}
\newcommand{\prospino}{\textsf{Prospino}\xspace}
\newcommand{\prosp}{\textsf{prosp}\xspace}

\newcommand{\madgraph}{\textsf{MadGraph}\xspace}
\newcommand{\resummino}{\textsf{Resummino}\xspace}
\newcommand{\nllfast}{\textsf{NLL-fast}\xspace}
\newcommand{\nnllfast}{\textsf{NNLL-fast}\xspace}
\newcommand{\deepxs}{\textsf{DeepXS}\xspace}
\newcommand{\pyslha}{\textsf{PySLHA}\xspace}
\newcommand\python{\textsf{Python}\xspace}
\newcommand\numpy{\textsf{NumPy}\xspace}
\newcommand\scipy{\textsf{SciPy}\xspace}
\newcommand\scikit{\textsf{scikit-learn}\xspace}
\newcommand\pip{\textsf{pip}\xspace}
\newcommand\lhapdf[1]{\textsf{LHAPDF\,#1}\xspace}
\newcommand\softsusy{\textsf{SoftSUSY}\xspace}
\newcommand\blas{\textsf{BLAS}\xspace}
\newcommand\lapack{\textsf{LAPACK}\xspace}
\newcommand\mkl{\textsf{Intel MKL}\xspace}
\newcommand\openblas{\textsf{OpenBLAS}\xspace}
\newcommand\atlas{\textsf{ATLAS}\xspace}
\newcommand\colliderbit{\textsf{ColliderBit}\xspace}

\newcommand{\eV}{\text{e\kern-0.1ex V}\xspace}

\newcommand{\GeV}{\text{G\eV}\xspace}
\newcommand{\TeV}{\text{T\kern-0.1ex\eV}\xspace}

\def\BibTeX{{\rm B\kern-.05em{\sc i\kern-.025em b}\kern-.08em
    T\kern-.1667em\lower.7ex\hbox{E}\kern-.125emX}\xspace}

\newcommand\mcM{\mathcal{M}}
\newcommand\mcD{\mathcal{D}}
\newcommand\mcDtest{\mathcal{D}_\textsf{test}}
\newcommand\mcDtesttb{\mathcal{D}_{\textsf{test-tb}}}
\newcommand\mcDMSSM{\mathcal{D}_\textsf{MSSM-24}}
\newcommand*\vecell{\boldsymbol{\ell}}
\renewcommand\vec{\mathbf}

\makeatletter
\newcommand{\preprintnumber}[1]{\gdef\@preprintnumber{\begin{flushright}{#1}\end{flushright}}}
\makeatother

\definecolor{solarized@base01}{HTML}{586e75}
\definecolor{solarized@base2}{HTML}{EEE8D5}
\definecolor{solarized@orange}{HTML}{CB4B16}
\definecolor{solarized@red}{HTML}{DC322F}
\definecolor{solarized@violet}{HTML}{6C71C4}
\definecolor{solarized@green}{HTML}{859900}
\definecolor{darkred}{HTML}{550003}
\lstdefinestyle{python}{
  language=Python,
  basicstyle=\small\ttfamily,
  basewidth={0.53em,0.44em},
  numbers=none,
  tabsize=2,
  breaklines=true,
  escapeinside={@}{@},
  showstringspaces=false,
  numberstyle=\tiny\color{solarized@base01},
  keywordstyle=\color{blue},
  stringstyle=\color{orange}\ttfamily,
  identifierstyle=\color{darkred},
  commentstyle=\color{purple},
  emphstyle=\color{green},
  frame=single,
  rulecolor=\color{solarized@base2},
  rulesepcolor=\color{solarized@base2},
  literate = {~}{\customtilde}1
  }
  \lstdefinestyle{terminal}{
    language=bash,
    basicstyle=\small\ttfamily,
    numbers=none,
    tabsize=2,
    breaklines=true,
    escapeinside={@}{@},
  frame=single,
  showstringspaces=false,
  numberstyle=\tiny\color{solarized@base01},
  keywordstyle=\color{solarized@orange},
  stringstyle=\color{solarized@red}\ttfamily,
  identifierstyle=\color{black},
  commentstyle=\color{solarized@violet},
  emphstyle=\color{solarized@green},
  frame=single,
  rulecolor=\color{solarized@base2},
  rulesepcolor=\color{solarized@base2},
  morekeywords={python, pip, git, xsec},
  deletekeywords={test},
  literate = {.git}{{.}{\color{black}}git}4
             {--help}{{--}{\color{black}}help}6
             {-download}{{\color{solarized@orange}}\!\!\!{-}download{\color{black}}}9
             {-gprocs}{{\color{solarized@orange}}\!\!\!{-}gprocs{\color{black}}}7
             {~}{\customtilde}1
}

\newcommand\py[1]{{\lstset{style=python}\lstinline!#1!}}
\newcommand\term[1]{{\lstset{style=terminal}\lstinline!#1!}}

\newcommand{\metavar}[1]{\textit{\color{OliveGreen}\texttt{#1}}}

\usepackage{todonotes}
\usepackage{xpatch}
\makeatletter
\xpretocmd{\todo}{\@bsphack}{}{}
\xapptocmd{\todo}{\@esphack}{}{}
\makeatother

\journalname{Eur. Phys. J. C}

\begin{document}

\preprintnumber{SAGEX-20-17-E}

\title{\Smoking:  the cross-section evaluation code}

\author{
  Andy~Buckley\thanksref{addr1}
  \and
  Anders~Kvellestad\thanksref{addr3,addr4}
  \and
  Are~Raklev\thanksref{addr4}
  \and
  Pat~Scott\thanksref{addr3,addr5}
  \and
  Jon~Vegard~Sparre\thanksref{addr6}
  \and
  Jeriek~Van~den~Abeele\thanksref{addr4,e1}
  \and
  Ingrid~A.~Vazquez-Holm\thanksref{addr2}
  }

  \institute{School of Physics and Astronomy, University of Glasgow, Glasgow, G12 8QQ, UK \label{addr1}
  \and
  Department of Physics, Imperial College London, South Kensington, SW7 2AZ, UK \label{addr3}
  \and
  Department of Physics, University of Oslo, N-0316 Oslo, Norway \label{addr4}
  \and
  School of Mathematics and Physics, The University of Queensland, St.\ Lucia, Brisbane, QLD 4072, Australia \label{addr5}
  \and
  The Norwegian Labour and Welfare Administration, N-0557 Oslo, Norway \label{addr6}
  \and
  Institut de Physique Th\'{e}orique, Universit\'{e} Paris-Saclay, CEA, CNRS, F-91191 Gif-sur-Yvette, France \label{addr2}
}

\thankstext{e1}{email: jeriekvda@fys.uio.no}

\date{Received: date / Accepted: date}

\maketitle

\begin{abstract}
The evaluation of higher-order cross-sections is an important component in the search for new physics,
both at hadron colliders and elsewhere. For most new physics processes of interest, total cross-sections are known at next-to-leading order (NLO) in the strong coupling $\alpha_s$, and often beyond, via either higher-order terms at fixed powers of $\alpha_s$, or multi-emission resummation. However, the computation time for such higher-order cross-sections is prohibitively expensive, and precludes efficient evaluation in parameter-space scans beyond two dimensions.
Here we describe the software tool \smoking, which allows for fast evaluation of cross-sections based on the use of machine-learning regression, using distributed Gaussian processes trained on a pre-generated sample of parameter points. This first version of the code provides all NLO Minimal Supersymmetric Standard Model strong-produc\-tion cross-sections at the LHC, for individual flavour final states, evaluated in a fraction of a second. Moreover, it calculates regression errors, as well as estimates of errors from higher-order contributions, from uncertainties in the parton distribution functions, and from the value of $\alpha_s$.
While we focus on a specific phenomenological model of supersymmetry, the method readily generalises to any process where it is possible to generate a sufficient training sample.
\end{abstract}

\tableofcontents

\section{Introduction}
\label{sec:intro}
The determination of cross-sections beyond leading order (LO) is typically very computationally expensive because of the evaluation of tensorial loop integrals.  This is especially so for hadronic interactions, where the loop integrals must themselves be numerically integrated over the relevant parton distribution functions (PDFs).

\begin{figure}
\includegraphics[width=0.5\textwidth]{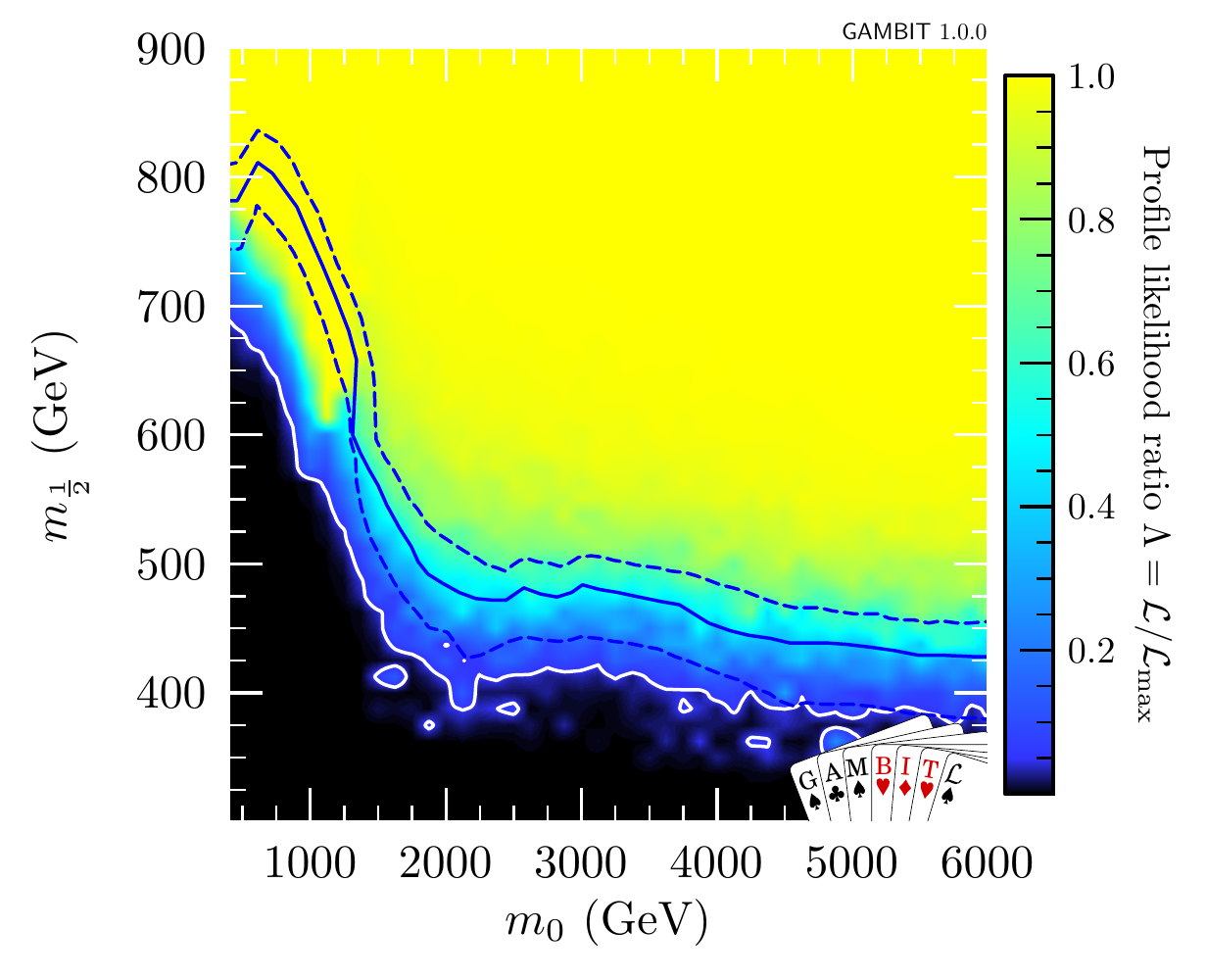}
\caption{Colour map showing the profile likelihood ratio $\mathcal{L}/\mathcal{L}_\mathrm{max}$ for a \colliderbit scan over the CMSSM model defined by $m_0$ and $m_{1/2}$, with $\tan\beta=30$, $A_0=-2m_0$ and $\mu>0$, using the ATLAS zero-lepton supersymmetry search likelihood from $\sqrt{s}=8$~TeV data~\cite{ATLAS:0LEP_20invfb}.
The solid white line indicates the 95\%~CL exclusion contour found with {\textsf GAMBIT~1.0}~\cite{Athron:2017ard} using LO cross-sections. The solid blue line shows the corresponding ATLAS 95\%~CL observed exclusion limit with NLO+NLL cross-sections, with dashed blue lines showing the $\pm 1 \sigma$ theoretical cross-section uncertainty. Reproduced from Fig.~3 in Ref.~\cite{Balazs:2017moi}.}
\label{fig:NLOcomp}
\end{figure}

The computational cost of evaluation per parameter point restricts the usage of next-to-leading order (NLO) cross-sections to (simplified) new physics models with only one or two relevant parameters. However, higher-order contributions can be very important in many other models.  This is especially true for strong interactions, where NLO contributions can be of comparable size to the LO contribution. The physics impact of this restriction is quite dramatic.  In Fig.~\ref{fig:NLOcomp}, reproduced from Ref.~\cite{Balazs:2017moi}, we show as an example the significant differences between the limits resulting from a parameter scan of the Constrained Minimal Supersymmetric Standard Model (CMSSM) using LO rather than NLO cross-sections. This also highlights the importance of propagating theory uncertainties, \eg from the PDFs and the scale dependence, through to the physical observables. As the uncertainties on LO hard-scattering cross-sections are typically very large, owing in part to the missing higher-order terms, such theory error propagation to the cross-sections can only really be considered representative from NLO accuracy onwards. Even at NLO, we can see in Fig.~\ref{fig:NLOcomp} that the impact is considerable, and needs to be taken into account when fitting models.

In this work we present \smoking, an attempt at a general solution to the speed problem irrespective of the type of cross-section being evaluated, the size of the parameter space of the model, and the underlying precision of the calculation. We achieve this by performing machine-learning regression on a pre-generated training dataset consisting of cross-sections sampled from a model. In this first version of \smoking, we will be focusing on strong-production cross-sections in the MSSM, but the selection of cross-sections and models will be extended in future versions.

The increasing sparsity of training data in models with larger numbers of free parameters means that a lot of care must go into performing the regression, and that reliable regression errors can only be determined on a point-by-point basis.
We choose to use Gaussian process (GP) regression \cite{Rasmussen:2006:GPM:1162254}, a highly flexible Bayesian method for modelling functions. It is not restricted to pre-determined functional shapes with a set number of parameters, and is instead directly informed by the data together with some prior understanding of the underlying correlation structure of the function that it is used to model.
Moreover, a point-specific regression uncertainty follows naturally from the
posterior predictive distribution, which is straightforward to compute in closed form.

Training GP models involves the inversion of a matrix of the size of the number of training points. This means that while GPs are a very powerful regression tool, single GPs scale very badly with increasing training data, and training becomes infeasible beyond ${\mathcal O}(10^4)$ data points. To overcome this difficulty we use a model with factorised and distributed training, where the data are split into manageable subsets each assigned to a single GP.  The regression prediction is subsequently determined by a combination of the predictions of the individual GPs \cite{2015arXiv150202843D,2018arXiv180600720L}.

Current codes that can evaluate a broad collection of MSSM cross-sections at NLO include \prospino~\textsf{2.1}~\cite{Beenakker:1996ch,Beenakker:1996ed,Beenakker:1997ut,Beenakker:1999xh,Spira:2002rd,Plehn:2004rp} and \madgraph~\textsf{5}~\cite{Frixione:2019fxg}. However, the evaluation time per parameter point on a modern CPU is in the range of minutes for a single final state, far too slow for large parameter scans. This does not include the calculation of scale, $\alpha_s$, nor PDF errors, which in \prospino increases the evaluation time by orders of magnitude.
In the \madgraph implementation, this can be done more efficiently through reweighting techniques.
However, the NLO functionality for the MSSM has only recently been made publicly available in \madgraph, and after the computationally expensive sample generation campaign for the current version of \smoking was completed.

A test of the calculation time per parameter point for all strong-production cross-sections in  \prospino, including the evaluation of scale, $\alpha_s$, and PDF errors, when \prospino has been heavily optimised with the \textsf{ifort} compiler, takes around 2.5 hours on a single Intel Xeon-Gold 6138 (Skylake) core running at 2.0 GHz.

Fast evaluation of strong-production cross-sections in the MSSM at NLO, with next-to-leading (NLL) and next-to-next-to-leading (NNLL) logarithmic resummation, already exists in the form of the \nllfast and \nnllfast codes~\cite{Kulesza:2008jb,Kulesza:2009kq,Beenakker:2009ha,Beenakker:2010nq,Beenakker:2011fu,Beenakker:2011sf,Beenakker:2013mva,Beenakker:2014sma,Beenakker:2015rna,Beenakker:2016gmf,Beenakker:2016lwe}, which add NLL and NNLL corrections to existing NLO results from \prospino. However, these results are restricted to interpolation in a two-dimensional subspace spanned by the gluino mass and a common squark mass, and results are given as a sum over outgoing squark flavours.

In addition to the processes found in \prospino, cross-sections for gaugino and slepton pair production, as well as gaugino-gluino production, can be evaluated at NLO+NLL precision by the \resummino code \cite{Bozzi:2006fw,Bozzi:2007qr,Bozzi:2007tea,Debove:2009ia,Debove:2010kf,Debove:2011xj,Fuks:2012qx,Fuks:2013vua,Fuks:2016vdc}. Here, the total running time for all chargino and neutralino production processes at NLO (again on a single 2.0\,GHz Intel Xeon-Gold 6138 Skylake core) is around three hours, with no evaluation of scale, $\alpha_s$, nor PDF errors. Combined NLO+NLL precision takes four days.

Recently, the \deepxs package \cite{Otten:2018kum} appeared, which performs a fast evaluation of neutralino and chargino pair production using neural networks. This is currently limited to a small set of the most phenomenologically relevant processes in the MSSM-19, and does not include uncertainties arising from PDFs, or the value of~$\alpha_s$.

The \smoking~\textsf{1.0} code is currently designed to reproduce all the NLO results of \prospino~\textsf{2.1} for strong production in the MSSM in a small fraction of the time, including scale errors, and with the addition of PDF and $\alpha_s$ errors based on a modern PDF set. It does not include NNLL or even NLL resummation as can be found in \nnllfast and \nllfast, however, unlike those codes it provides reliable results for non-degenerate squark masses, within the inherent limitations of the training set generated with \prospino. We intend to extend the code to also include NLL resummation in future releases.

Unlike \nnllfast, \smoking performs a separate evaluation of the cross-section for all distinct flavour combinations of the first two generations of squarks. The \smoking code also treats sbottom and stop pair-production processes separately, but this has limited impact as the two cross-sections are the same at LO (for the same masses),\footnote{Considering only QCD corrections, corrections that  depend on the mixing angles contribute only through $t\tilde t_i\tilde g$ and $\tilde t_1\tilde t_1\tilde t_2\tilde t_2$ vertices (and similarly for sbottoms)  at NLO and are typically small.} as \prospino\xspace --- and therefore our training set --- is limited to light quark initial states. A complete list of available processes can be found in Table~\ref{tab:processes}.

The code is written in \python with heavy use of the \numpy~\cite{NumPy} package for numerical calculations, and is compatible with both \python \textsf{2} and \textsf{3}. It can be installed using the \pip package manager and run as a set of functions from a self-contained module. We also provide interfaces through SLHA files~\cite{Skands:2003cj} and a command-line tool.

The rest of this paper is structured as follows.
We begin in Sec.~\ref{sec:regression_framework} with an introduction to the GP regression framework that we employ. In Sec.~\ref{sec:training} we describe our GP training regime, including how we compute the required training sets of NLO cross-sections. In Sec.~\ref{sec:validation}, we perform a thorough validation of the results from \smoking, with a comparison to results from existing codes. Section~\ref{sec:structure} covers the structure of the code and its interfaces, and we conclude in Sec.~\ref{sec:conclusion}.

\section{Regression framework}
\label{sec:regression_framework}

\subsection{Gaussian process regression}
\label{sec:GP}

The basic objective in regression is to estimate an unknown function value $f(\vec x)$ at some new $\vec x$ point, given that we know the function values at some other $\vec x$ points.
A Bayesian approach to this task is provided by Gaussian process regression, in which we express our degree of belief about any set of function values as a joint Gaussian pdf.
We underline here that this pdf should be understood in a purely Bayesian sense --- it does not imply any randomness in the true function we are approximating.

We begin by defining some notation and terminology, indicating in italics the terminology commonly used in the GP and machine learning literature.
Each input point $\vec x$ has $m$ components (\textit{features}), which in our case will correspond to masses and mixing angles from the MSSM squark and gluino sector.
We denote by $\vec x_*$ the new input point (\textit{test point}) for which we will estimate the unknown, true function value $f_* \equiv f(\vec x_*)$, here an NLO production cross-section.
We let $\vec x_i$ with $i=1,\ldots,n$ denote the $n$ input points (\textit{training points}) at which we know the function values $f_i \equiv f(\vec x_i)$ (\textit{targets}).
The combined set $\mcD = \{\vec x_i, f_i\}_{i=1}^{n}$ is referred to as the \textit{training set}.
The complete set of input components in our training set can be expressed as an $n \times m$ matrix $X = [\vec x_1, \ldots, \vec x_n]^\mathrm{T}$.
Similarly, the complete set of known function values can be collected in a vector $\vec f = [f_1, \ldots, f_n]^\mathrm{T}$. Thus, our training set can also be expressed as $\mcD = \{X, \vec f\}$.

The starting point for GP regression is the formulation of a joint Gaussian prior pdf\footnote{We always treat the input $\vec x$ points in a dataset $\mcD$ as known. Thus, the pdf $p(\mcD)$ should be understood as the pdf $p(f_1,\ldots,f_n | \vec x_1,\ldots, \vec x_n) = p(\vec f | X)$, and similarly for other pdfs involving $\mcD$.}
\begin{align}
  p(\mcD, f_* \big| \vec x_*) \equiv p(\vec f, f_* \big| X, \vec x_*),
\end{align}
which formally describes our degree of belief for possible function values at both the training points $X$ and the test point $\vec x_*$, before we look at the training data. This prior is chosen indirectly by choosing a \textit{mean function} $m(\cdot)$ and a \textit{covariance function} or \textit{kernel} $k(\cdot,\cdot)$, defined to specify the following expectation values for arbitrary input points:
\begin{align}
  & m(\vec x) = \mathbb{E}[f(\vec x)] \equiv \int_{-\infty}^\infty f p(f|\vec x) \, \mathrm{d} f \label{eq:m_f}\\
  & k(\vec x, \vec x') = \mathbb{E}[(f(\vec x)-m(\vec x)) (f(\vec x') - m(\vec x'))] \label{eq:k_f}\,.
\end{align}
We note that while the mean function and kernel are defined as functions of inputs in $\vec x$ space, the function values represent mean and covariance values in $f$ space. Our joint Gaussian prior can then be expressed as
\begin{align}
  \label{eq:GP_prior}
  p\Bigg(
  \begin{bmatrix}
    \vec f\\
    f_*
  \end{bmatrix}
  \Bigg|
  \begin{bmatrix}
    X\\
    \vec x_*
  \end{bmatrix}
  \Bigg)
  &= \nonumber\\
  &\mathcal{N} \Bigg(
    \begin{bmatrix}
      m(X)\\
      m(\vec x_*)
    \end{bmatrix},
  \begin{bmatrix}
  \Sigma & k(X, \vec x_*)\\
  k(\vec x_*, X) & k(\vec x_*, \vec x_*)
  \end{bmatrix}
   \Bigg),
\end{align}
where
\begin{align}
m(X)          &\equiv [m(\vec x_i),\ldots,m(\vec x_n)]^\mathrm{T}\\
k(\vec x_*,X) &\equiv [k(\vec x_*,\vec x_i),\ldots,k(\vec x_*,\vec x_n)]\\
k(X,\vec x_*) &\equiv k(\vec x_*,X)^\mathrm{T}\\
\Sigma        & \equiv k(X, X)\mathrm{,\ i.e.\ }\Sigma_{ij} = k(\vec x_i, \vec x_j) \label{eq:covar_matrix_def}.
\end{align}
The choice and optimisation of the kernel and mean function constitute the main challenge in GP regression, and we will discuss these aspects in detail in the next sections.

Our goal is to obtain a predictive posterior pdf for the unknown function value $f_*$ at $\vec x_*$. From the fully specified GP prior we can now find this simply by ``looking at'' the training data $\vec f$, \ie by deriving from the GP prior $p(\mcD, f_* | \vec x_*)$ the conditional pdf
\begin{align}
  \label{eq:GP_predicted_f}
  p(f_* \big| \mcD, \vec x_*) = \mathcal{N} (\mu_*, \sigma_*^2).
\end{align}
The mean and variance of this univariate Gaussian can be expressed in closed form as
\begin{align}
  \label{eq:GP_mean_predicted_f}
  &\mu_*=m(\vec x_*) + k(\vec x_*, X)\Sigma^{-1} (\vec f - m(X)) \\
  \label{eq:GP_variance_predicted_f}
  &\sigma_*^2=k(\vec x_*, \vec x_*)- k(\vec x_*, X)\Sigma^{-1}k(X, \vec x_*).
\end{align}
The prediction $\mu_*$ for $f_*$ is thus simply the prior mean $m(\vec x_*)$ plus a shift given by a weighted sum of the shifts of the known function values from their corresponding prior means, $\vec f - m(X)$. The weights are proportional to the covariances between the prediction at $\vec x_*$ and the known function values at the training points $X$, as set by the kernel $k(\vec x_*, X)$. The prediction variance $\sigma_*^2$ is given as the prior variance $k(\vec x_*, \vec x_*)$ reduced by a term representing the additional information provided by the training data about the function value at $\vec x_*$. This naturally depends only on the kernel. We will refer to the width $\sigma_*$ simply as the regression error or GP prediction error, keeping in mind that it should be interpreted in a Bayesian manner.

\subsection{Kernel choice and optimisation}
\label{sec:kernel-choice}

Choosing the kernel, Eq.\ (\ref{eq:k_f}), is the main modelling step in GP regression. It effectively determines what types of functional structure the GP will be able to capture.
In particular, it encodes the smoothness and the periodicity (if applicable) of the function that is being modelled, as it controls the expected correlation between function values at two different points. The choice of prior mean function, Eq.\ (\ref{eq:m_f}), is typically much less important, as we discuss at the end of this section.

The question of the optimal kernel choice is covered in more detail in Refs.~\cite{Rasmussen:2006:GPM:1162254,duvenaud2014}.
The squared-exponential kernel
\begin{align}
  \label{eq:squared-exp}
  k\left(\vec{x}, \vec{x}'; \sigma_f^2, \ell\right) = \sigma_f^2 \exp\left(-(\vec{x} - \vec{x}')^2/(2\ell^2)\right),
\end{align}
is the standard choice. It results in an exponentially decreasing correlation as the Euclidean distance between two input points increases with respect to a length-scale hyperparameter $\ell$. The signal variance $\sigma_f^2$ is a hyperparameter containing information about the amplitude of the modelled function. This is a universal kernel \cite{micchelli2006universal}, which means that it is in principle capable of approximating any continuous function given enough data. The infinite differentiability and exponential behaviour of this kernel typically result in a very smooth posterior mean.

However, for our purposes, the squared-exponential has some problems. Its sensitivity to changes in the function means that the length scale $\ell$ is usually determined by the smallest `wiggle' in the function~\cite{duvenaud2014}.
We hence consider also the Mat{\'e}rn kernel family:
like the squared-exponential, these are universal and stationary, \ie, only functions of the relative positions of the two input points,
but additionally incorporate a smoothness hyperparameter $\nu$ following the basic form
\begin{align}
  \label{eq:matern}
  k_\mathrm{M} (\vec{x}, &\vec{x}'; \nu, \ell) \nonumber \\
    &= \frac{2^{1-\nu}}{\Gamma (\nu) } \Big( \sqrt{2 \nu}\frac{|\vec{x} - \vec{x}'|}{\ell} \Big)^{\nu} K_{\nu} \Big( \sqrt{2 \nu}\frac{|\vec{x} - \vec{x}'|}{\ell} \Big) \, ,
\end{align}
where $\Gamma (\nu)$ is the gamma function and $K_{\nu}$ is a modified Bessel function of the second kind. For the modelling of cross-section functions, we adopt the Matérn kernel class on the basis of its superior performance.  This has followed significant testing and cross-validation across a number of different problems \cite{Ingrid, Iza, Nick}. During testing we found $\nu=\frac{3}{2}$ to be optimal for our purposes, in which case Eq.\ (\ref{eq:matern}) simplifies to
\begin{align}
  k_\mathrm{M} (\vec{x}, &\vec{x}'; \nu=\tfrac{3}{2}, \ell) \nonumber \\
    & = \Big(1 + \sqrt{3}\frac{|\vec{x} - \vec{x}'|}{\ell}\Big) \exp\Big(-\sqrt{3}\frac{|\vec{x} - \vec{x}'|}{\ell}\Big).
\end{align}

To account for the fact that some directions in the input space of masses and mixing angles may have more impact on the cross-section values than others, we use an anisotropic, multiplicative Mat{\'e}rn kernel,
\begin{align}
  k_\mathrm{AMM} (\vec x, \vec x'; \nu, \sigma_f^2, \vecell) = \sigma_f^2 \prod_{d=1}^m k_M (x^{(d)}, x'^{(d)}; \nu, \ell_d),
\end{align}
where we have also included a signal variance hyperparameter $\sigma_f^2$, similar to the one in Eq.\ (\ref{eq:squared-exp}). Here $x^{(d)}$ denotes the $d$th component of the input vector $\vec x$, and $\vecell$, with components $\ell_d$, is a vector containing one length scale per $\vec x$ component.
The product over the dimensions of the parameter space results in points only being strongly correlated if in each dimension, their distance is small with respect to the relevant length scale.

So far we have focused on the ``noise-free'' case, in which the training targets $\vec f$ are the exact values of the true function at the training points. In this case the predictive posterior $p(f_* | \mcD, \vec x_*)$ collapses to a delta function when $\vec x_*$ equals a training point.
In theory this is a reasonable approach, since what we seek is a surrogate model for an expensive, but precise and deterministic numerical computation. In practice, however, allowing for some uncertainty also at the training points typically results in a more well-behaved and stable regression model. The main reason for this is that the additional wiggle-room in the modelling can ease the challenging matrix numerics of GP regression, as we will discuss in some detail in Sect.\ \ref{sec:regularisation}.
We therefore add a ``white-noise'' term,
\begin{align}
  \label{eq:white-kernel}
  k_\mathrm{WN} (\vec x, \vec x'; \sigma_{\epsilon}^2) = \delta_{\vec x \vec x'} \sigma_{\epsilon}^2 ,
\end{align}
to our kernel, where $\sigma_{\epsilon}^2$ is the hyperparameter that sets the amount of ``noise''. The effect of this term is simply to add $\sigma_{\epsilon}^2$ along the diagonal of the covariance matrix $\Sigma$, as well as to the prior variance at the test point, $k(\vec x_*, \vec x_*)$. It is known as \textit{homoscedastic} noise, as it is the same for all data points.

In GP terminology, to include this additional variance term corresponds to going from the noise-free case to a scenario with noisy training data. The targets are then considered measurements $y_i \equiv y(\vec x_i) = f(\vec x_i) + \epsilon_i$, with the noise $\epsilon_i$, introduced in the process of performing the $i$th measurement, modelled by a Gaussian distribution $\mathcal{N}(0, \sigma_\epsilon^2)$. However, we remind the reader that for our case this Gaussian pdf represents an adopted effective Bayesian degree of belief in the accuracy of the training data, rather than an expression of actual random noise.

Conceptually we should then make the substitution $f \rightarrow y$ in our definitions from Sect.\ \ref{sec:GP}. Our training set becomes $\mcD = \{X, \vec y\}$, with $\vec y = [y_1, \ldots, y_n]^\mathrm{T}$, and the GP prior becomes a joint pdf for $\vec y$ and $y_*$:
\begin{align}
  p(\mcD, y_* \big| \vec x_*) = p(\vec y, y_* \big| X, \vec x_*).
\end{align}
The prior mean function and kernel now specify expectation values in $y$ space,
\begin{align}
  & m(\vec x) = \mathbb{E}[y(\vec x)] \label{eq:m}\\
  & k(\vec x, \vec x') = \mathbb{E}[(y(\vec x)-m(\vec x)) (y(\vec x') - m(\vec x'))] \label{eq:k}\,,
\end{align}
where we note that $\mathbb{E}[y(\vec x)] = \mathbb{E}[f(\vec x)]$ since the Gaussian noise term has zero mean. Likewise, the predictive posterior pdf becomes
\begin{align}
  \label{eq:GP_predicted}
  p(y_* \big| \mcD, \vec x_*) = \mathcal{N} (\mu_*, \sigma_*^2),
\end{align}
with mean and variance
\begin{align}
  \label{eq:GP_mean_predicted}
  &\mu_*=m(\vec x_*) + k(\vec x_*, X)\Sigma^{-1} (\vec y - m(X)) \\
  \label{eq:GP_variance_predicted}
  &\sigma_*^2=k(\vec x_*, \vec x_*)- k(\vec x_*, X)\Sigma^{-1}k(X, \vec x_*).
\end{align}

\noindent Our complete kernel is then given by
\begin{align}
  \label{eq:full_kernel}
  k(\vec x, \vec x') = k_\mathrm{AMM}(\vec x, \vec x'; \nu, \sigma_f^2, \vecell)
    + k_\mathrm{WN}(\vec x, \vec x'; \sigma_{\epsilon}^2).
\end{align}

Fixing $\nu = \frac{3}{2}$, as discussed above, we are left with the set $\vec\theta=\{\sigma_f^2,\vecell,\sigma_\epsilon^2\}$ of undetermined hyperparameters. To be fully Bayesian, one would introduce a prior pdf $p(\vec\theta)$ for the hyperparameters and obtain the GP posterior $p(y_* | \mcD, x_*)$ by marginalising over $\vec\theta$,
\begin{align}
  p(y_* | \mcD, x_*) &= \int p(y_*, \vec\theta | \mcD, x_*) \,\mathrm{d}\vec\theta \nonumber\\
    &= \int p(y_* | \vec\theta, \mcD, x_*) p(\vec\theta | \mcD)\,\mathrm{d}\vec\theta \nonumber\\
    &\propto \int p(y_* | \vec\theta, \mcD, x_*) p(\mcD | \vec\theta) p(\vec\theta)\,\mathrm{d}\vec\theta.
\end{align}
In our high-dimensional case with large datasets, such integration would come at a steep computational expense, even with MCMC methods. We therefore follow the common approach of using a point estimate for the hyperparameters, found by maximising the log-likelihood function \cite{Rasmussen:2006:GPM:1162254}
\begin{eqnarray}
  \label{eq:log-likelihood}
  \log p(\mcD | \vec\theta) &=&\log p(\vec y | X, \vec\theta) \nonumber\\
    &=&-\frac{1}{2}{(\mathbf y - m(X))}^\mathrm{T}\Sigma^{-1}{(\mathbf y - m(X))}\nonumber\\
    &&-\frac{1}{2}\log|\Sigma| -\frac{n}{2}\log 2\pi.
\end{eqnarray}

Finding an adequate set of hyperparameters constitutes the model training step in the GP approach. It is complicated by the fact that each optimisation step requires the computation of the inverse and determinant of the $n\times n$ covariance matrix $\Sigma$, which scales poorly with the number of training points $n$. To increase speed and numerical stability, $\Sigma$ is generally not directly inverted in practice, and its Cholesky decomposition is used instead. In an attempt to avoid local optima, we employ the \scipy implementation of the differential evolution method~\cite{StornPrice95,SciPy}, rather than performing a gradient-based search.

Recent work has demonstrated that the theoretical prediction error $\sigma_*^2$ in Eq.~(\ref{eq:GP_variance_predicted}) systematically underestimates the mean-squared prediction error when the hyperparameters are learned from the data \cite{pmlr-v54-wagberg17a}. As proposed there, we account for the uncertainty on the point estimate of the hyperparameter by adding a correction term to $\sigma_*^2$, derived from the Hybrid Cram{\'e}r-Rao Bound. In our case, with a constant prior mean function, this extra term amounts to
\begin{align}
  \Delta\sigma_*^2 = \left(1-\vec 1\Sigma^{-1}k(X, \vec x_*)\right)^2 \left/ \sum_{i,j=1}^n [\Sigma^{-1}]_{ij}\right.,
\label{eq:add_variance_from_hyperpar_uncertainty}
\end{align}
where $\vec 1 \equiv [1,\ldots,1]$. In particular, this increases the prediction error at test points far from the training data.

Compared to the choice of kernel, the choice of the prior mean function, Eq.\ (\ref{eq:m}), is typically less important. Following conditioning on a sufficiently large training set, the prior gets overpowered and the posterior mean is primarily influenced by the training data through the second term in Eq.\ (\ref{eq:GP_mean_predicted}). For this reason the prior mean function is commonly taken to be zero everywhere. Nevertheless, it is sensible to incorporate our knowledge of the mean, and we therefore use the sample mean of the target values $\vec y$ as a prior mean function that is constant in $\vec x$.

\subsection{Regularisation of the covariance matrix}
\label{sec:regularisation}

A practical challenge when training GPs is to ensure numerical stability when inverting the covariance matrix $\Sigma$. The precision of the result is controlled by the condition number $\kappa$ of $\Sigma$, which can be considered a measure of the sensitivity of the inversion to roundoff error. It is computed as the ratio $\lambda_\mathrm{max}/\lambda_\mathrm{min}$ between the highest and lowest eigenvalues of $\Sigma$, and becomes infinite for a singular matrix. The loss of numerical precision at high $\kappa$ becomes most obvious when the predictive variance, computed according to Eq.~(\ref{eq:GP_variance_predicted}), evaluates to a negative number.
In order to prevent this problem, it is essential to understand how to control $\kappa$.

When the target values of training points are strongly correlated, their corresponding rows and columns in $\Sigma$ are nearly identical. This leads to eigenvalues close to zero and a very large condition number. It has been shown that in the worst case, $\kappa$ can grow linearly with the number of training points and quadratically with the signal-to-noise ratio $\mathrm{SNR} = \sigma_f/\sigma_\epsilon$~\cite{PracticalGuideGPs}.

Increasing the noise level improves numerical stability, as a larger diagonal contribution $\sigma_\epsilon^2$ to $\Sigma$ enhances the difference between otherwise similar rows and columns.
Therefore, we add a term to the log likelihood in Eq.~(\ref{eq:log-likelihood}) that penalises hyperparameter choices with extremely high signal-to-noise ratios, as suggested in Ref.~\cite{PracticalGuideGPs}. Our objective function for training GPs then becomes
\begin{align}
  \log p({\mathbf y} | X, \vec\theta) - \left(\frac{\log(\sigma_f/\sigma_\epsilon)}{\log(\mathrm{SNR}_\mathrm{max})}\right)^{50}.
\label{eq:likelihood_penalty}
\end{align}\footnotetext[50]{This is a power, not a footnote.}%
The large exponent guarantees that situations where $\mathrm{SNR} > \mathrm{SNR}_\mathrm{max}$ are the only ones where the penalty term has a significant effect. We use $\mathrm{SNR}_\mathrm{max}=10^4$.

In some cases the likelihood penalty in Eq.~(\ref{eq:likelihood_penalty}) does not decrease the condition number sufficiently to stabilise the inversion. However, choosing a lower overall value for $\mathrm{SNR}_\mathrm{max}$ dilutes the information in the training data to an extent that it can sometimes be fitted by noise, even when unnecessary. We therefore check the condition number after the optimisation with the penalty term, and proceed to increase the homoscedastic noise $\sigma_\epsilon^2$ just for the inversion step until the condition number drops below a reasonable value $\kappa_\mathrm{max}$~\cite{mohammadi2016analytic}:
\begin{align}
  \Delta \sigma_\epsilon^2 = \frac{\lambda_\mathrm{max} - \kappa_{\mathrm{max}} \lambda_{\mathrm{min}}}{\kappa_{\mathrm{max}} - 1}.
\end{align}
We set $\kappa_\mathrm{max} = 10^9$, roughly corresponding to a maximal loss of nine digits accuracy from the total of 16 in a 64-bit double-precision floating-point number.

These measures may seem to deteriorate the performance of our regression model, but they are necessary to ensure the numerical stability. The underlying reason is that we have essentially noiseless data, and are hitting the limits of floating-point precision in the process of calculating the GP predictions. In comparison to the scale and PDF uncertainties on the cross-sections, the resulting regression errors nevertheless remain small, as we demonstrate in Sec.~\ref{sec:validation}.

\subsection{Distributed Gaussian processes and prediction aggregation}
\label{sec:DGP}

With $n$ training points, the complexity of the matrix inversion operations in Eqs.~(\ref{eq:GP_mean_predicted}) and (\ref{eq:GP_variance_predicted}) scales as $n^3$, making standard GP regression unsuitable for problems that require large training sets. To overcome this challenge we construct a regression model based on \textit{distributed Gaussian processes} (DGPs) \cite{2015arXiv150202843D}: We partition the total training set $\mcD$ into $d$ manageable subsets $\mcD_i$, and for each $\mcD_i$ we train a new GP $\mcM_i$. These GPs are referred to as \textit{experts}. The prediction from our regression model is obtained by aggregating the predictions from the individual experts. For this prediction aggregation we follow the approach know as the \textit{Generalised Robust Bayesian Committee Machine} (GRBCM) \cite{2018arXiv180600720L}, for which we summarise the main steps below.

First we construct a data subset $\mcD_1 \equiv \mcD_c$, randomly chosen from $\mcD$ without replacement, which will be used to train a single \textit{communication expert} $\mcM_c$. Next, we partition the remaining data into subsets $\{\mcD_i\}_{i=2}^d$, each of which will serve to train one expert $\mcM_i$. Following Refs.~\cite{2015arXiv150202843D,2018arXiv180600720L}, all experts are then trained simultaneously, such that they share a common set of hyperparameters.

The GRBCM approach places no restrictions on how to partition the data to form the subsets $\{\mcD_i\}_{i=2}^d$. However, empirical studies have shown that some clustering of the data can help the experts to become sensitive to local, short-scale variability of the target function~\cite{2016arXiv160705432R,2018arXiv180600720L}.
Compared to using a simple random partition, we have noticed minor improvements with a disjoint partition, where the data is split into local subsets based on the mass parameter with the smallest length-scale hyperparameter. Tests with $k$-means clustering did not indicate further improvements in our case, nor did tests with sorting on less dominant features.

The special role of the communication expert $\mcM_c$ becomes evident at the prediction stage. For each of the experts $\{\mcM_i\}_{i=2}^d$, we construct an improved expert $\mcM_{+i}$ by replacing the corresponding dataset $\mcD_i$ with the extended set $\mcD_{+i} = \{\mcD_i, \mcD_c\}$. That is, for prediction the communication dataset $\mcD_{c}$ is shared by all the experts $\mcM_{+i}$. The communication expert $\mcM_c$ serves as a common baseline to which the experts $\mcM_{+i}$ can be compared. In the final combination, the prediction from expert $\mcM_{+i}$ will be weighted according to the differential entropy difference between its predictive distribution and that of $\mcM_{c}$.

The central approximation that allows for computational gains in DGPs and related approaches is an assumption that the individual experts can be treated as independent, which corresponds to approximating the kernel matrix of the combined problem, i.e. without partition into experts, as block-diagonal. In the GRBCM approach, this approximation is expressed as the conditional independence assumption $\mcD_i \perp \mcD_j | \mcD_c, y_*, \vec x_*$ for $2 \leq i \neq j \leq d$, which enables the approximation $p(\mcD_i|\mcD_j,\mcD_c,y_*, \vec x_*) \approx p(\mcD_i|\mcD_c, y_*, \vec x_*)$. That is, when the information contained in the communication set $\mcD_c$ is known, we assume that the predictive distribution for points in subset $\mcD_i$ should not be strongly influenced by the additional information contained in subset $\mcD_j$.

Using Bayes' theorem and the above independence assumption, the exact predictive distribution $p(y_* | \mcD, \vec x_*)$ can now be approximated as
\begin{align}
\label{eq:GBRCM_eq12}
p(& y_* | \mcD, \vec x_*) \nonumber \\
  &\propto\, p(y_* | \vec x_*) p(\mcD_c | y_*,\vec x_*) \prod\limits_{i=2}^d p(\mcD_i | \mcD_1,\ldots\mcD_{i-1},y_*,\vec x_*) \nonumber \\
  &\approx\, p(y_* | \vec x_*) p(\mcD_c | y_*,\vec x_*) \prod\limits_{i=2}^d p^{\beta_i}(\mcD_i | \mcD_c,y_*,\vec x_*) \nonumber \\
  &=\, p(y_* | \vec x_*) p(\mcD_c | y_*,\vec x_*) \prod\limits_{i=2}^d \frac{p^{\beta_i}(\mcD_i, \mcD_c | y_*,\vec x_*)}{p^{\beta_i}(\mcD_c | y_*,\vec x_*)} \nonumber \\
  &=\, \frac{p(y_* | \vec x_*) \prod\limits_{i=2}^d p^{\beta_i}(\mcD_{+i} | y_*,\vec x_*)}{p^{\beta_1}(\mcD_c | y_*,\vec x_*)},
\end{align}
where we have introduced the weights $\beta_i$ for the predictions from different experts, and defined $\beta_1 \equiv -1 + \sum_{i=2}^d \beta_i$.  By applying Bayes' theorem again, we can express our approximation for $p(y_* | \mcD, \vec x_*)$ in terms of the corresponding predictive distributions from the individual experts, $p_{+i} (y_* | \mcD_{+i}, \vec x_*)$ and $p_c (y_* | \mcD_c, \vec x_*)$. Leaving out normalisation factors, the distribution for the aggregated prediction becomes
\begin{equation}
\label{eq:GBRCM_eq13}
  p_A(y_* | \mcD, \vec x_*) \propto \frac{\prod\limits_{i=2}^d p_{+i}^{\beta_i}(y_* | \mcD_{+i}, \vec x_*)}{p_c^{\beta_1}(y_* | \mcD_c, \vec x_*)},
\end{equation}
with mean $\mu_\mathrm{DGP}$ and variance $\sigma_\mathrm{DGP}^2$ at $\vec x_*$ given by
\begin{eqnarray}
\label{eq:GBRCM_eq14b}
\sigma_\mathrm{DGP}^{-2}(\vec x_*) &=& - \beta_1 \sigma_c^{-2}(\vec x_*) + \sum\limits_{i=2}^d \beta_i \sigma_{+i}^{-2}(\vec x_*)\\
\frac{\mu_\mathrm{DGP}(\vec x_*)}{\sigma_\mathrm{DGP}^2(\vec x_*)} &=& - \beta_1 \sigma_c^{-2}(\vec x_*) \mu_c(\vec x_*) \nonumber\\
&&+ \sum\limits_{i=2}^d \beta_i \sigma_{+i}^{-2}(\vec x_*) \mu_{+i}(\vec x_*).
\end{eqnarray}%
Following Ref.~\cite{2018arXiv180600720L}, we set the weights $\beta_i$ to
\begin{align}
\label{eq:GBRCM_eq5}
  &\beta_2 = 1, \\
  &\beta_{i\geq3} = 0.5 \left[\log\sigma_c^2(\vec x_*) - \log\sigma_{+i}^2(\vec x_*)\right].
\end{align}
The reason for assigning weight $\beta_2 = 1$ for expert $\mcM_{+2}$ is that the transition
\begin{align}
  p(\mcD_i | \mcD_1,\ldots\mcD_{i-1},y_*,\vec x_*) \rightarrow p^{\beta_i}(\mcD_i | \mcD_c,y_*,\vec x_*)
\end{align}
in Eq.~(\ref{eq:GBRCM_eq12}) is exact for $i=2$, $\beta_2=1$. For each remain\-ing expert $\mcM_{+i\geq3}$, the weight is taken to be the difference in differential entropy between the baseline predictive distribution of the communication expert, $p_c(y_* | \mcD_c, \vec x_*)$, and that of the given expert, $p_{+i}(y_* | \mcD_{+i}, \vec x_*)$. Thus, if an expert $\mcM_{+i}$ provides little additional predictive power over $\mcM_{c}$, its relative influence on the aggregated prediction is low.

Requiring the experts to share a common set of hyper\-parameters effectively disfavours overfitting of individual experts. Moreover, the risk of overfitting is alleviated by the fact that after training, each expert is extended with the communication dataset $\mcD_c$ that it did not see during training, and its weight in the prediction aggregation is regularised through the comparison to the communication expert.

The GRBCM split of the dataset into $d$ experts reduces the complexity of training from $n^3$ to $\mathcal{O}(d (n/d)^3 = n^3d^{-2})$. The memory, storage space, and evaluation all depend directly on the size of the matrix, and scale as $\mathcal{O}(n^2)$ for a regular GP, but as $\mathcal{O}(n^2/d)$ in the GRBCM approach.

\section{Training}
\label{sec:training}

\subsection{Sample generation}
\label{sec:sample generation}
We generate the inputs for our training data calculations by sampling the physical gluino and squark masses and (for third-generation squarks) the angles describing mass mixing between gauge eigenstates. The only other parameters involved in the production of gluinos and squarks to NLO QCD are the strong coupling $\alpha_s$ and the SM quark masses.
The cross-sections depend on $\alpha_s$ both through the matrix element and the PDF. To capture the cross-section variation due to the uncertainty on $\alpha_s$, we generate separate input points with $\alpha_s$ set to 0.1180 (central value), 0.1165 ($1\sigma$ lower value) and 0.1195 ($1\sigma$ upper value), using the corresponding PDF sets, and train separate GPs on the ratio of cross-sections obtained with the central and $\pm1\sigma$ values. For the SM masses we use a fixed value for the bottom and top quark masses, and assume the other quark masses to be zero.

In the sample generation we do not simply sample over a regular grid of parameter values, for three reasons:
\begin{enumerate}
\item Grid sampling is inefficient when one parameter is more important than the others: too many evaluations are spent on varying the less influential parameters whilst keeping the important one at a fixed value.
\item The curse of dimensionality renders this sampling technique infeasible for processes that depend on more than three or four parameters.
\item The complexity of sampling and cross-section calculation for the large number of processes that we consider (in terms of final-state squark flavours) means that it is more efficient to evaluate multiple cross-sections for every parameter combination than to generate separate samples for each final state.
\end{enumerate}

We sample individual baseline masses for the gluino, $m_{\tilde g}$, for the first and second-generation (gauge eigenstate) squarks, $m_{\tilde u_L}$, $m_{\tilde d_L}$, $m_{\tilde c_L}$, $m_{\tilde s_L}$, $m_{\tilde u_R}$, $m_{\tilde d_R}$, $m_{\tilde c_R}$, and $m_{\tilde s_R}$, and for the third-generation (mass eigenstate) squarks $m_{\tilde b_1}$, $m_{\tilde t_1}$, $m_{\tilde b_2}$, and $m_{\tilde t_2}$. We do this in two different ways: either drawing from a uniform distribution on the interval $[50, 3500]$\,\GeV, or from a hybrid distribution uniform on the interval $[50, 150]$\,\GeV and logarithmic on the interval $[150, 3500]$\,\GeV. We order the third-generation squarks in mass after sampling, so that ${\tilde t_1}$ and ${\tilde b_1}$ are by definition the lightest. We intentionally choose these sampling ranges to be slightly beyond our final claimed region of validity for the regression --- typically 200--3000\,\GeV\xspace--- in order to try to avoid large regression errors at the edges.  The regions where our regression has been validated more than cover the ranges of masses of interest for the LHC. We sample the cosines of the sbottom and stop mixing angles, $\cos\theta_{\tilde b}$ and $\cos\theta_{\tilde t}$, uniformly on the interval $[-1, 1]$.

On top of our baseline sampling, we employ further sub-sampling of the particle masses in order to properly include cross-section resonances within our training set.  This ensures that our training dataset is more densely sampled where the cross-sections of interest vary the most.  To do this, we generate and then combine six different training sets:%
\begin{itemize}%
\item[i)] All masses sampled from the uniform prior.
\item[ii)] All masses sampled from the hybrid prior.
\item[iii)] Employing the uniform prior for both the mass of the gluino and for a common squark mass scale, and taking a Gaussian prior with width 50\,\GeV for the difference between the common squark mass scale and the masses of the individual squarks.
\item[iv)] Employing the uniform prior for the mass of the gluino, the hybrid prior for a common squark mass scale, and the same Gaussian prior as in iii) to draw values for the squark masses around the common scale.
\item[v)] Employing the uniform prior for a common mass scale, and using the same Gaussian prior as in iii) to draw a gluino mass and the individual squark masses around the common scale.
\item[vi)] Varying the gluino mass and a common squark mass independently on a two-dimensional grid spanning from 60 to 3000\,\GeV,  with steps of 60\,\GeV.\footnote{This set was only used to improve the regression results for the stop and sbottom pair production cross-sections.}
\end{itemize}%

By joining these samples, we are able to achieve both good sampling of low masses (where cross-sections are large) via the logarithmic mass prior, and acceptable sampling of large masses via the uniform prior.

Because $R$-parity is assumed in the MSSM, there are no new $s$-channel resonances in the mass range where we train our GPs, and SM resonances are too light to have any impact. Therefore, we do not have to worry directly about sampling densely in the region of resonances. However, at LO there are potential effects for gluino pair production near threshold from destructive interference between diagrams with $s$-channel gluons and those with $t$-channel squarks.
For squark production, the chirality of the final-state squarks affects the contribution from $t$-channel gluino exchange, so that for example in squark--squark production for equal chiralities the matrix element has a zero at $m_{\tilde g}\to0$, and a maximum around $m_{\tilde g}\simeq\bar m_{\tilde q}$, where $\bar m_{\tilde q}$ is the average mass of the first and second-generation squarks. Both these effects can lead to non-monotonous behaviour of the cross-section as a function of masses (bumps and dips).
In Sec.~\ref{sec:validation} we shall see this particularly for gluino pair production. This necessitates the separate samples with (near) degenerate masses, where such effects are larger.

Even though all non-degenerate squark masses enter into the LO cross-sections, to the precision of our training sample (see Sec.~\ref{sec:NLO}), only two mass parameters enter into the NLO corrections to gluino and first/second-generation squark production: the gluino mass $m_{\tilde g}$ and the averaged first and second-generation squark mass $\bar m_{\tilde q}$.  Therefore, any additional interference structures present in the NLO corrections to these processes must be visible in the $(m_{\tilde g},\bar m_{\tilde q}) $ slice of the parameter space.  For stop/sbottom production, such structures should be similarly visible in the $(m_{\tilde g}, m_{\tilde t_1/\tilde b_1})$ slice.  As a result, we shall spend some time below in validation (Sec.~\ref{sec:validation}) looking at gluino and first/second-generation squark production cross-sections in terms of $m_{\tilde g}$ and $\bar m_{\tilde q}$, and third-generation squark production in terms of $m_{\tilde g}$ and $m_{\tilde t_1}$.

Table~\ref{tab:processes} gives the list of cross-sections available from \smoking, and their parameter dependencies. The reader may wonder at this point why we train our GPs on the total cross-section in terms of all the non-degenerate masses, rather than simply on the NLO corrections (in terms of just $m_{\tilde g}$ and $\bar m_{\tilde q}$ for gluino/first/second-generation squark production, supplemented with the three relevant third-generation parameters each for stop and sbottom production).  One reason is that we have designed \smoking to be more general than \prospino; in future releases we intend to make use of training data from other tools able to move beyond the degenerate-squark-mass approximation. The other reason is that we feel it is more convenient for a user to simply obtain full LO+NLO (and future +NLL+NNLL+\ldots) cross-sections from \smoking, rather than needing to install the correct LO cross-section calculator and PDF set, call them, and then combine the results.  Similarly, training on and returning the full cross-section will make simultaneously including cross-sections from different calculators (chosen e.g.\ as most appropriate for different theories or processes) far more straightforward.

\begin{table}
\caption{List of all available cross-sections in \smoking.  Explicit first and second-generation squarks $\tilde q_i$ can be $\tilde u_L,\tilde d_L,\tilde c_L,\tilde s_L,\tilde u_R,\tilde d_R,\tilde c_R$ or $\tilde s_R$, whilst $\bar m_{\tilde q}$ denotes the average over all eight of these masses. Where the charge-conjugate process is distinct from the original, \smoking returns the sum of the cross-section for the process and its conjugate (as indicated; for $q_i\tilde q_j^*$, the conjugate is only distinct when $i\ne j$).
\label{tab:processes}}
\centering
\begin{tabular}{ll}
\hline\noalign{\smallskip}
Final state       & Variables  \\
\noalign{\smallskip}\hline\noalign{\smallskip}
$\tilde g\tilde g$    & $m_{\tilde g}, \bar m_{\tilde q}, m_{\tilde u_L},m_{\tilde d_L},m_{\tilde c_L},m_{\tilde s_L}, $ \\ & $m_{\tilde u_R},m_{\tilde d_R},m_{\tilde c_R},m_{\tilde s_R}$  \\
$\tilde g\tilde q_i + \textrm{c.c.}$  & $m_{\tilde g}, \bar m_{\tilde q}, m_{\tilde q_i}$  \\
$\tilde q_i\tilde q_j + \textrm{c.c.}$  & $m_{\tilde g}, \bar m_{\tilde q}, m_{\tilde q_i}, m_{\tilde q_j}$ \\
$\tilde q_i\tilde q_j^* (+\, \textrm{c.c.})$ & $m_{\tilde g}, \bar m_{\tilde q}, m_{\tilde q_i}, m_{\tilde q_j}$ \\
$\tilde b_i\tilde b_i^*$  & $m_{\tilde g}, \bar m_{\tilde q}, m_{\tilde b_1}, m_{\tilde b_2}, \cos\theta_{\tilde b}$ \\
$\tilde t_i\tilde t_i^*$    & $m_{\tilde g}, \bar m_{\tilde q}, m_{\tilde t_1}, m_{\tilde t_2}, \cos\theta_{\tilde t}$  \\
\noalign{\smallskip}\hline
\end{tabular}
\end{table}

\subsection{Calculation of NLO training cross-sections}
\label{sec:NLO}

We use \prospino~\textsf{2.1} to generate cross-sections for our training samples. This calculates, amongst other things, NLO cross-sections for strong production processes in the MSSM for proton-proton collisions at a given centre-of-mass (CoM) energy, and for a choice of renormalisation/factorisation scales.  We have modified the code to set $\alpha_s$ accordingly, and to accept generic PDF sets from \lhapdf{6.2}~\cite{Buckley:2014ana}. For the current version of \smoking we have used the \textsf{PDF4LHC15\_nlo\_30\_pdfas} symmetric Hessian NLO PDF set with 30 eigenvector members and two members with varied strong coupling $\alpha_s(m_Z)=0.1180\pm0.0015$~\cite{Butterworth:2015oua}.

\prospino performs the PDF integral of the partonic process using the VEGAS \cite{1978JCoPh..27..192L} importance sampling algorithm for Monte Carlo integration. The convergence criterion leaves some numerical noise in the result, typically of the order of $10^{-3}$ relative to the central cross-section value.

In order to obtain $K$-factors for gluino and first\slash{}second-generation squark production, \prospino first calculates the LO and NLO cross-sections using a single squark mass, obtained as the average over the masses of all first and second-generation squarks, i.e.\ eight in total; this mass is employed even for any internal third-generation squark propagators.  The ratio of the LO and NLO results gives the $K$-factor for the process in question. \prospino then recomputes the cross-section at LO, without the assumption of an average mass, and the corresponding NLO value is found by multiplying this LO result by the $K$-factor calculated for the average squark mass.  The calculation proceeds similarly for stop and sbottom production, except that the third-generation squark masses are kept non-degenerate for all steps of the calculation, meaning that first and second-generation masses are still averaged in the $K$-factor calculation.  The final cross-sections should thus be viewed as an approximation to a fully non-degenerate squark mass NLO calculation. The effect of this assumption was investigated in Ref.~\cite{GoncalvesNetto:2012yt} and found to be relatively small in most parts of parameter space.

In total, \prospino allows 141 final states containing gluinos and different flavour squarks. However, due to charge conjugation relations and NLO QCD identities in the cross-section (relating certain combinations of left and right-handed final-state squarks under the exchange of masses), only 49 distinct final states are needed to represent all processes calculated by \prospino in our training sample. These relationships are handled internally in \smoking, and therefore need not directly concern the user. A list of the available cross-sections from the user's side can be found in Table~\ref{tab:processes}.

For every parameter point and every process in the training sample, we calculate 34 different cross-section values. These include: i) a central value, ii) 30 values using the PDF eigenvectors, iii) one value where $\alpha_s$ is taken to its lower value, iv) one where $\alpha_s$ is taken to its upper value,  v) a value where the renormalisation/factorisation scale is doubled, and, vi) a value where the renormalisation/factorisation scale is halved.

From the values in i) and ii) we follow the PDF4LHC guidelines to calculate the symmetric PDF uncertainty on the central value, understood to be a 68\% confidence level bound.\footnote{Due to the extreme calculational cost, we do not use the recommended Monte Carlo PDF sets with 100 replicas, but use the symmetric Hessian set with 30 eigenvector members.}
Similarly, we follow the PDF4LHC prescription to calculate the 68\% confidence level bound from varying $\alpha_s$ alone, using the results from iii) and iv). We do not add the PDF and $\alpha_s$ errors, but train different DGPs for the two errors. We note that these errors should be added in quadrature after evaluation if the user wishes to obtain a single 68\% confidence bound incorporating both effects.

Further, we follow the standard lore of estimating an uncertainty associated with missing higher-order corrections by calculating the spread in cross-section values under scale variation coming from v) and vi). In \smoking this is referred to as the scale uncertainty. Choosing exactly how to interpret this asymmetric uncertainty as a probability distribution (flat, Gaussian or otherwise), and whether/how to combine it with the (Gaussian) $\alpha_s$ and PDF uncertainties, is left to the user.

In total this leaves us with seven cross-sections: the central value and the two-sided PDF, $\alpha_s$ and scale uncertainties.


\subsection{Training implementation details}
\label{sec:training_implementation}

For each final state, we train one large DGP with multiple experts on the central cross-section value, and five smaller regular GPs: one each on the upper and lower values arising from regularisation and factorisation scale variation, one each on the cross-sections for the variation of $\alpha_s$ (and corresponding PDFs) to its upper and lower limit in its 68\% confidence interval (these values are later symmetrised), and a single GP on the symmetric 68\% confidence level PDF variation.

To improve the training, we try to simplify the target model. The span of some ten orders of magnitude in cross-sections across the sampled parameter space means that it is numerically challenging to train the DGPs on the raw cross-section numbers. Before training, we scale out part of the dependence on the final-state masses by first multiplying the central input cross-section by the square of the (average) final-state mass, and then take the logarithm of the result. For the remaining five `error' values, we train on the logarithm of the ratio to the central cross-section, such that after the reverse transformation, the final predictions for the ratios are strictly positive. As noted in section~\ref{sec:kernel-choice}, we choose a constant prior mean equal to the sample average of the transformed target values, which results in GPs effectively modelling the deviation from this mean value. We also rescale all input parameters to the interval $[0,1]$ before training, to improve numerical stability and precision.

The transformations during training are automatically recorded and reversed at the time of evaluation. As the GPs are trained on the logarithm of the (mass-rescaled) cross-section, the resulting predictive distribution for the absolute cross-section is a log-normal distribution. Due to the positive skew of this distribution, we base the central cross-section estimate on the median. Specifically, let
$\mathcal N(\mu_\text{DGP}(\vec x_*), \sigma^2_\text{DGP}(\vec x_*))$ denote the aggregated DGP predictive distribution for the log-transformed cross-section at point $\vec x_*$, after accounting for the constant prior shift and the mass scaling applied to the training data. The central cross-section value $y_\smoking$ returned by \smoking is then
\begin{align}
  y_\smoking(\vec x_*) = \exp[\mu_\text{DGP}(\vec x_*)],
\label{eq:xsec_return_value_definition}
\end{align}
with associated asymmetric regression uncertainties $\Delta_\smoking^-$ and $\Delta_\smoking^+$. These are constructed from the bounds of the $1\sigma$~credible interval $\exp(\mu_\text{DGP}\pm\sigma_\text{DGP})$:\footnote{As detailed in Sec.~\ref{sec:structure}, \smoking returns these regression errors in the form of the signed, relative errors $(-\Delta_\smoking^-/y_\smoking)$ and $(\Delta_\smoking^+/y_\smoking)$.}
\begin{align}
\begin{split}
  \Delta^-_\smoking(\vec x_*) = &\exp[\mu_\text{DGP}(\vec x_*)]\\
   &- \exp[\mu_\text{DGP}(\vec x_*)-\sigma_\text{DGP}(\vec x_*)],\\
  \Delta^+_\smoking(\vec x_*) = & \exp[\mu_\text{DGP}(\vec x_*)+\sigma_\text{DGP}(\vec x_*)] \\
  &- \exp[\mu_\text{DGP}(\vec x_*)].
\label{eq:xsec_regression_error_definition}
\end{split}
\end{align}
More generally, the range
$y_\smoking \pm m \Delta^\pm_\smoking$
corresponds to an $n \sigma$ credible interval, where $n$ is given by
\begin{align}
  n = \frac{\log\left(1 \pm \frac{m \Delta^\pm_\smoking}{y_\smoking}\right)}{\log\left(1 \pm \frac{\Delta^\pm_\smoking}{y_\smoking}\right)},
\label{eq:xsec_credible_interval_relation}
\end{align}
so that $n = m$ to first order in $(m \Delta^\pm_\smoking / y_\smoking)$ and $(\Delta^\pm_\smoking / y_\smoking)$.

We train the GPs with the union of the training sets discussed in Sec.~\ref{sec:sample generation}.
In order to choose the optimal training parameters (relative sizes of the five different training samples, total number of training points, and experts per process), we have carried out a long programme of testing and cross-validation, in which we optimised training parameters separately for each process type, e.g.\ separately for gluino pair production and gluino--squark production.
Our goal was to achieve good performance in terms of the regression error estimated by the DGP, while keeping disk size, memory footprint, training and evaluation time down to acceptable levels.

The processor time spent on training the GPs varies for each process and depends critically on the number of training points per expert because of the GP scaling relations, as well as on optimiser settings like the population size and convergence thresholds for the differential evolution.  In practice, the total time we spent on training a single process ranges from 4.5 CPU hours for a simple case with only three parameters, such as $\tilde c_L^* \tilde c_L$ production, to 141 CPU hours for gluino pair production, which has ten parameters at NLO. In these most extreme examples, the use of parallel processing for the matrix operations, in particular for the different experts, reduced the actually elapsed wall-clock time by a factor 8 to 12, compared to the CPU time.

\section{Validation}
\label{sec:validation}

To validate the \smoking cross-section results we generate three main test sets: $\mcDtest$, $\mcDtesttb$ and $\mcDMSSM$. The set $\mcDtest$ is used for testing all cross-sections except for the stop/sbottom pair-production processes, which are tested with the set $\mcDtesttb$. The sets $\mcDtest$ and $\mcDtesttb$ contain 10\,000 and 5\,000 points, respectively, and in both cases the input points are sampled in the same way as for the corresponding training sets. The third set, $\mcDMSSM$, contains 19\,000 points.  We use this in the validation of all cross-sections. Here we draw the input points from the MSSM-24, as defined at $Q=1$\,\TeV, using uniform priors for the MSSM parameters. Our definition of the MSSM-24 follows that in Ref.~\cite{Athron:2017ard}, except that we parameterise the Higgs sector using the higgsino mass parameter $\mu$ and the tree-level mass of the $A^0$ boson $m_{A^0}$, instead of the soft-breaking mass parameters $m_{H_u}^2$ and $m_{H_d}^2$. For the samples in $\mcDMSSM$ we use \softsusy~\textsf{4.0} to calculate the physical mass spectrum from the MSSM input parameters \cite{Allanach:2001kg,Allanach:2016rxd}. In addition to the three main test sets, we also generate a number of process-specific two-dimensional parameter grids for further validation.

As part of the validation we will discuss two simple error measures and their distributions for our sets of test points. The first measure is just the relative error,
\begin{align}
\epsilon(\vec x_*) = \frac{y_\prosp(\vec x_*) - y_\smoking(\vec x_*)}{y_\prosp(\vec x_*)},
\label{eq:xsec_relative_error}
\end{align}
which measures the relative deviation of the \smoking result $y_\smoking$ from the true \prospino value $y_\prosp$. By assuming some sampling prior $\pi(\vec x_*)$ for the test points $\vec x_*$ and looking at the resulting distribution of $\epsilon(\vec x_*)$ values, we can get a \textit{global} picture of how well \smoking performs for the different supersymmetric production processes included in \smoking.

One of the strengths of GP regression is that the method directly provides \textit{point-wise} uncertainty estimates, here in terms of $\Delta^-_\smoking(\vec x_*)$ and $\Delta^+_\smoking(\vec x_*)$, which are based on the width $\sigma_\text{DGP}(\vec x_*)$ of the DGP predictive distribution (see Eq.~\ref{eq:xsec_regression_error_definition}). As discussed in Sec.~\ref{sec:regression_framework}, the point-wise regression uncertainty is connected to a Bayesian degree of belief regarding the unknown function value at $\vec x_*$, given the particular training set $\mcD$ and the modelling choices made in constructing and optimising the kernel.

An interesting question is then how this point-wise uncertainty compares to the \textit{actual} deviation between $y_\smoking$ and the true \prospino value $y_\prosp$ across the input feature space. That is, we should also investigate the distribution of the standardised residual
\begin{align}
  z(\vec x_*) = \frac{y_\prosp(\vec x_*) - y_\smoking(\vec x_*)}{\Delta^\pm_\smoking(\vec x_*)},
\label{eq:xsec_residual}
\end{align}
in our sets of test points. Here the notation $\Delta^\pm_\smoking$ is shorthand for using $\Delta^+_\smoking$ when $y_\prosp - y_\smoking > 0$ and $\Delta^-_\smoking$ when $y_\prosp - y_\smoking < 0$. By studying the distribution of $z(\vec x_*)$ values
for our test sets, and comparing to the unit normal distribution,
we will obtain a global picture of the extent to which the point-wise regression errors are conservative or not, compared to the true deviations.

In cases where the main source of GP regression uncertainty is actual random noise in the training data, and we learn this noise level from the data by including a white-noise term, Eq.\ (\ref{eq:white-kernel}), in the GP kernel, we expect the residual in Eq.\ (\ref{eq:xsec_residual}) to be distributed as $\mathcal{N}(0,1)$.\footnote{That is, at least up to log-normal corrections implied by Eq.~(\ref{eq:xsec_credible_interval_relation}) for $n\ne1$.} We can understand this from a simple example with single-component input points $x$. Assume that the target data are noisy measurements of the form $y(x) = f(x) + \delta$,
where the noise $\delta$ is distributed as $p(\delta) = \mathcal{N}(0,\sigma^2_{\textrm{noise}})$. Let the GP posterior predictive distribution for $y(x_*)$ be given by $\mathcal{N}(\mu_\textrm{GP}(x_*),\sigma^2_\textrm{GP}(x_*))$. If the noise is the dominant uncertainty we have $\sigma_\textrm{GP}(x_*) \approx \sigma_{\textrm{noise}}$, and we can express the residual as
\begin{align}
\begin{split}
  z(x_*) &= \frac{y(x_*) - \mu_\textrm{GP}(x_*)}{\sigma_\textrm{GP}(x_*)}\\
         &\approx \frac{y(x_*) - f(x_*)}{\sigma_\textrm{noise}}
            + \frac{f(x_*) - \mu_\textrm{GP}(x_*)}{\sigma_\textrm{GP}(x_*)}.
\label{eq:residual_normal_dist_example}
\end{split}
\end{align}
Given the generative model for the data, the first term will follow an $\mathcal{N}(0,1)$ distribution. The second term is the number of standard deviations (as measured by the GP's own uncertainty) by which the GP prediction $\mu_\textrm{GP}$ differs from the true value of the underlying function $f$. While this term can in general not be expected to be normally distributed, the assumption of noise-dominated uncertainty implies that its contribution to the residual is $\ll 1$, and hence, that the $z$ distribution is close to $\mathcal{N}(0,1)$.

On the other hand, if the true noise level is tiny and some other source of uncertainty dominates $\sigma_\textrm{GP}$, \ie, if $y \approx f$ and $\sigma_\textrm{GP} \gg \sigma_\textrm{noise}$, we get
\begin{align}
  z(x_*)  \approx \frac{f(x_*) - \mu_\textrm{GP}(x_*)}{\sigma_\textrm{GP}(x_*)}.
\end{align}
Thus, in this limit it is the generally small, and potentially non-Gaussian, second term from Eq.\ (\ref{eq:residual_normal_dist_example}) that dominates the residual.

The latter scenario is most similar to the case we have for the \smoking residual in Eq.\ (\ref{eq:xsec_residual}). The actual noise level in the \prospino training data is very small, as typically is the additional error in Eq.~(\ref{eq:add_variance_from_hyperpar_uncertainty}) accounting for uncertainty in the hyperparameter choice. Assuming some reasonably uninformative sampling prior $\pi(\vec x_*)$, this means that for most points the widths $\Delta^\pm_\smoking(\vec x_*)$ are dominated by the homoscedastic error contribution that we include to stabilise the numerics (see Sec.~\ref{sec:regularisation}). As this is a global error contribution, we can expect the resulting regression error for most test points to be larger than the actual error, $y_\prosp(\vec x_*) - y_\smoking(\vec x_*)$. We therefore in general expect the $z(\vec x_*)$ distributions to be narrow compared to $\mathcal{N}(0,1)$, and not necessarily Gaussian. For comparison, we will include a graph of $\mathcal{N}(0,1)$ in all our plots of residual distributions.

In the coming subsections, much of our focus will be on the regression errors and the related relative error $\epsilon$ and residual $z$. However, we remind the reader that the regression errors that we find are typically far subdominant to the cross-section uncertainties coming from the scale and PDF errors.

\subsection{Gluino pair production}

The gluino pair-production cross-section to NLO in QCD depends on the gluino mass and all the other squark masses. Naturally, the gluino mass is the dominant feature of the DGP after training. The mass-averaging approximation that \prospino uses (see Sec.\ \ref{sec:NLO}) means that the average first/second-generation squark mass $\bar m_{\tilde q}$ is a strong predictor of the NLO contribution.  We therefore provide it to the GPs as a separate feature, i.e. as if it were part of the vector of parameters $\vec x$. The importance of the individual first and second-generation squark masses roughly follows the PDF contributions from their corresponding quarks, due to LO $t$-channel squark exchange diagrams.

Given that the mass range of parameters (features) in the training samples is $[100, 3500]$\,\GeV, we validate the cross-section on the sub-interval $[200,3000]$\,\GeV for both gluino and squark masses, where the cross-section regression has solid support from training data and the regression error is small.

\begin{figure*}[t!]
\begin{minipage}[t]{0.485\textwidth}
\centering
\includegraphics[width=\textwidth]{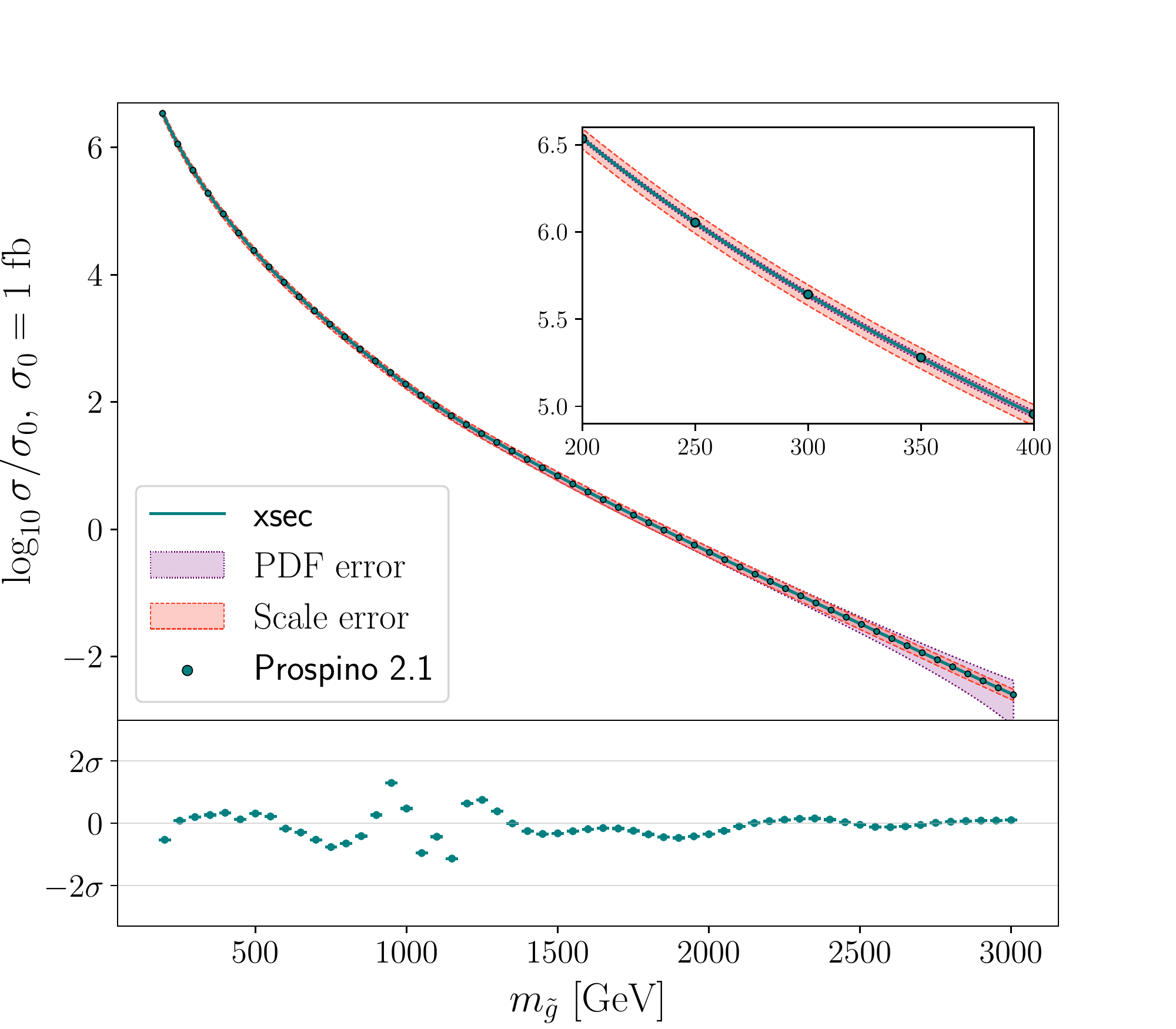}
\caption{Gluino pair-production cross-section as a function of gluino mass, with all squark masses fixed at 1\,\TeV. The central value is shown in light green, the scale error in pink, and the PDF error in violet. The $\alpha_s$ error is too small to be visible on the scale of the plot.
Superimposed on the prediction are the \prospino values (dots). Inset is a close-up of the region at low gluino mass, and below we show the residual between the \smoking prediction and the \prospino values.}
\label{fig:gg_gmass}
\end{minipage}%
\hspace{0.03\textwidth}
\begin{minipage}[t]{0.485\textwidth}
\centering
\includegraphics[width=\textwidth]{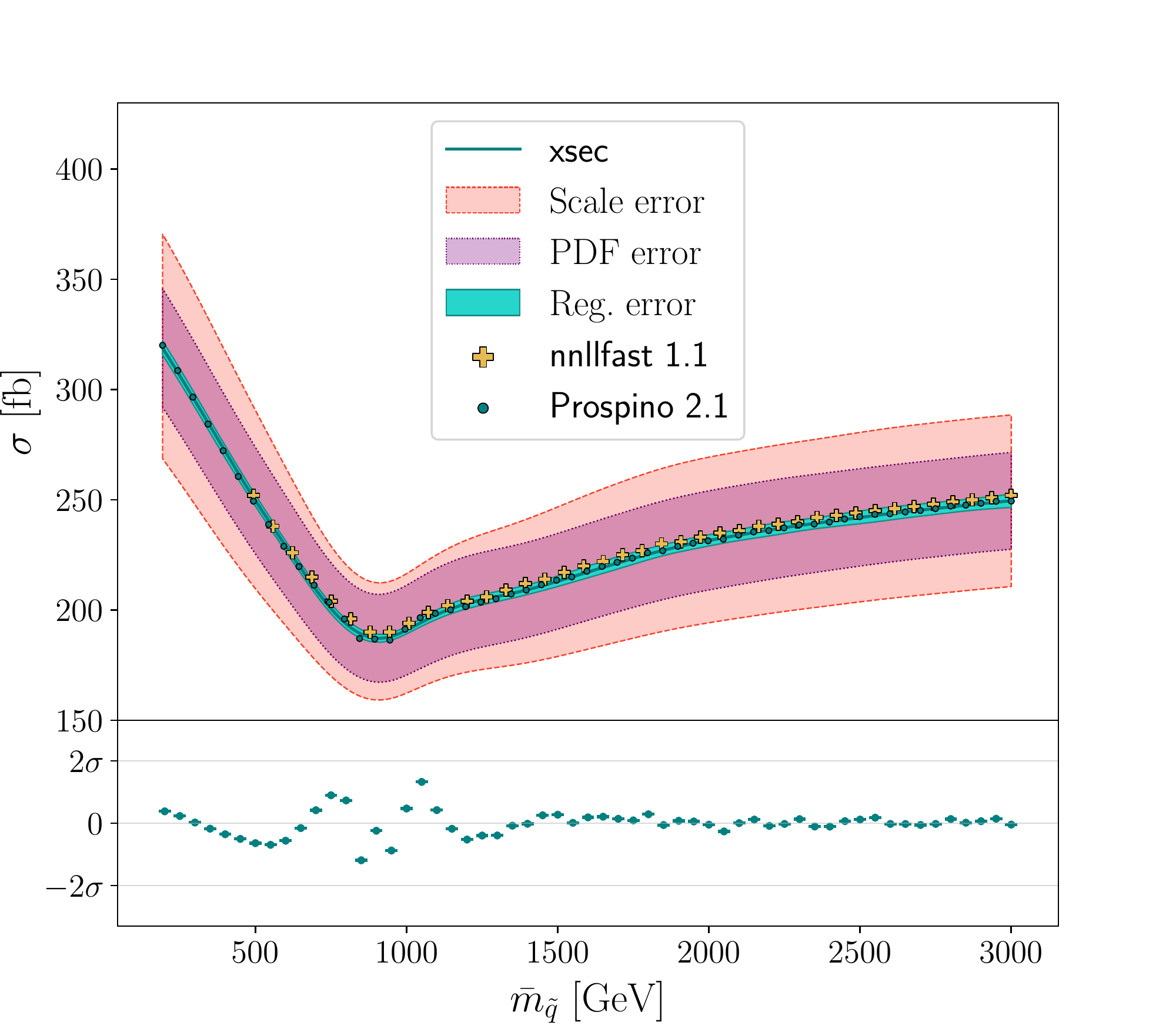}
\caption{Gluino pair-production cross-section as a function of average first and second-generation squark mass. The gluino mass is fixed at 1\,\TeV. Shown are the central \smoking prediction (solid line), the $1\sigma$ regression error band (light green), the scale error (pink), the PDF error (violet), the \prospino values (dots) and the corresponding \nnllfast NLO result (crosses).
\label{fig:gg_qmass}}
\end{minipage}
\end{figure*}

In Fig.~\ref{fig:gg_gmass} we compare the gluino pair-production cross-sections predicted by \smoking, presented as a function of the gluino mass, with values taken directly from \prospino (but not in the training set).
For this comparison we fix the squark masses to a common value of 1\,\TeV.
We also show the associated \smoking-predicted uncertainty from the renormalisation scale and the PDFs as bands.
The uncertainties from the regression provided by \smoking and from $\alpha_s$ are too small to be visible on the logarithmic scale. However, below the plot we show the residual between the \smoking prediction and the \prospino value (Eq.\ \ref{eq:xsec_residual}), as a multiple of the \smoking regression uncertainty.
We observe good agreement overall with the \prospino result.  As expected, the scale error dominates at low gluino masses, and the PDF error at high masses.

\begin{figure*}[p]
\begin{minipage}[c]{\textwidth}
\centering
\includegraphics[width=\textwidth]{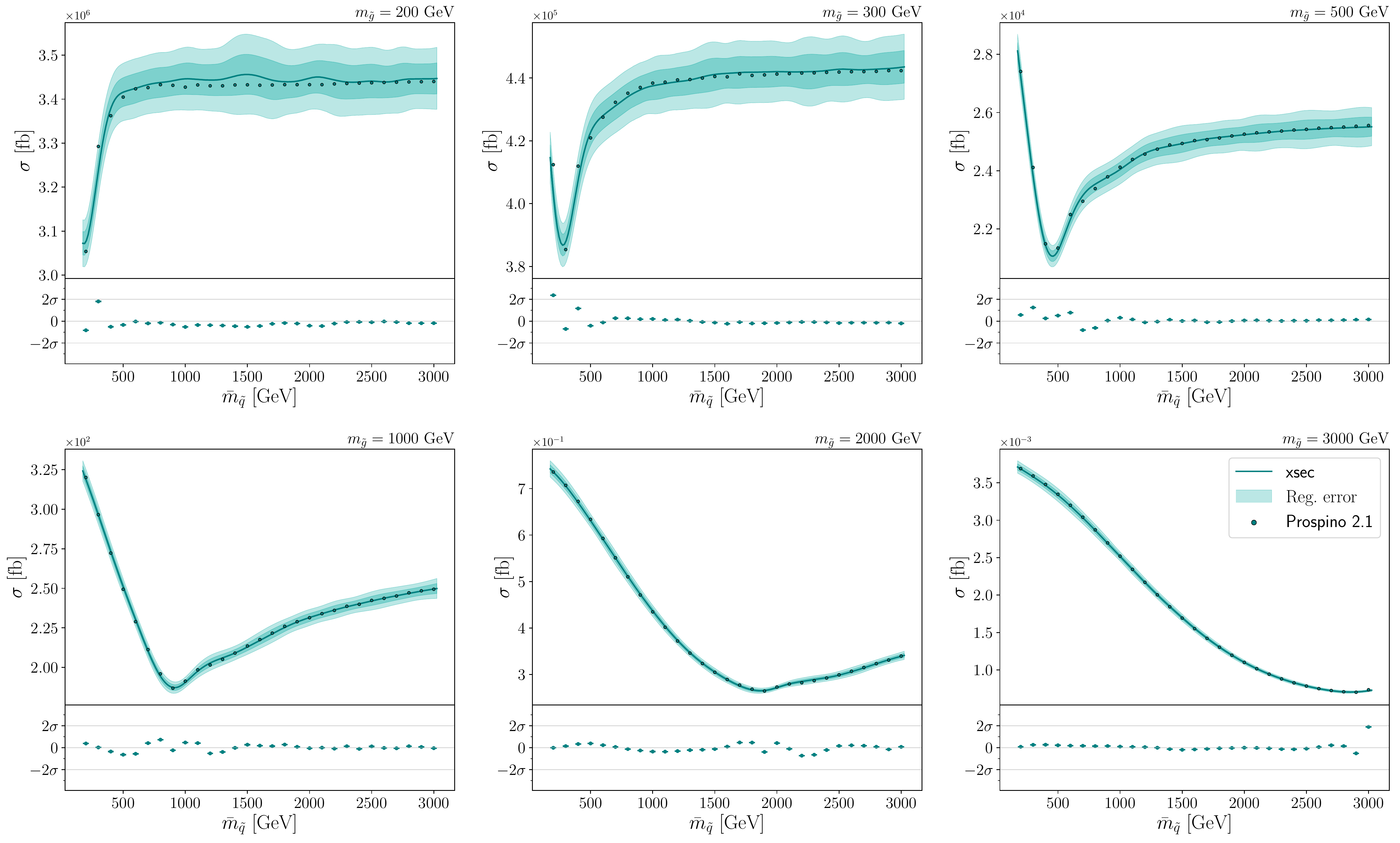}
\caption{Gluino pair-production cross-section as a function of average first and second-generation squark mass for a set of different gluino masses. Shown is the central \smoking prediction (solid line), the $1\sigma$ and $2\sigma$ regression error bands (shaded regions), and the \prospino values (dots). We also show the residuals of the comparison to \prospino.
\label{fig:gg_qmass_multiplot}}
\end{minipage} %
\begin{minipage}[t]{0.485\textwidth}
\centering
\includegraphics[width=\textwidth]{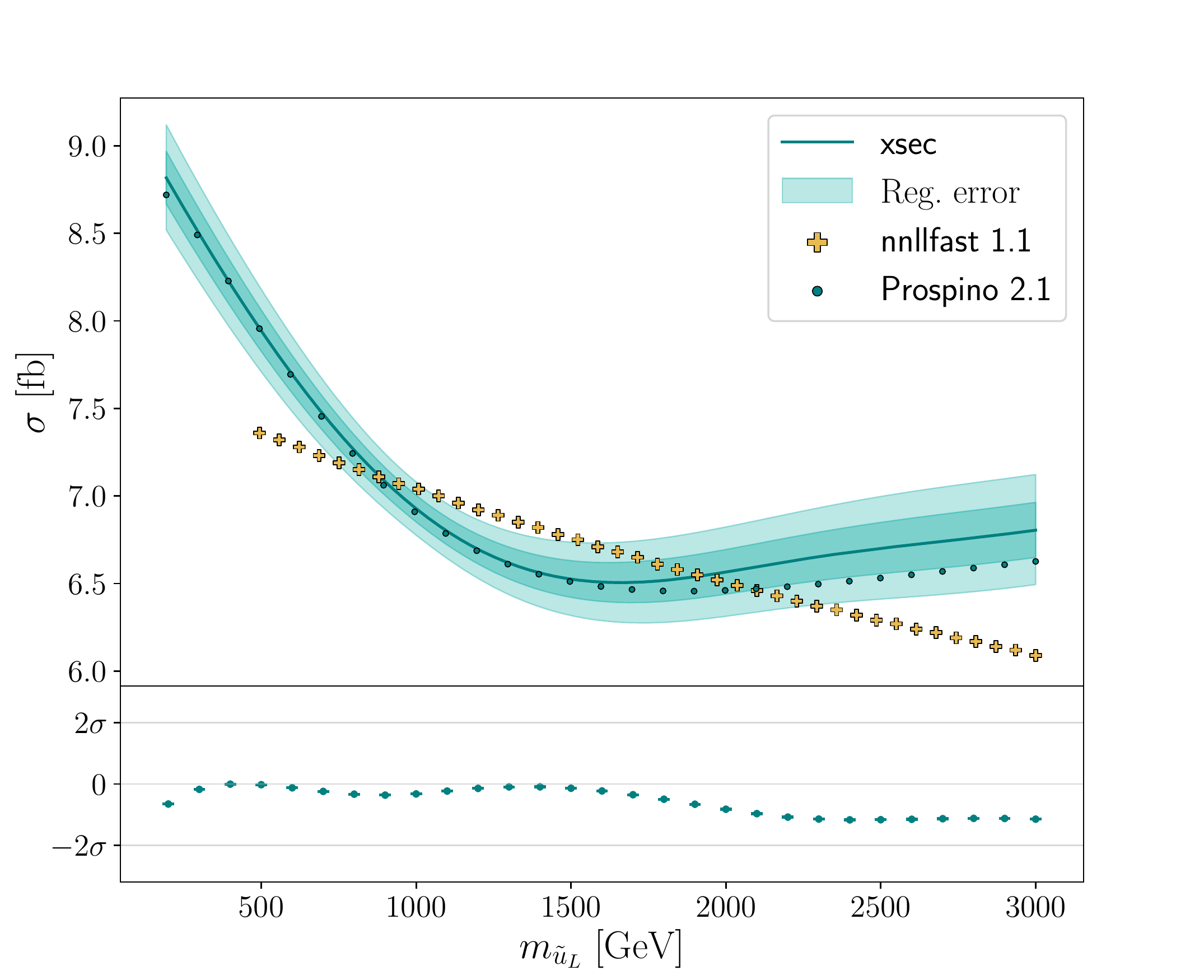}
\caption{Gluino pair-production cross-section as a function of the ${\tilde u}_L$  mass. All other squark masses are fixed at 1\,\TeV and the gluino mass is set to 1.5\,\TeV. Shown is the central \smoking prediction (solid line), the $1\sigma$ and $2\sigma$ regression error bands (shaded regions), the \prospino values (dots) and the corresponding \nnllfast NLO result (crosses).}
\label{fig:gg_uLmass}
\end{minipage}%
\hspace{0.03\textwidth}
\begin{minipage}[t]{0.485\textwidth}
\centering
\includegraphics[width=\textwidth]{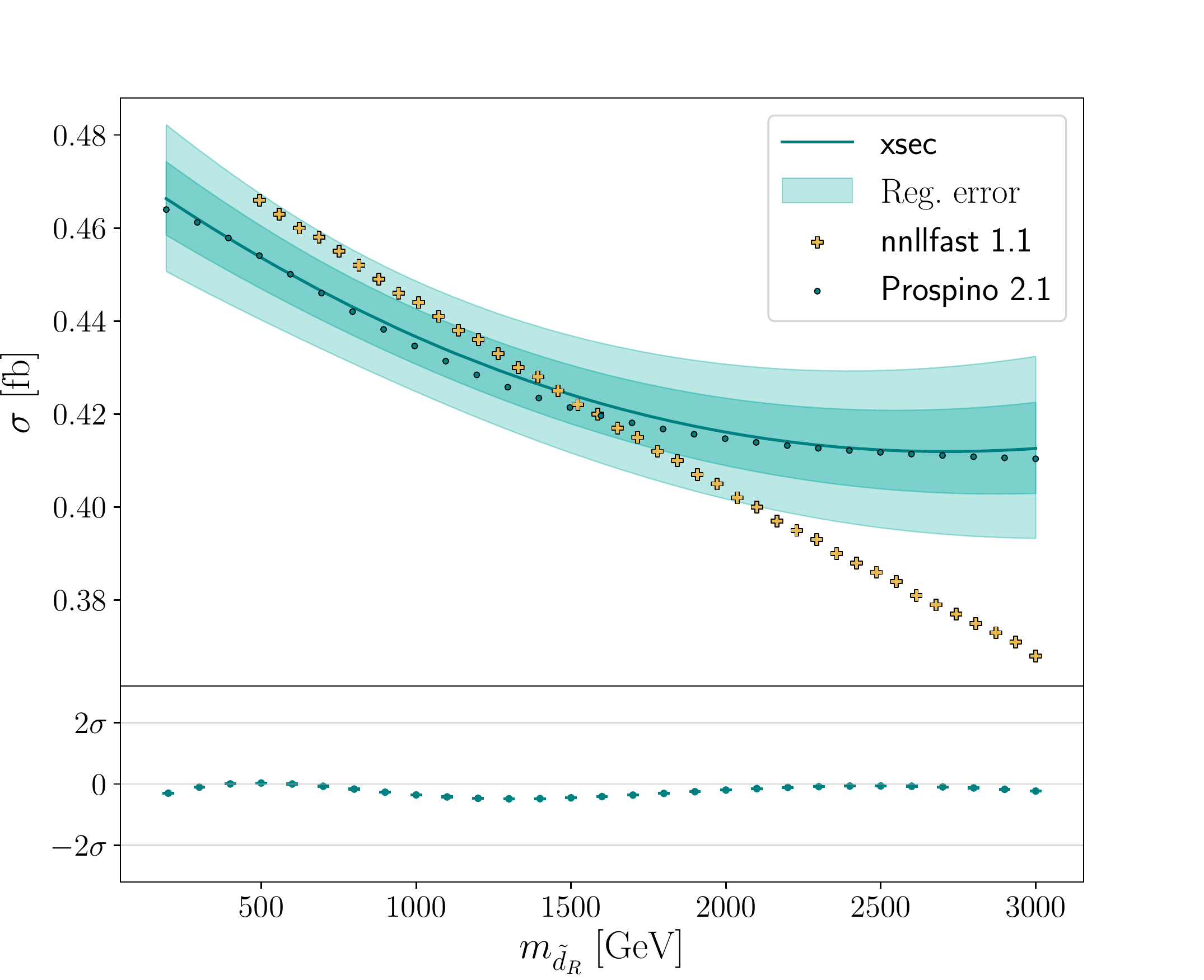}
\caption{Gluino pair-production cross-section as a function of the ${\tilde d}_R$  mass. All other squark masses are fixed at 1\,\TeV and the gluino mass is set to 2\,\TeV. Shown is the central \smoking prediction (solid line), the $1\sigma$ and $2\sigma$ regression error bands (shaded regions), the \prospino values (dots) and the corresponding \nnllfast NLO result (crosses).}
\label{fig:gg_dRmass}
\end{minipage}
\end{figure*}


\begin{figure*}[t]
\includegraphics[width=0.5\textwidth]{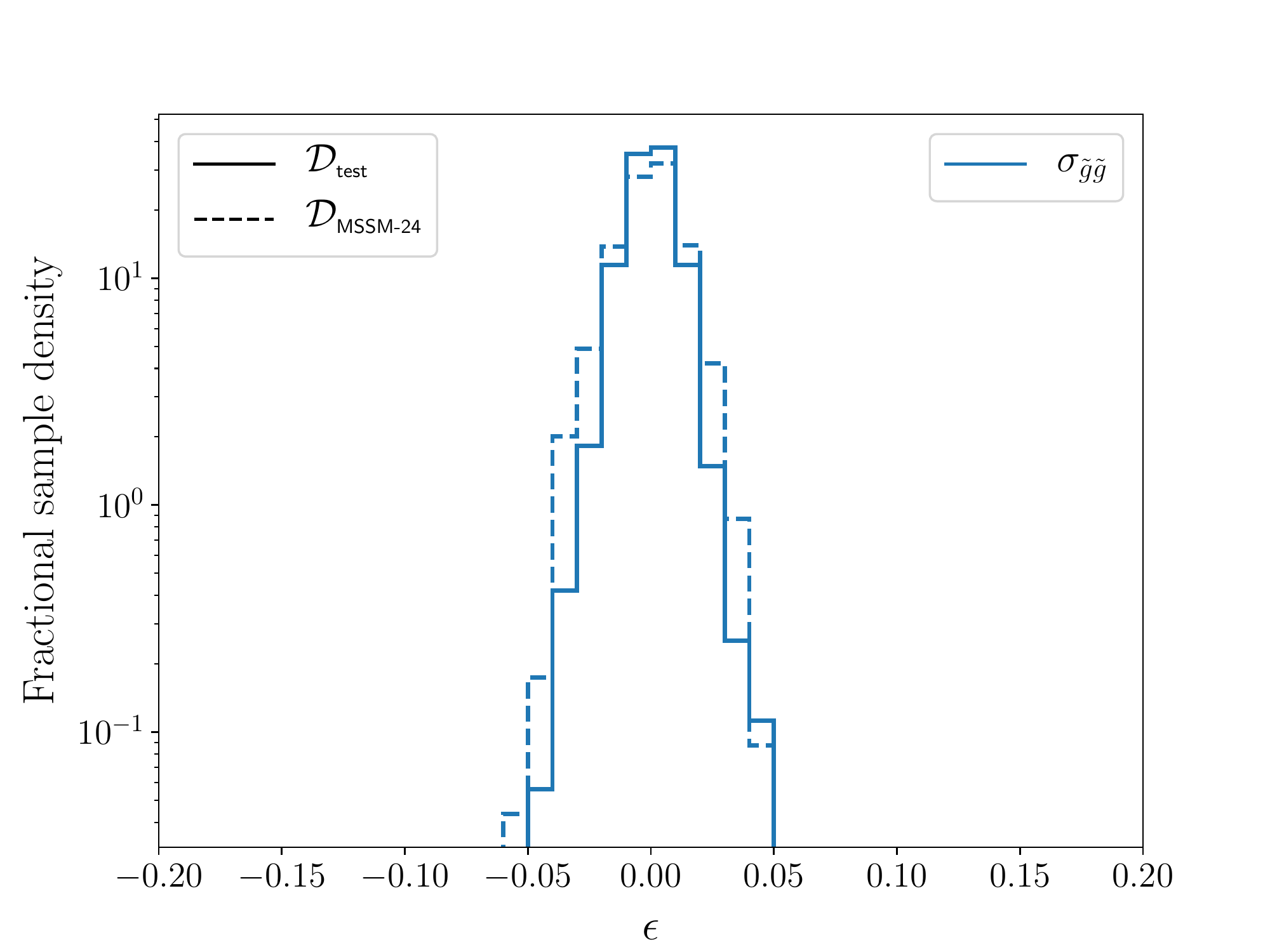}
\includegraphics[width=0.5\textwidth]{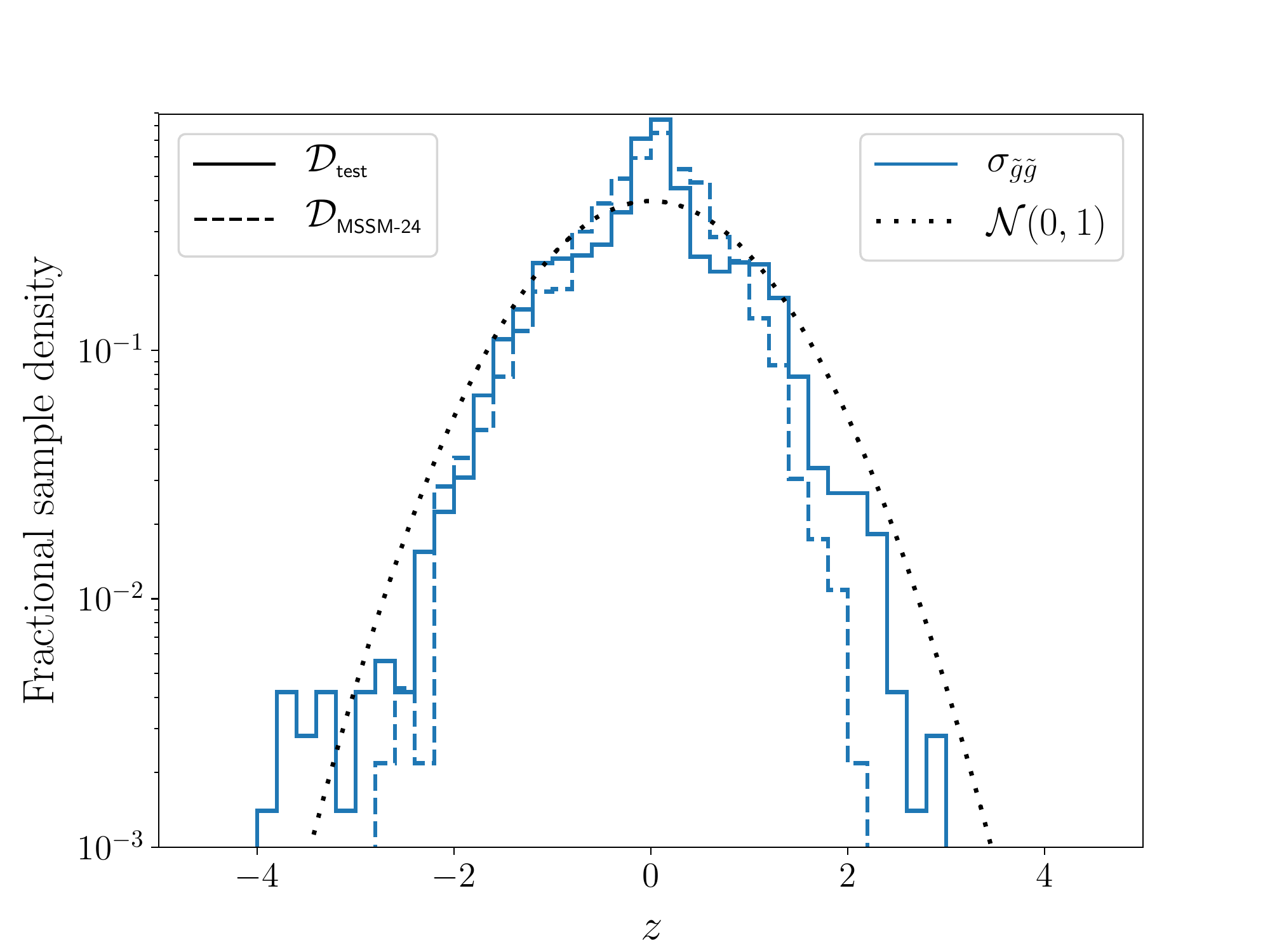}
\caption{The relative error (left) and residual (right) distributions for the gluino pair-production cross-section in the test sets $\mcDtest$ (solid) and $\mcDMSSM$ (dashed). The input points in $\mcDtest$ are sampled from the same distribution as the training set, while the points in $\mcDMSSM$ are sampled from the MSSM-24 using flat priors for the MSSM parameters. All distributions are normalised to unity. The unit normal distribution is shown for comparison as a dotted black line.}
\label{fig:gg_rd}
\end{figure*}
A particular phenomenon occurs in gluino pair production due to destructive interference between LO diagrams when $m_{\tilde g}\approx m_{\tilde q}$, resulting in a vanishing partonic cross-section at threshold~\cite{Beenakker:1996ch}. This can be found as a significant dip in the total pair-production cross-section when one or more of the squark masses become degenerate with the gluino. We show that \smoking reproduces this behaviour to very good precision  in Fig.~\ref{fig:gg_qmass}. Here the gluino mass is fixed to 1\,\TeV while the squark masses are run together as a common squark mass $\bar m_{\tilde q}$. This figure also clearly demonstrates how subdominant the regression error is compared to the scale and PDF errors.
Finally, Fig.~\ref{fig:gg_qmass} compares the results of \smoking to the corresponding NLO result from \nnllfast based on the same PDF set. We observe a slight systematic difference at the $\sim$1\% level.  This is at the level of the interpolation error quoted by \nnllfast.

The same gluino pair-production cross-section as a function of a common squark mass for a selection of different gluino masses can be found in Fig.~\ref{fig:gg_qmass_multiplot}, showing that \smoking reproduces the feature across the whole assumed range of validity for the regression. As expected, the regions in which the cross-section changes most rapidly are the most difficult to predict, but we see that with one exception the prediction is always within $2\sigma$ of the \prospino value, over a large number of test points.

In Figs.~\ref{fig:gg_uLmass} and ~\ref{fig:gg_dRmass},
we show the potential importance of being able to deal with non-degenerate squark masses in \smoking. Here, we vary the ${\tilde u}_L$ and ${\tilde d}_R$ mass alone, respectively, with all other squarks fixed at 1\,\TeV and the gluino at 1.5\,\TeV for  ${\tilde u}_L$ and 2\,\TeV for the plot with ${\tilde d}_R$. While less pronounced, qualitatively the same dip feature due to the $t$-channel interference as discussed above can be seen here as well. In this example, the \nnllfast NLO result fails to reproduce the feature, as expected from its inherent assumption of degenerate squark masses.

Figure~\ref{fig:gg_rd} shows the distributions of the relative error (left) and the residual (right) for the central cross-section value using the test sets $\mcDtest$ and $\mcDMSSM$. All distributions are normalised to unity, and the $y$-axis range is set to show all bins with non-zero values. In both sets the relative error is well below 5\% for most points, and in fact there is only a single point, in the $\mcDMSSM$ set, with an error greater than 5\%. From the comparison to the unit normal distribution we see that, as expected, the \smoking regression error is somewhat larger than the true error, but with no apparent bias. The \smoking prediction is also robust under a change of the test sample to the MSSM-24.

\begin{figure*}[t]
\includegraphics[width=0.5\textwidth]{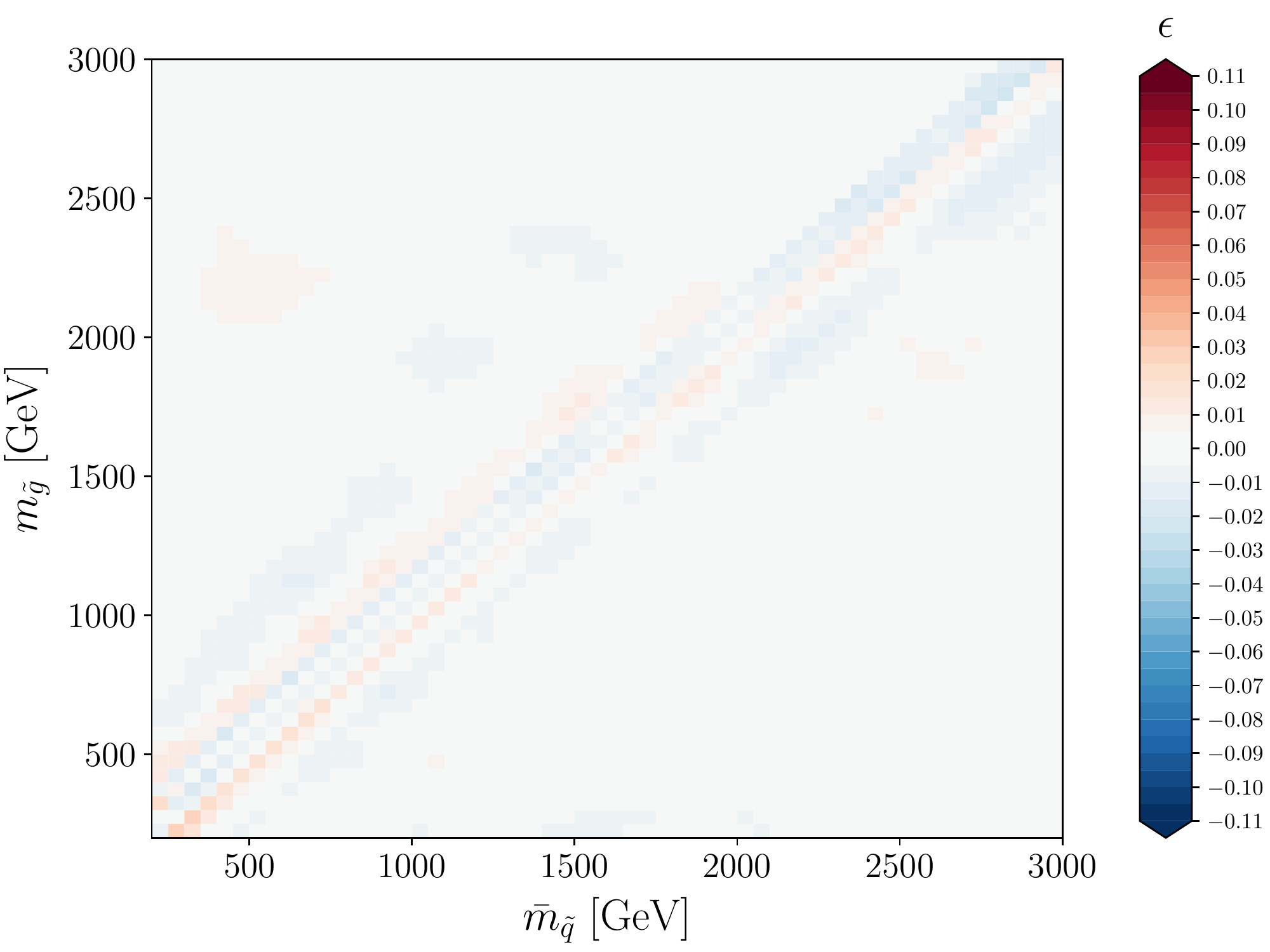}
\includegraphics[width=0.5\textwidth]{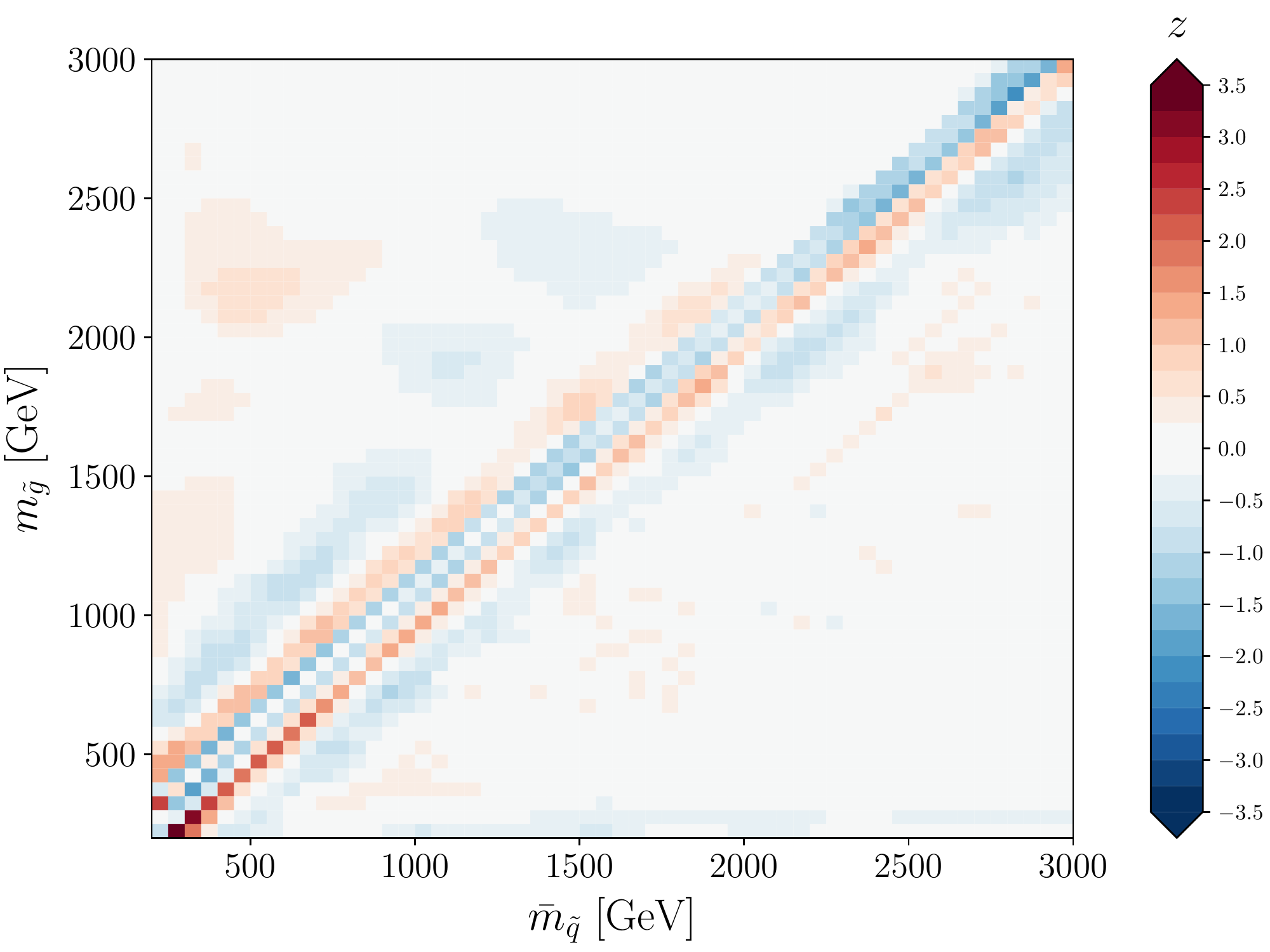}
\includegraphics[width=0.5\textwidth]{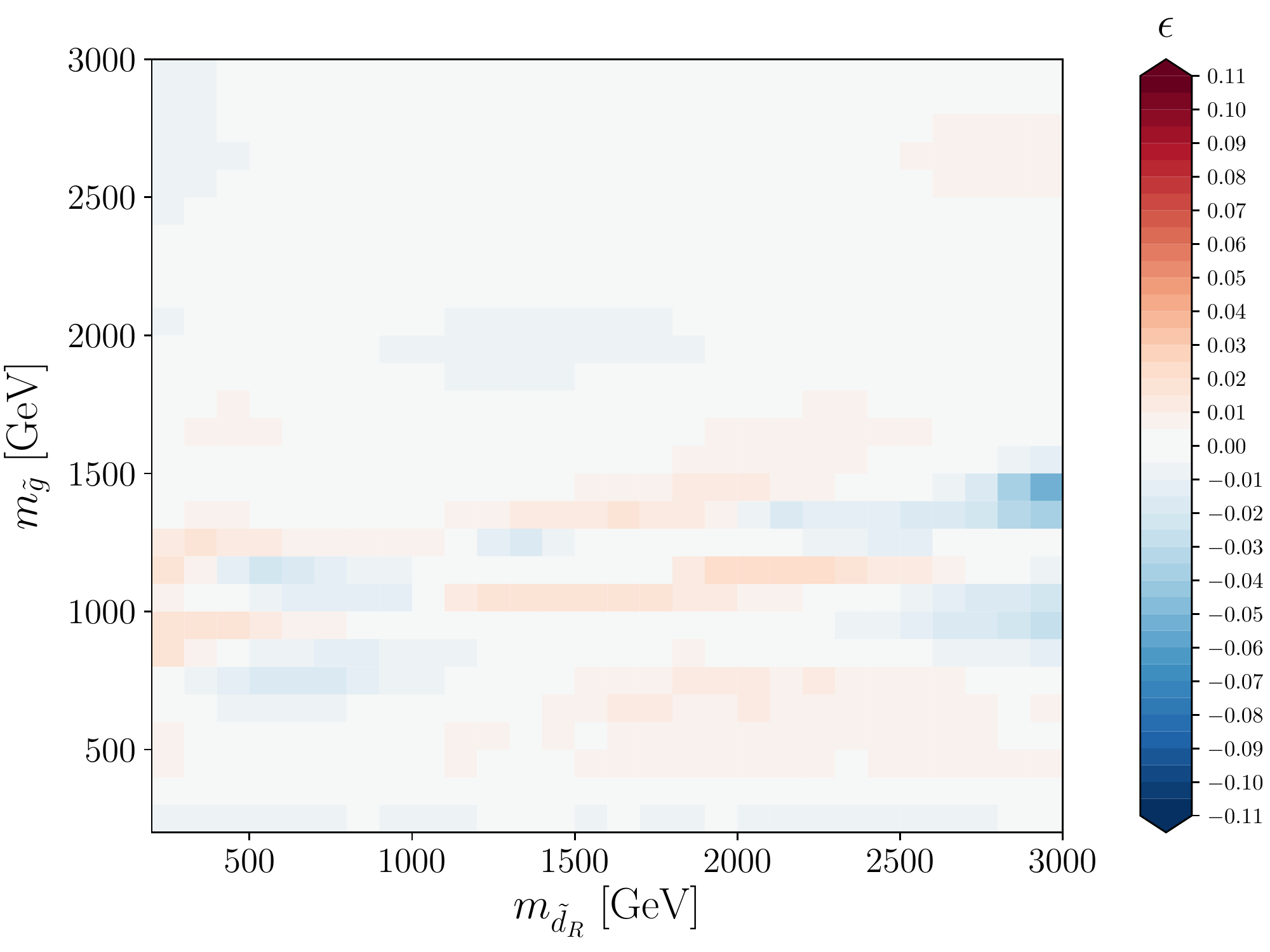}
\includegraphics[width=0.5\textwidth]{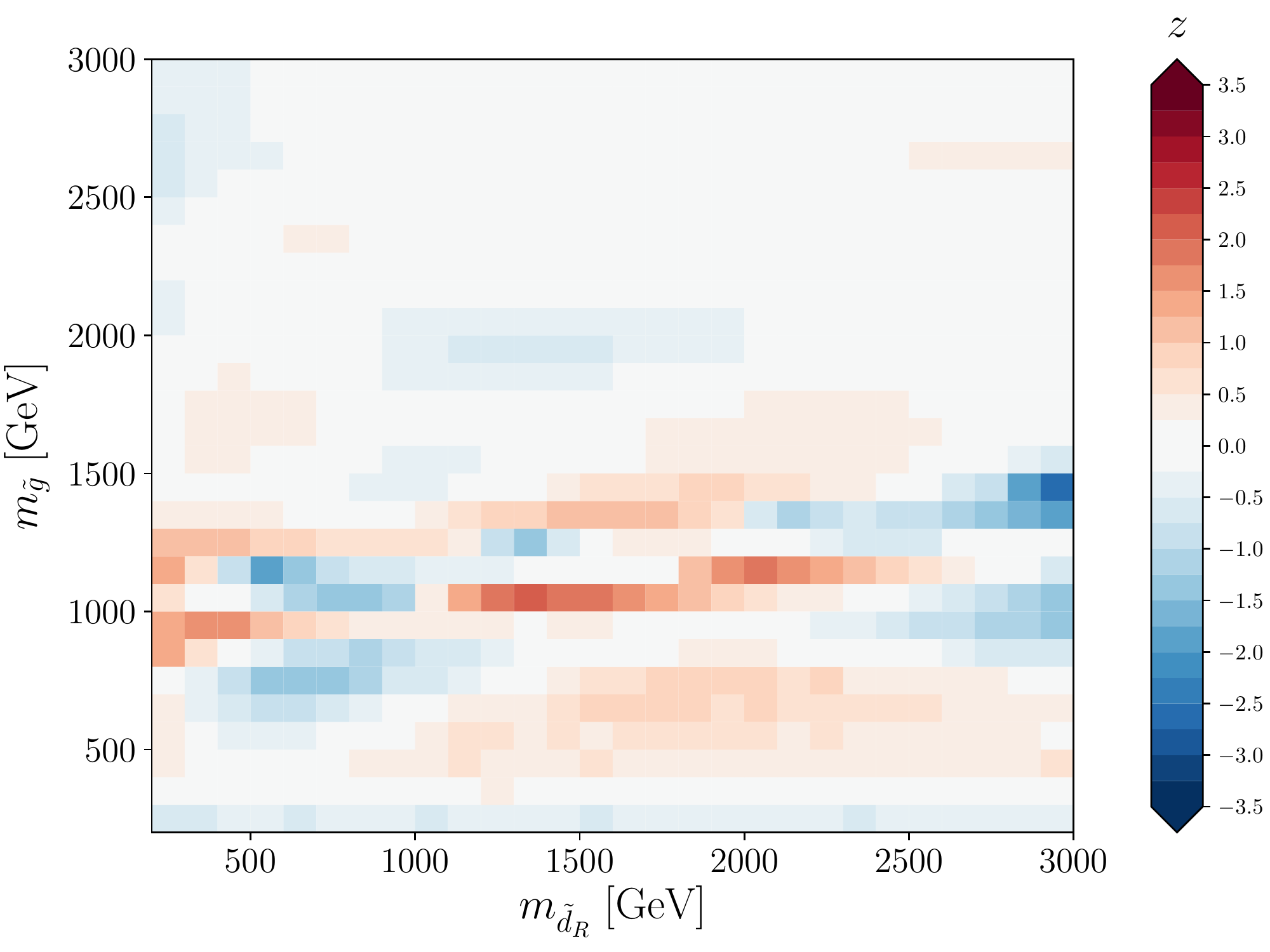}
\caption{The relative error (left) and residual (right) for the gluino pair-production cross-section as a function of the gluino mass versus a common squark mass (top) and the $\tilde d_R$ mass (bottom). All other masses are fixed at 1\,\TeV.
\label{fig:gg_relative_deviance_massplanes}}
\end{figure*}

It is also instructive to perform two-dimensional grid scans of mass planes to show the relative error and residual of the central cross-section value as a function of two of the features at a time. Two examples of this are found in Fig.~\ref{fig:gg_relative_deviance_massplanes}, where we show the result in the planes of $(m_{\tilde g}, \bar m_{\tilde q})$ and $(m_{\tilde g}, m_{\tilde d_R})$.
We see that the relative errors and residuals are correlated in the mass planes, as should be expected from Gaussian processes when the dominant uncertainty is not due to random noise in the training data, but rather due to the lack of information in regions where the function is changing quickly.
We also note that the regression uncertainty is largest when $m_{\tilde g}\approx m_{\tilde q}$. This shows that the destructive interference dip seen in Figs.~\ref{fig:gg_qmass}--\ref{fig:gg_uLmass} is the part of this cross-section function that is the most challenging to capture in the regression.
Further improvements could be made by adding extra training points, but only at significant cost to the evaluation speed, which seems unwarranted given the small regression errors compared to the other errors. Naive counting of the number of bins in the residual plots (right panels) above the 1, 2, and $3\sigma$ levels indicates that the quoted \smoking regression error is in general conservative compared to the actual error (i.e.\ fewer bins show large residuals than expected from Gaussian statistics).

\begin{figure}
\includegraphics[width=0.5\textwidth]{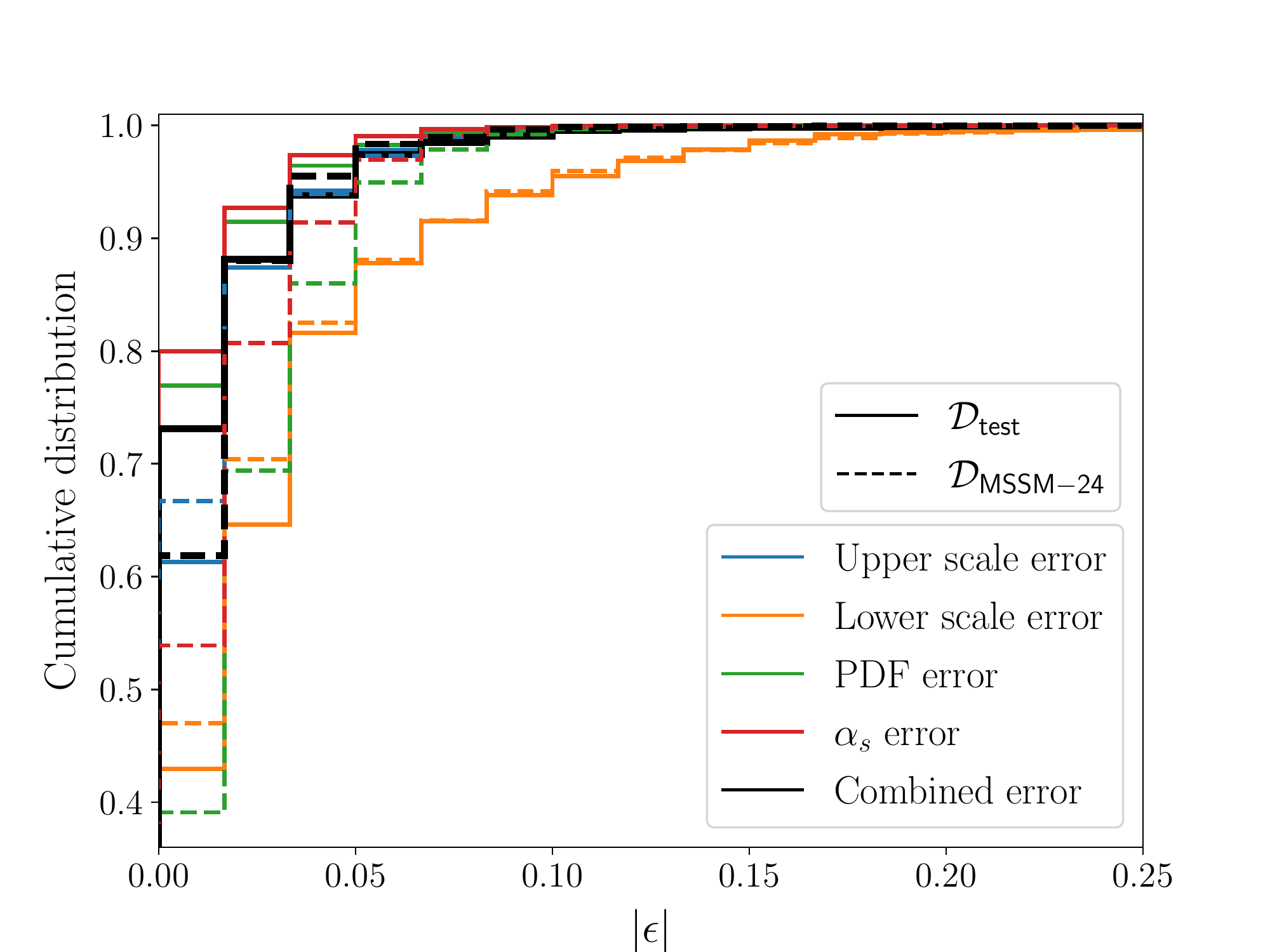}
\caption{Cumulative distribution of the relative error magnitude for the upper and lower scale errors, the PDF error and the $\alpha_s$ error for the gluino pair-production process.}
\label{fig:gg_errorerror}
\end{figure}

In addition to the regression errors, \smoking also predicts scale, PDF and $\alpha_s$ errors from separate GPs performing regression on the relative size of the respective error bands, see Sec.~\ref{sec:training_implementation}. The performance of these GPs in the case of gluino pair-production is shown in Fig.~\ref{fig:gg_errorerror}
in terms of the resulting relative error on the errors, compared to values calculated using \prospino for the $\mcDtest$ and $\mcDMSSM$ test sets. To be precise, we compute the analogue of Eq.~\ref{eq:xsec_relative_error} for each individual type of error:
\begin{align}
  \epsilon(\vec x_*) = \frac{\delta_\prosp(\vec x_*) - \delta_\smoking(\vec x_*)}{\delta_\prosp(\vec x_*)},
  \label{eq:xsec_relative_error_on_error}
\end{align}
where $\delta_\prosp(\vec x_*)$ represents the true width of the scale, PDF, or $\alpha_s$ error band at the test point $\vec x_*$, as computed with \prospino, and $\delta_\smoking(\vec x_*)$ is the corresponding \smoking estimate.

While PDF and $\alpha_s$ errors computed by \smoking are symmetric by default (see Sec.~\ref{sec:python-interface}), the scale errors are not. We therefore symmetrise these before adding them in quadrature to the PDF and $\alpha_s$ errors, to obtain a combined error. We find that the relative error on this combined error is below 10\% in over 90\% of the test points.
Given that the absolute magnitude of the cross-section is much larger than the absolute magnitude of the individual errors, these errors on errors are largely insignificant.
Very similar conclusions can be reached for the scale, PDF and $\alpha_s$ errors for the other processes included in \smoking, and for the sake of brevity we do not show the relative errors on the errors for other processes.

\begin{figure*}[t]
\includegraphics[width=0.5\textwidth]{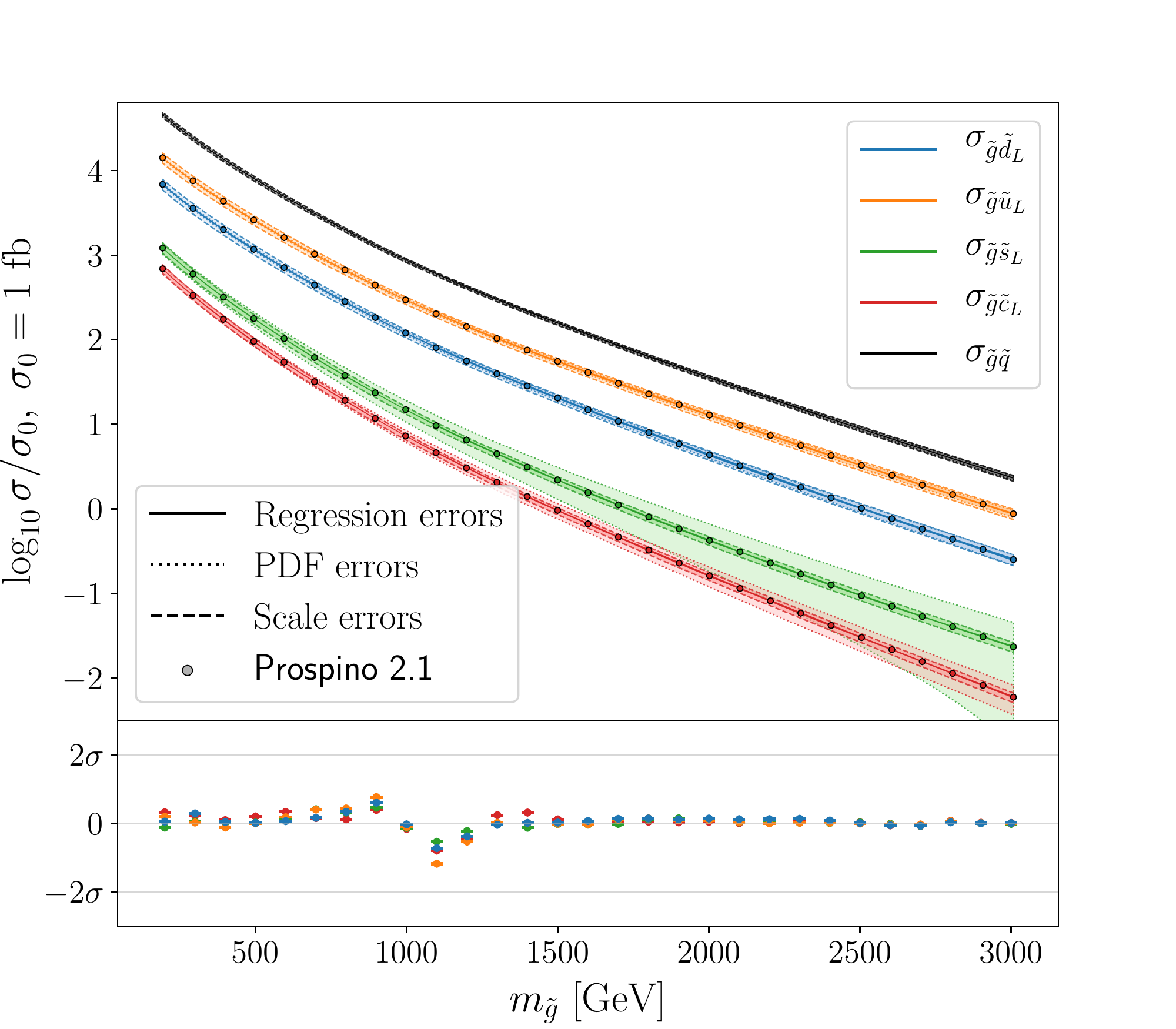}
\includegraphics[width=0.5\textwidth]{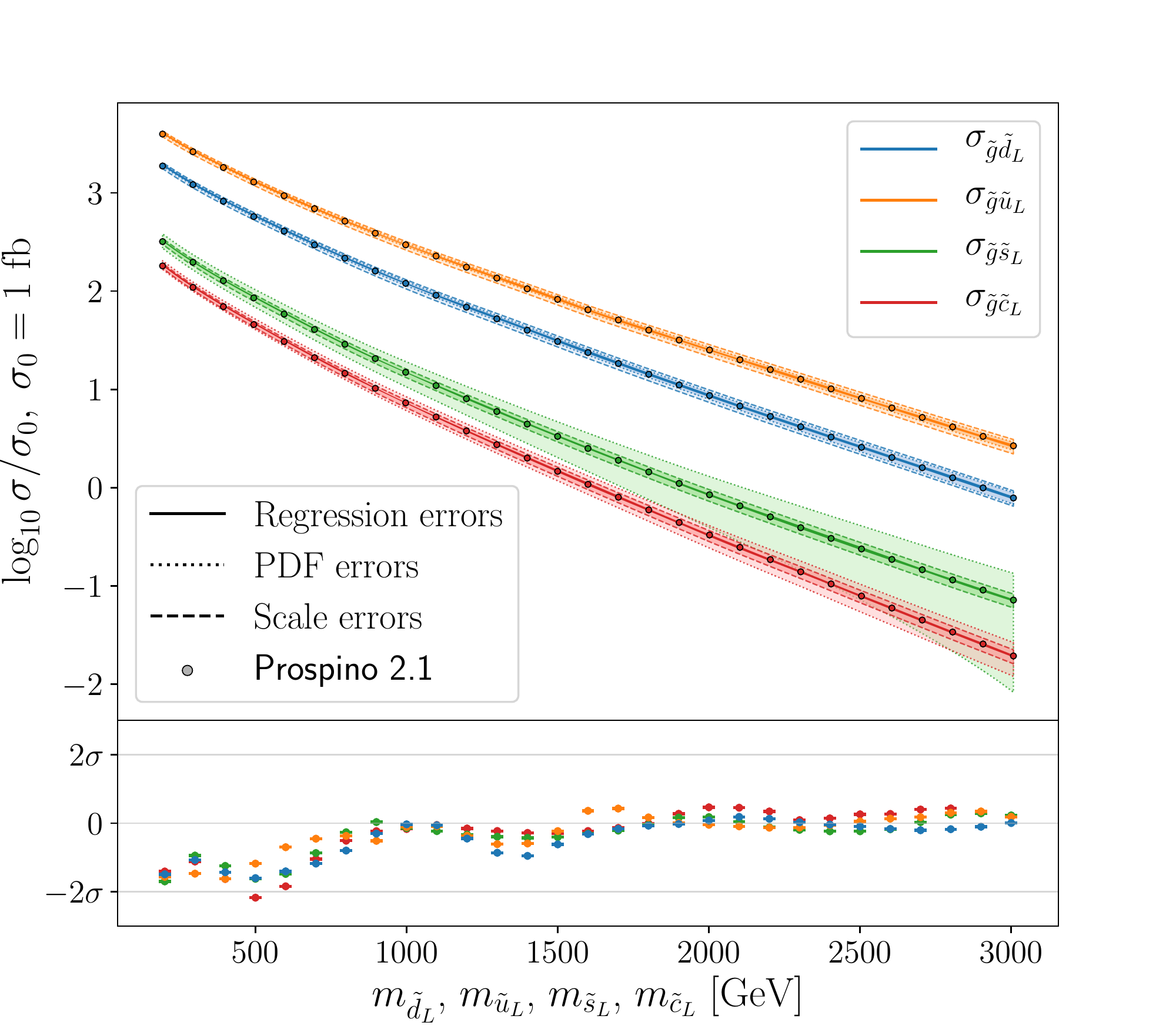}
\caption{Gluino--squark pair-production cross-section as a function of gluino mass (left) and squark masses (right), for production of first and second-generation squarks.  Shown are individual (left-handed) squark final states (colours) and the sum of all first and second-generation final states (black). In the left-hand plot all squark masses are fixed at 1\,\TeV. In the right-hand plot all masses except for the final-state squark mass are fixed at 1\,\TeV.  The central-value \smoking prediction is shown with error bands from regression (solid line), scale error (dashed) and PDF error (dotted). The $\alpha_s$ error is too small to be visible. Also shown are the \prospino values (dots).
}
\label{fig:gq_masses}
\end{figure*}


\subsection{Pair production of gluinos with first or second-generation squarks}

The data that we employ for training \smoking \textsf{1.0} are limited by the fact that \prospino assumes flavour conservation and neglects heavy quarks in the proton PDFs.  It therefore offers gluino--squark pair-production cross-sections only for processes with first and/or second-generation squarks in the final state. However, cross-sections for gluino--squark production with sbottoms or stops in the final state are expected to be very small. We also note that \prospino returns the cross-section for the sum over charge-conjugate final states, i.e. $\tilde g\tilde q_i+\tilde g\tilde q_i^*$, making it pointless to train \smoking separately on the two final states.  In addition to these limitations, at NLO QCD the numerical value of the cross-section is identical for left and right-handed squark final states, as long as their masses are identical. Although we use this fact internally in \smoking to reduce the total file size of the DGPs, the user can freely request any first or second-generation final-state squark.

\begin{figure*}[t!]
\includegraphics[width=0.5\textwidth]{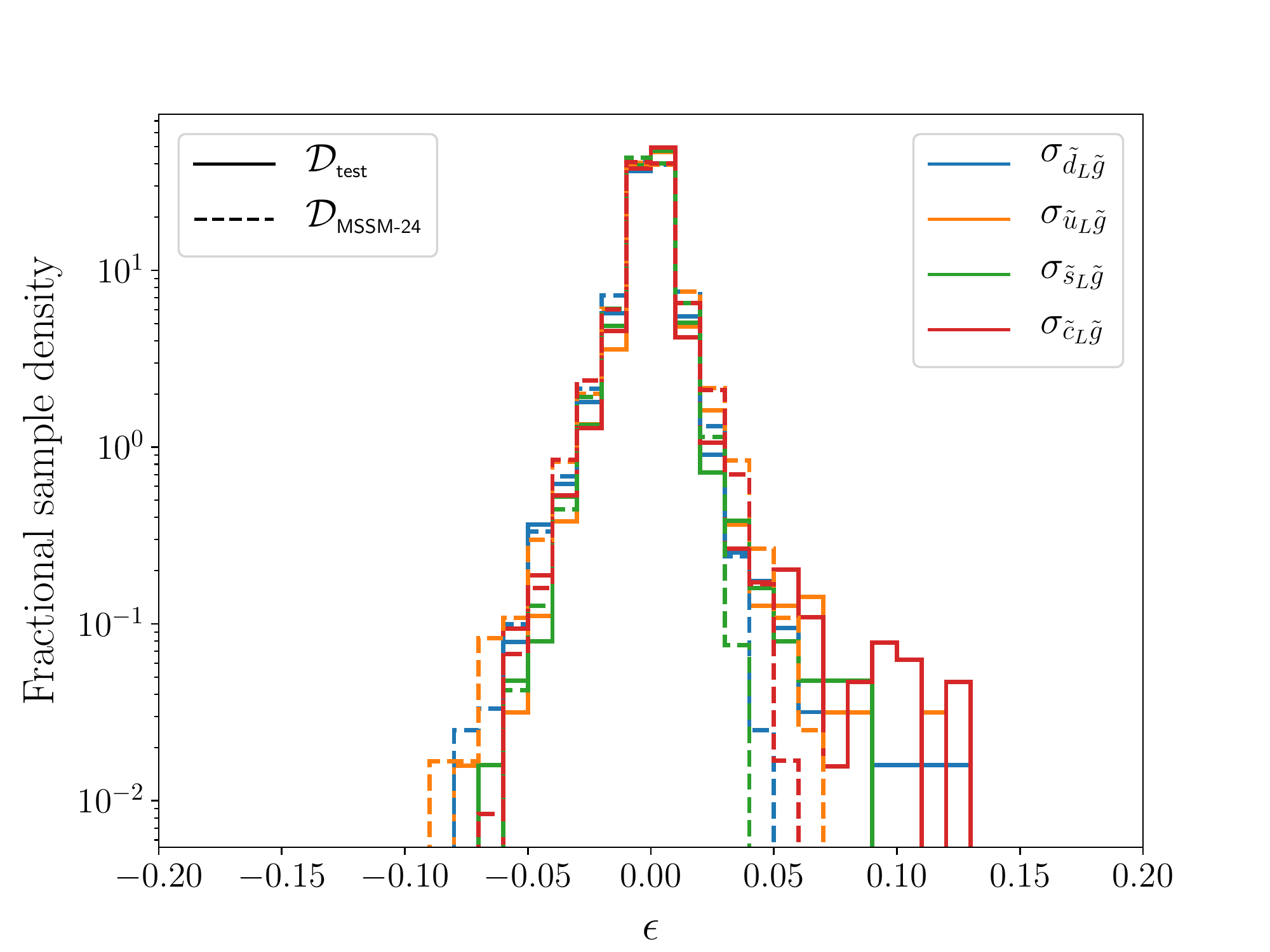}
\includegraphics[width=0.5\textwidth]{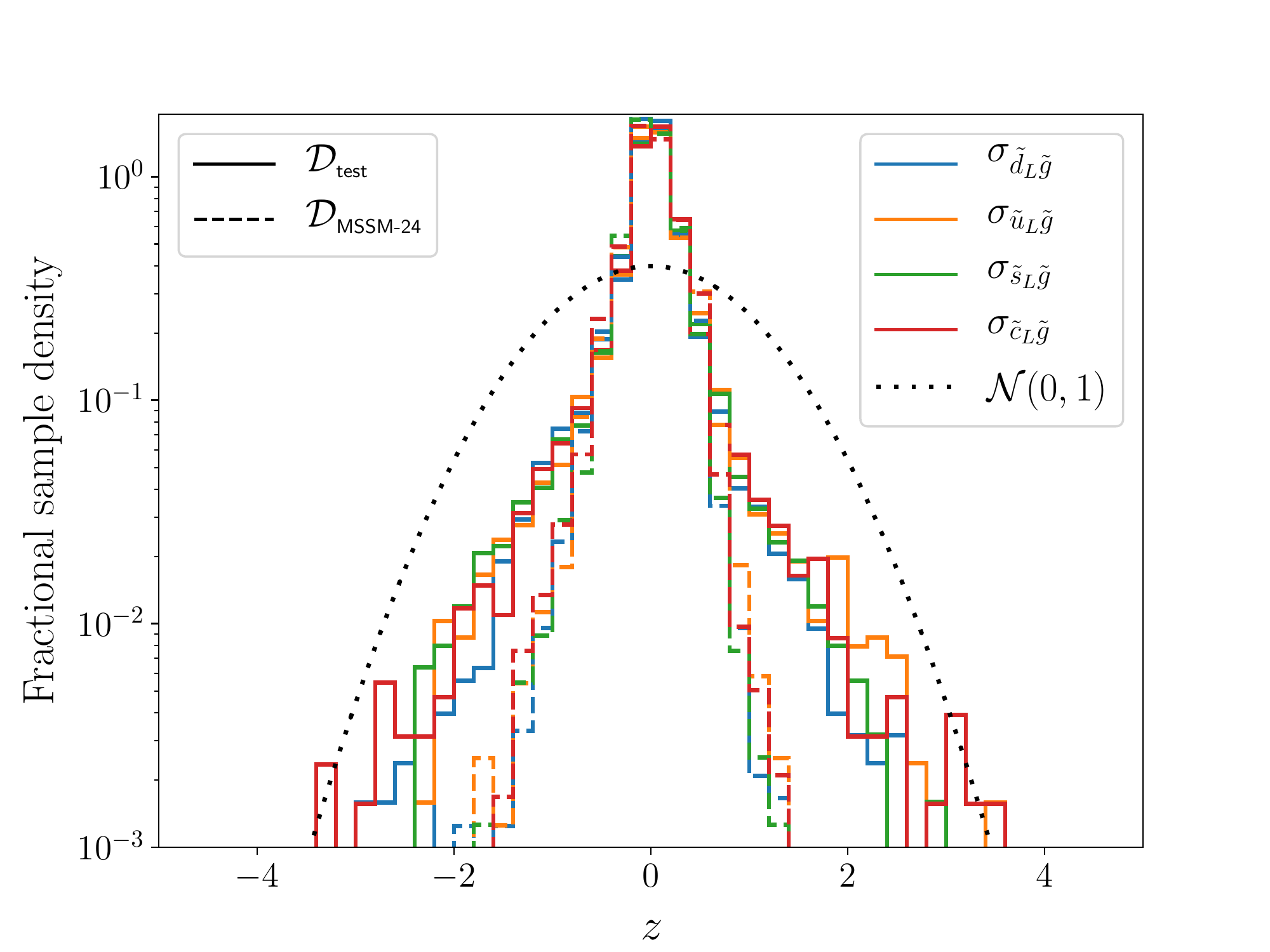}
\caption{The relative error (left) and residual (right) distributions for the first and second-generation gluino-squark cross-sections, for the test sets $\mcDtest$ (solid) and $\mcDMSSM$ (dashed). All distributions are normalised to unity. The unit normal distribution is shown as a dotted black line for comparison to the residual distributions.
}
\label{fig:gq_rd}
\end{figure*}

The sizes of the gluino--squark production cross-sections are naturally dominated by the gluino and final-state squark masses.  Because of flavour conservation, to the level of approximation used in \prospino's $K$-factor calculation, the only additional property of the model parameter space (i.e.\ feature) used by \smoking is the average squark mass $\bar m_{\tilde q}$.
The range of sparticle masses over which we assume this cross-section evaluation to be valid is the same as for gluino pair production, i.e. $[200,3000]$\,\GeV.

In Fig.~\ref{fig:gq_masses} we show the predicted gluino--squark production cross-sections as a function of the gluino mass (left) and the individual $\tilde q_L$ masses (right). For the individual squark masses we keep all other masses at 1\,\TeV and change only the mass of the final-state squark.
Also shown are the predicted regression, PDF and scale errors, and the residual between the \smoking predictions and the \prospino values calculated for the same parameters.
We see that \smoking reliably predicts the contribution from individual squark final-state flavours. This is even the case in the region of very low final-state squark masses, which tests \smoking on the arguably strange scenario in which the particular final-state squark is much lighter than the average mass of the first and second-generation squarks.

For the $\tilde g \tilde d_L $ and $\tilde g \tilde u_L$ processes, the scale error is the dominant uncertainty across the full mass range in both the gluino mass and the final-state squark mass. For the $\tilde g \tilde s_L $ and $\tilde g \tilde c_L$ processes it is generally the PDF error that dominates the uncertainty, except at low gluino masses, where the scale error is more important. The fact that the regression error residuals comparing the \smoking and \prospino values seem correlated between the four processes shown is due to the same training sample being used for all processes, causing the distances to the nearest, most influential training points to be the same in all cases.

In Fig.~\ref{fig:gq_rd} we show the distributions of the relative error between the \smoking prediction and the corresponding \prospino results, as well as the residual, for each individual flavour final state. We use the same two test sets, $\mcDtest$ and $\mcDMSSM$, as for the gluino pair-production cross-section in Fig.~\ref{fig:gg_rd}. All the distributions are normalised to unity.

The relative errors and residuals are similar across all squark flavours.  There are no obvious differences in performance with the two sets of test points. The relative error distributions show that for almost all points in the test sets, the true regression error is below 10\%, and \smoking tends to overestimate the \prospino cross-section by a few percent. Comparing the residual and $\mathcal{N}(0,1)$ distributions, we see that the predicted  \smoking regression uncertainty is conservative; indeed, notably more so than for the gluino pair-production cross-section (Fig.\ \ref{fig:gg_rd}).

\begin{figure*}[p]

\begin{minipage}[c]{\textwidth}
\includegraphics[width=0.5\textwidth]{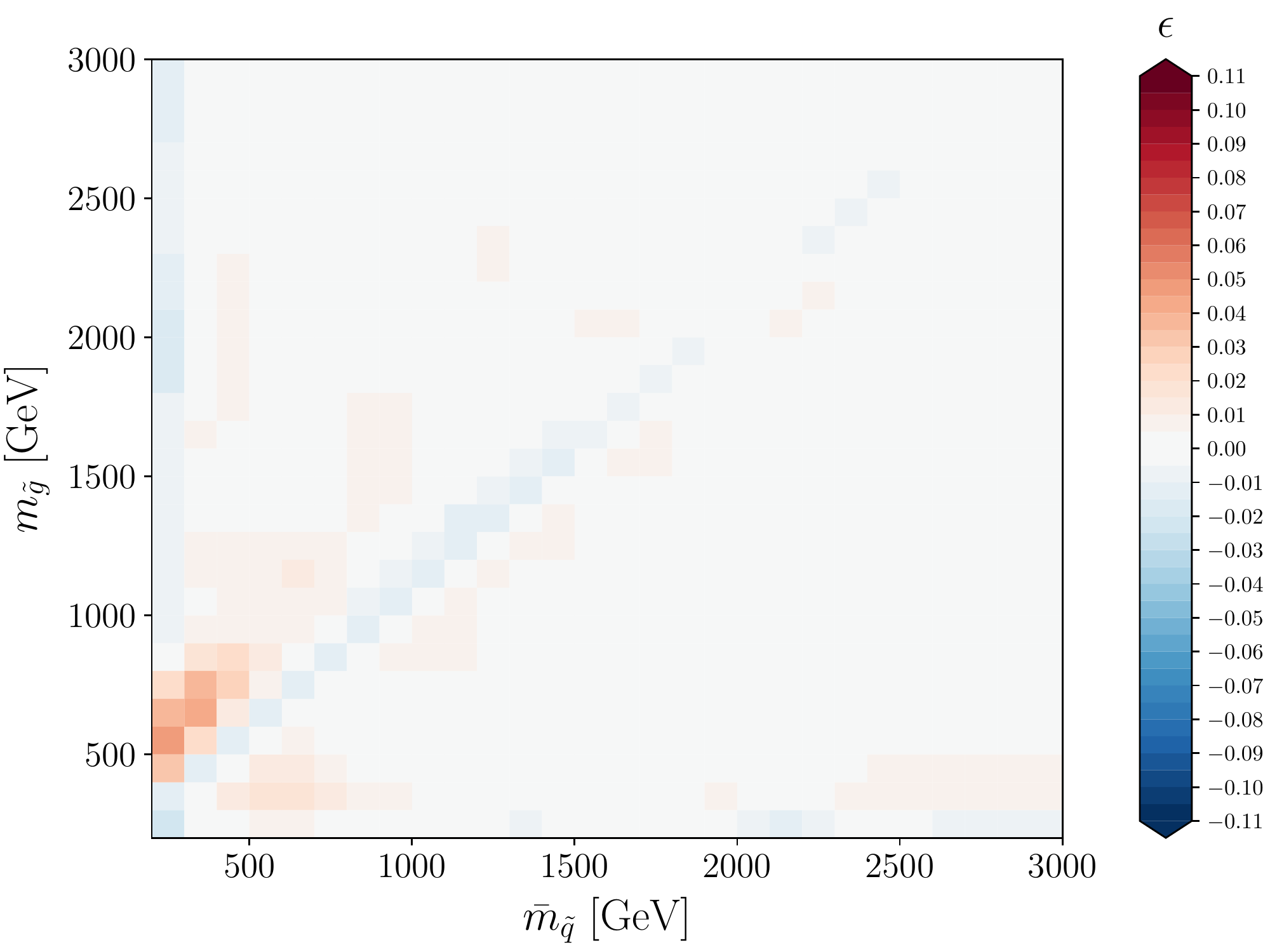}
\includegraphics[width=0.5\textwidth]{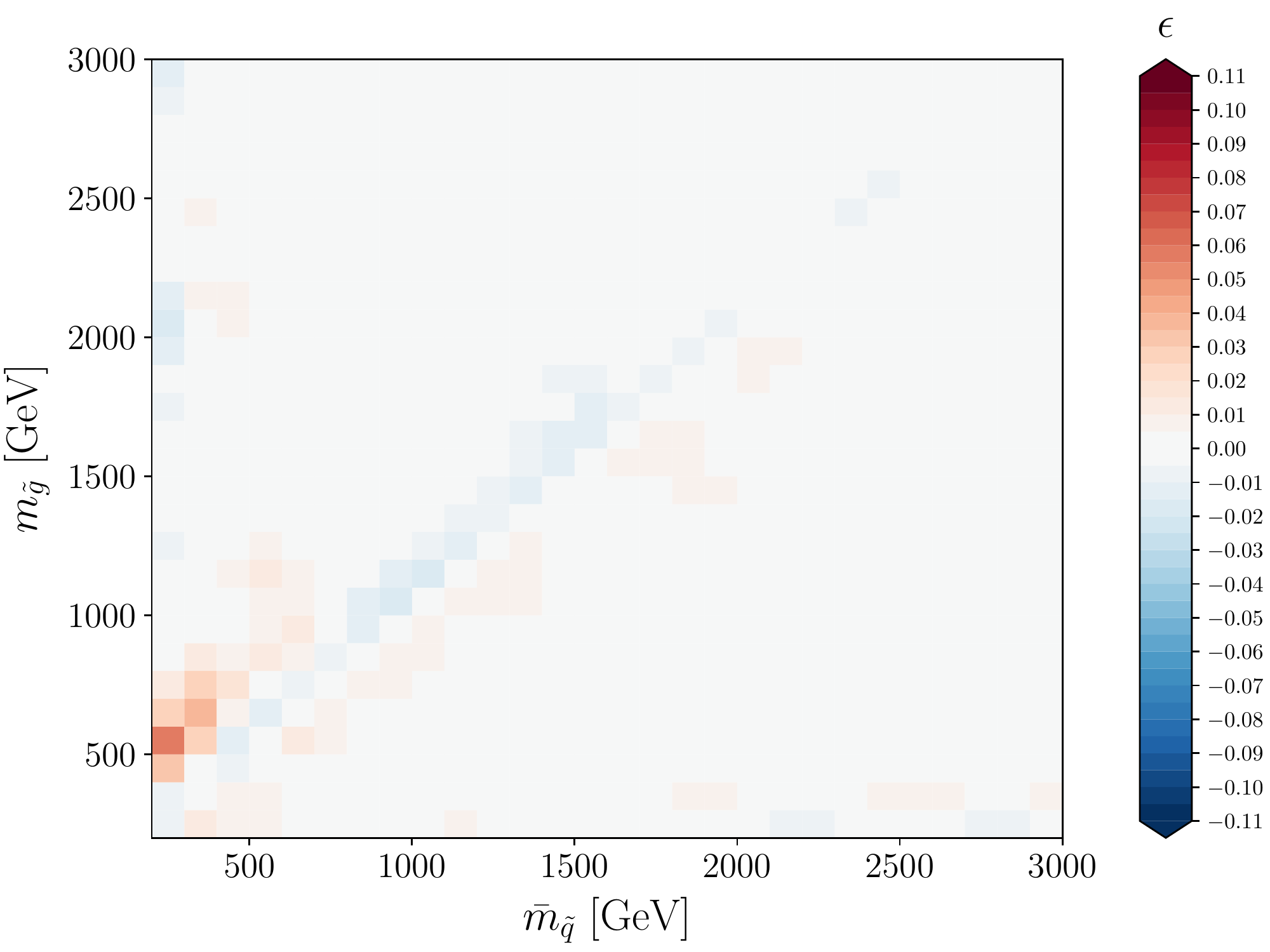}
\newline\includegraphics[width=0.5\textwidth]{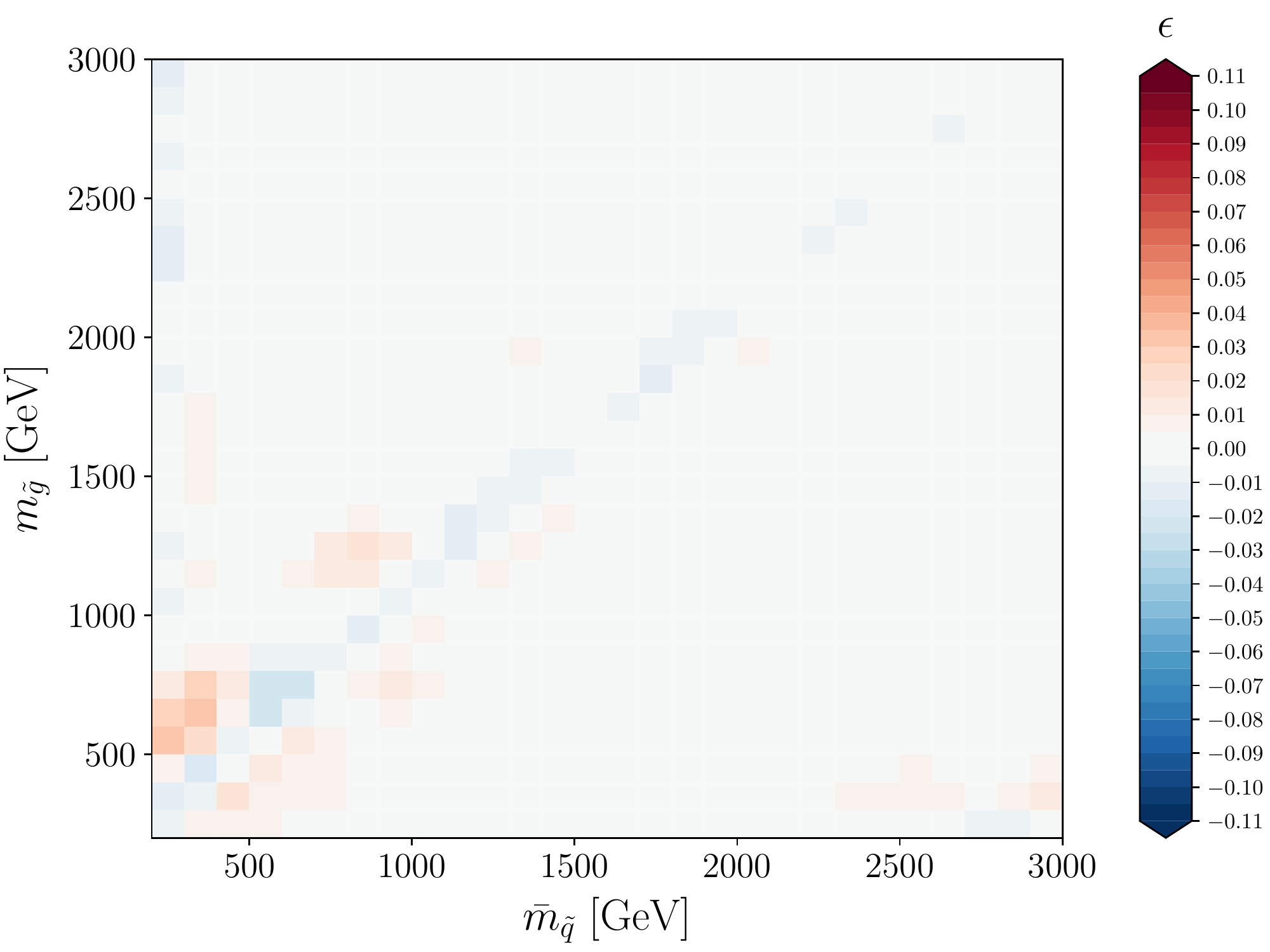}\includegraphics[width=0.5\textwidth]{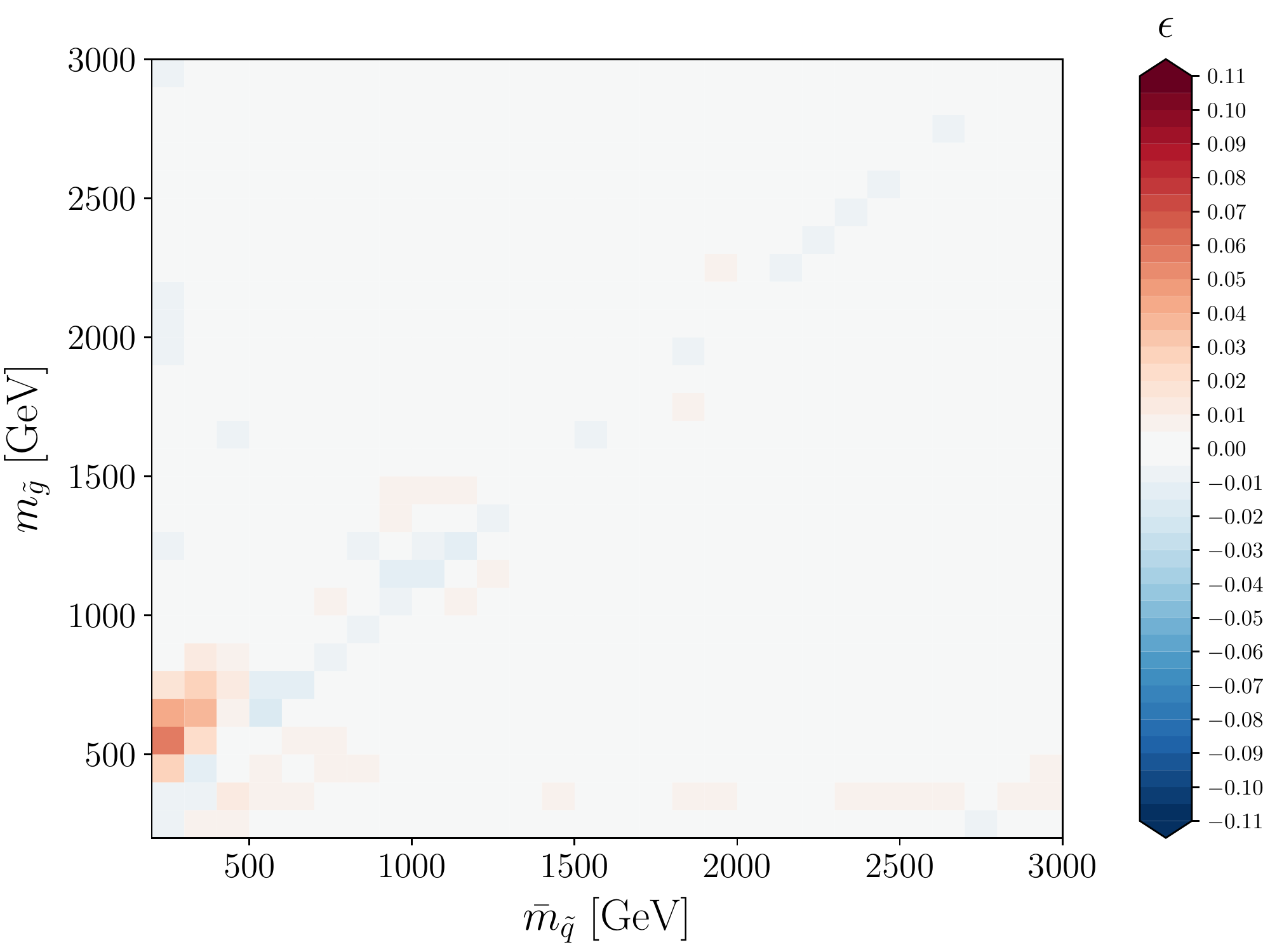}
\caption{The relative error of the production cross-section for gluinos and first or second-generation squarks, as a function of the average mass of first and second-generation squarks and the mass of the gluino. Shown are results for the production of $\tilde g \tilde d_L $ (top left), $\tilde g \tilde u_L $ (top right), $\tilde g \tilde s_L $ (bottom left), and $\tilde g \tilde c_L $ (bottom right). The final-state squark mass for each process is set equal to the average squark mass.}
\label{fig:gq_relative_error_massplanes}
\end{minipage}
\vspace{0.02\textheight}

\begin{minipage}[c]{\textwidth}
\includegraphics[width=0.5\textwidth]{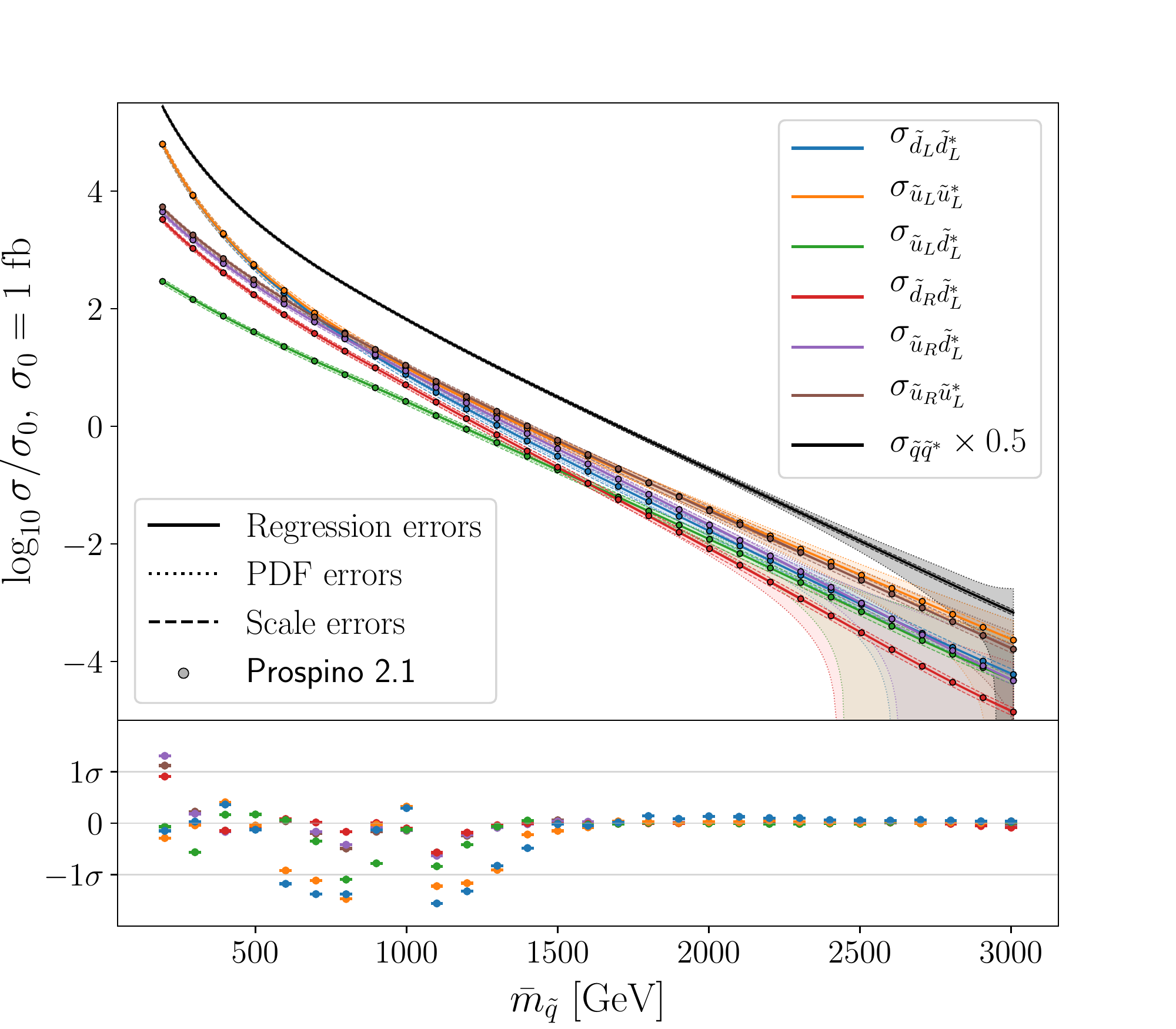}
\includegraphics[width=0.5\textwidth]{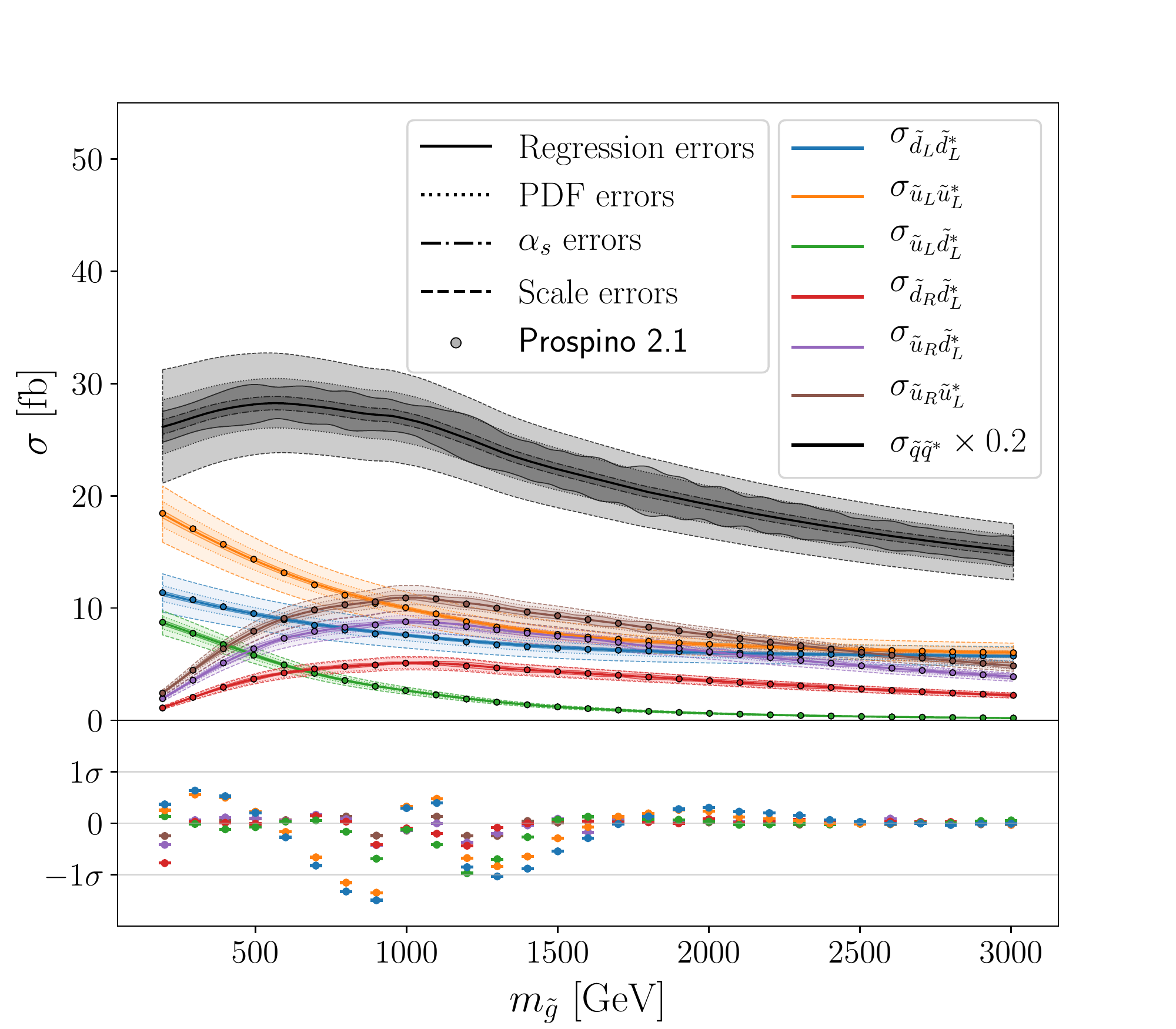}
\caption{Squark--anti-squark pair-production cross-sections for first and second-generation squarks.  Panels show cross-sections as a function of the average of all first and second-generation squark masses (left) and gluino mass (right), for a selection of final states (colours) and the total cross-section for production of first and second-generation squark--anti-squark pairs (black).}
\label{fig:qb_masses}
\end{minipage}

\end{figure*}

We can also compare the relative error across mass planes, which is shown in Fig.~\ref{fig:gq_relative_error_massplanes} separately for all $\tilde g\tilde q_L$ processes, in terms of the gluino mass and the average first and second-generation squark mass. Here the final-state squark mass for each plotted cross-section is set equal to the average squark mass. We see that the regression error is below 8\% across this plane for all four of the $\tilde g \tilde d_L $, $\tilde g \tilde u_L $, $\tilde g \tilde s_L $, and $\tilde g \tilde c_L $ production cross-sections.

\subsection{First and second-generation squark--anti-squark pair production}

We now look at the production cross-sections for squark--anti-squark pairs ${\tilde q}_{L/R} {\tilde q}^{(\prime)*}_{L/R}$. The flavours of the pair may be identical or different, and all four combinations of squark handedness are treated as separate processes. If the flavours of the two squarks are different, the process is assumed to include the charge-conjugate state. In this section we discuss only final states with first and second-generation (anti-)squarks, where the final-state flavour may come from first and second-generation (anti-)quarks sampled from the proton.

Within the limitations set by the training data from \prospino, the LO first and second-generation squark--anti-squark cross-sections depend on the masses of the gluino and the final-state squark(s), and the NLO corrections further on the mean mass of all first and second-generation squarks. Again, we validate the cross-section on the sub-interval $[200,3000]$\,\GeV of the training data (in both gluino and squark mass). For squark--anti-squark production, one should keep in mind that masses below the lower end of this range may be affected by resonant production through $Z$ and $W$, and while \smoking's reported regression error increases below 200\,\GeV, it cannot take these resonances into account, as they are not included in its training data.  The resulting cross-sections reported by \smoking must thus be seen as wholly unreliable for squark masses below 50\,\GeV.

In NLO QCD, cross-sections for the two sets of process pairs $({\tilde q}_L {\tilde q}_L^*, {\tilde q}_R{\tilde q}_R^*)$ and $({\tilde q}_R {\tilde q}_L^{\prime *},{\tilde q}'_R {\tilde q}_L^*)$ differ within each set only by an exchange of the appropriate squark masses. Removing also charge-conjugate states, an initial number of 64 independent processes $\tilde q_{L/R}\tilde q_{L/R}^{\prime *}$ therefore reduces to 20 unique cross-sections. To save training time and user disk space, we reuse the DGPs for the identical processes in \smoking simply by employing symbolic links and mapping the masses accordingly. This is however invisible to the user.

In Fig.~\ref{fig:qb_masses} we show the predicted first and second-generation squark--anti-squark production cross-sections as a function of the mean first and second-generation squark mass $\bar m_{\tilde q}$ (left) and the gluino mass (right).  We show results for a selection of sub-processes, and for the total cross-section (black, rescaled for readability). All other masses are kept at 1\,\TeV. We see that \smoking reliably predicts the contribution from individual squark final states, although at high squark masses the PDF error (dotted line) for some processes is consistent with zero cross-section. We also see that \smoking correctly captures the contribution of the gluino $t$-channel diagram, which controls the cross-section when the final-state squarks have different chirality, leading to the peak in Fig.~\ref{fig:qb_masses} (right) for L-R combinations when $m_{\tilde g}\simeq m_{\tilde q}$.

\begin{figure*}
\includegraphics[width=0.5\textwidth]{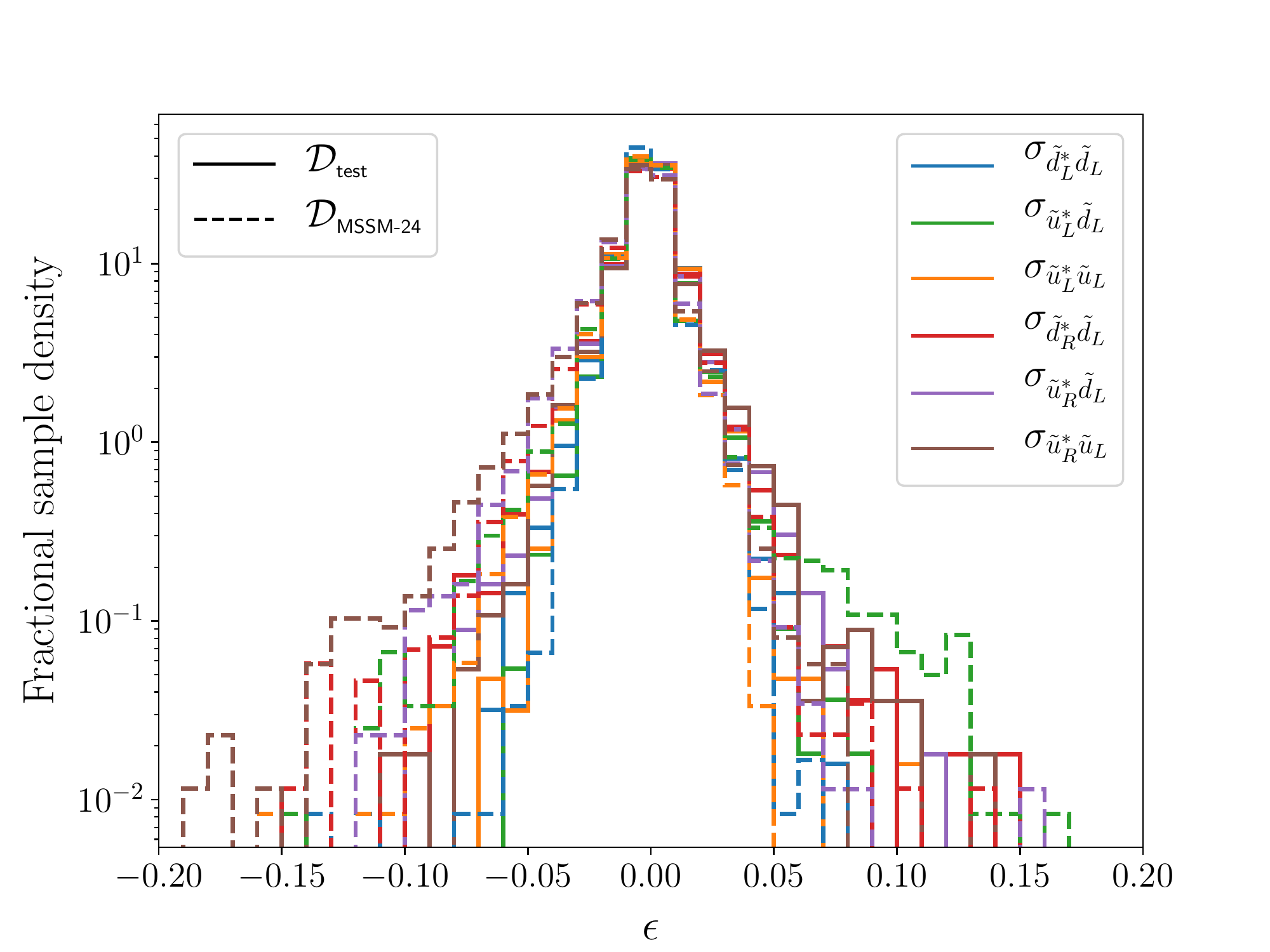}
\includegraphics[width=0.5\textwidth]{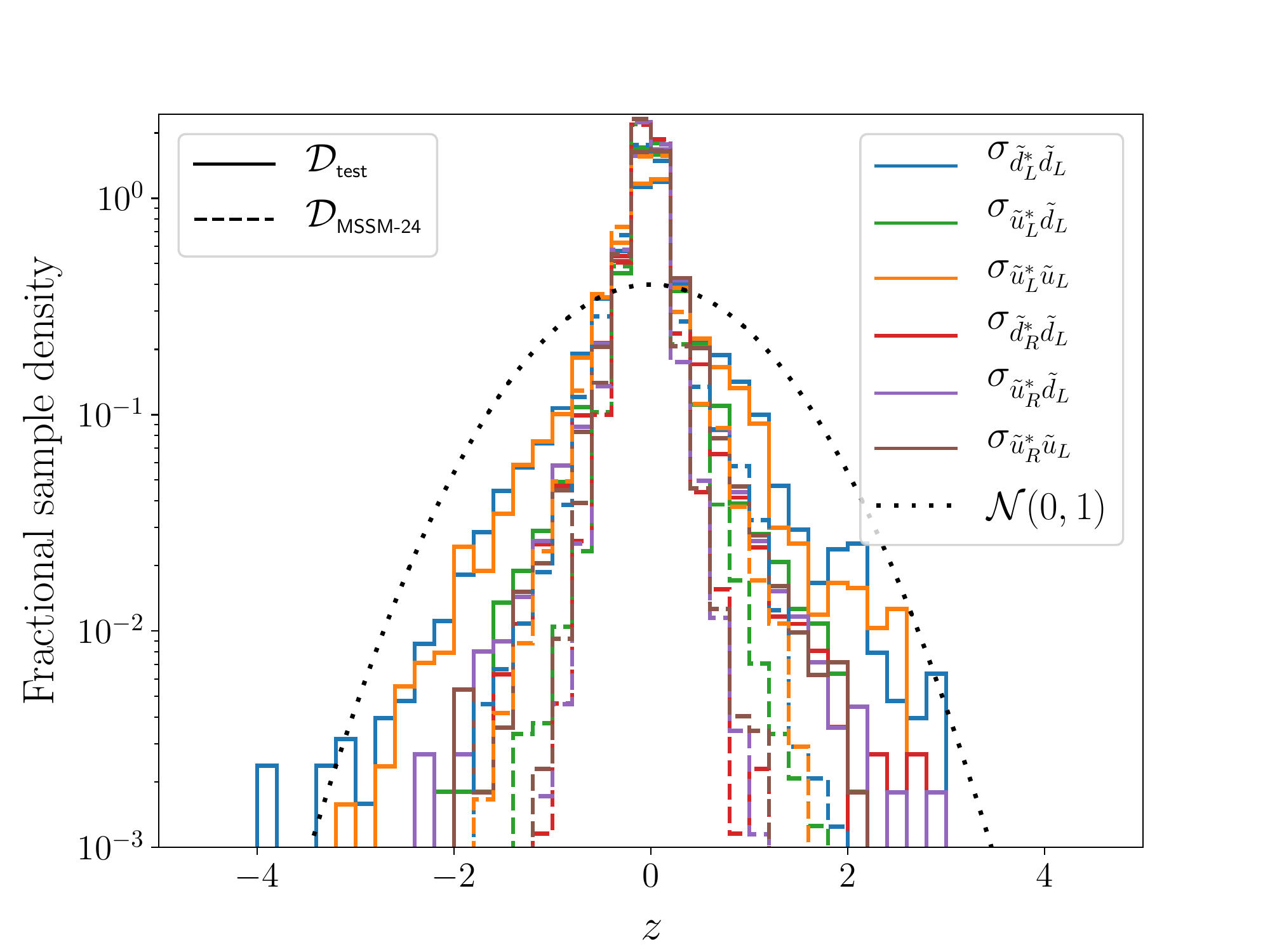}
\includegraphics[width=0.5\textwidth]{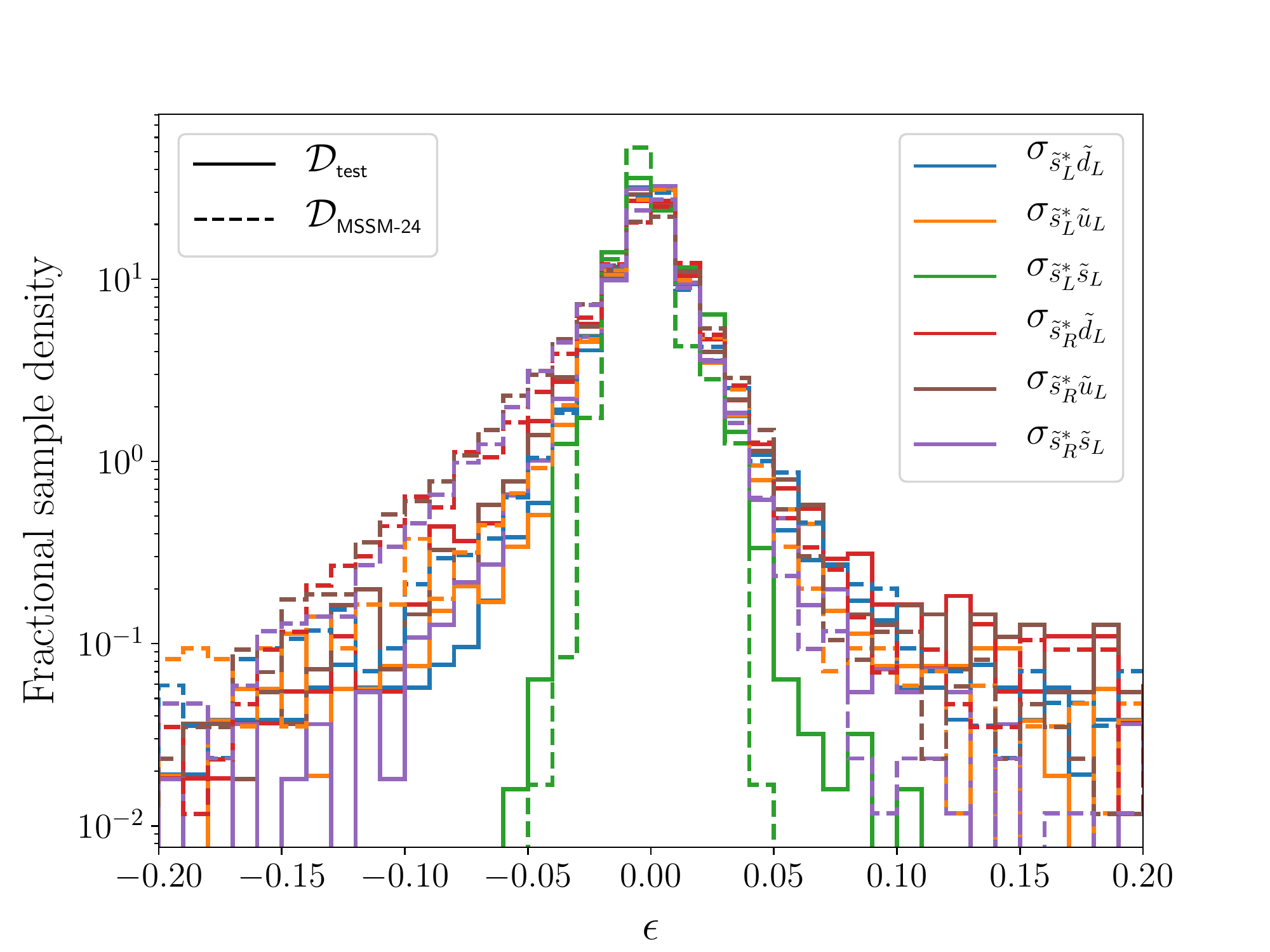}
\includegraphics[width=0.5\textwidth]{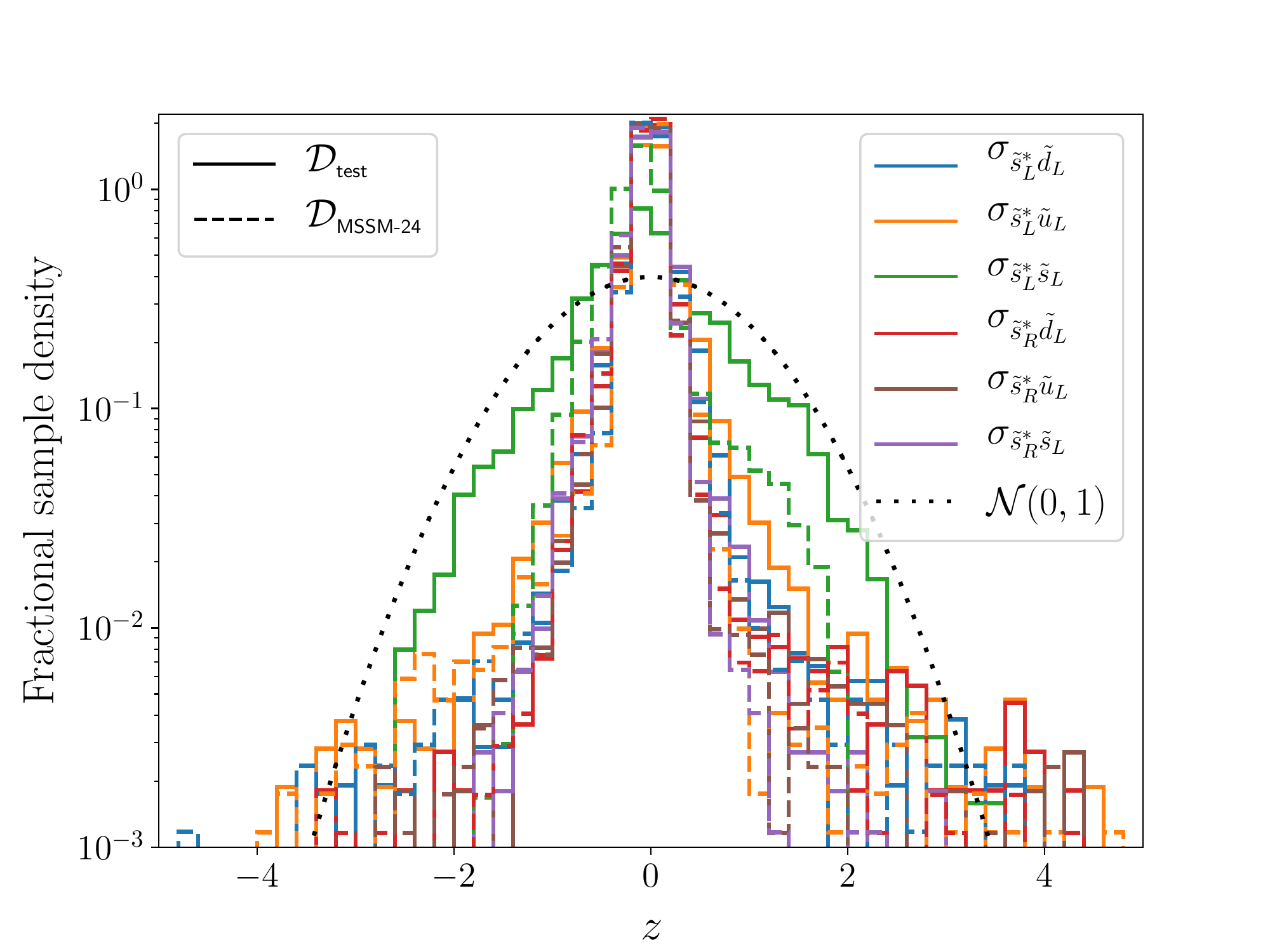}
\includegraphics[width=0.5\textwidth]{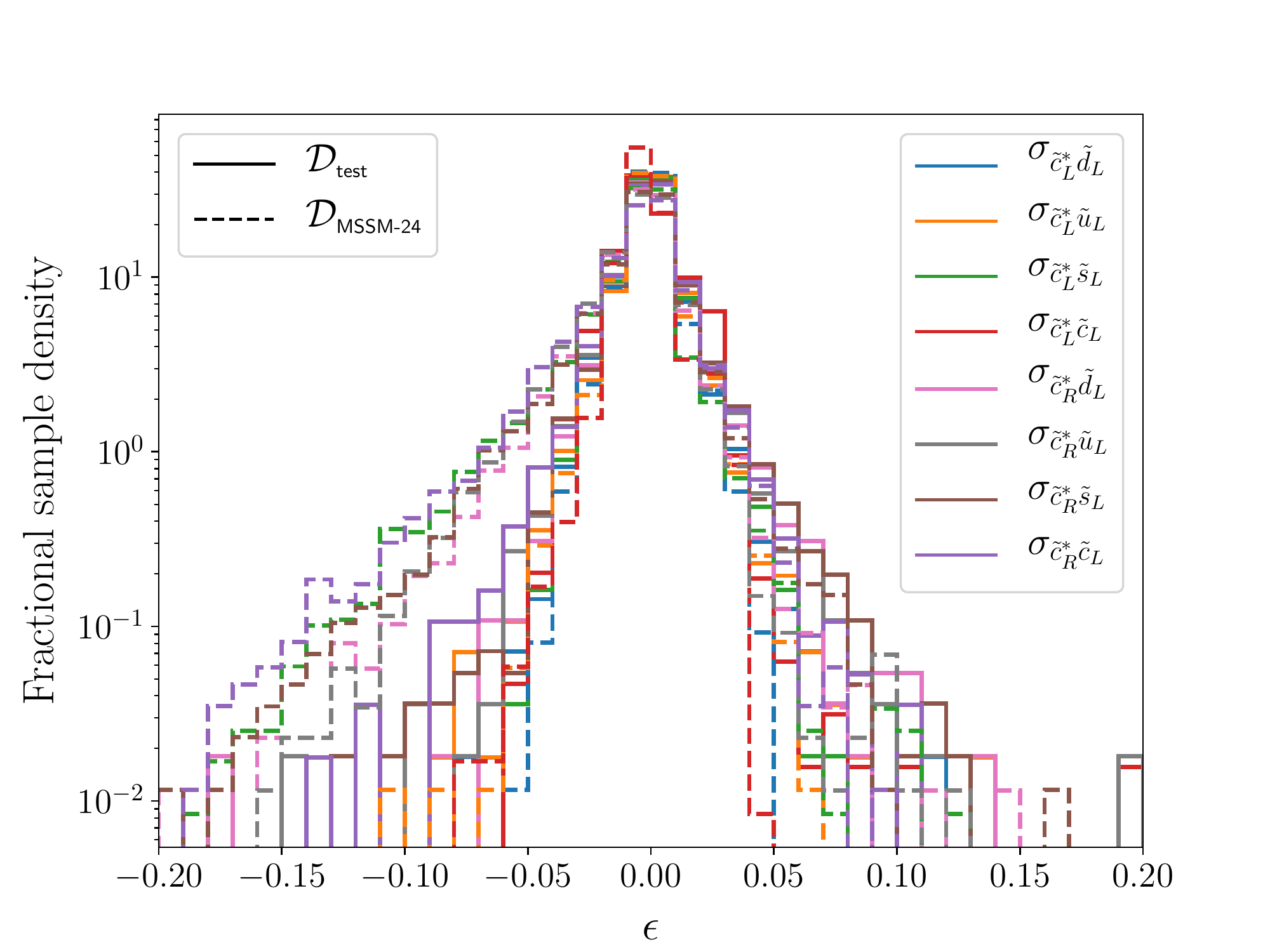}
\includegraphics[width=0.5\textwidth]{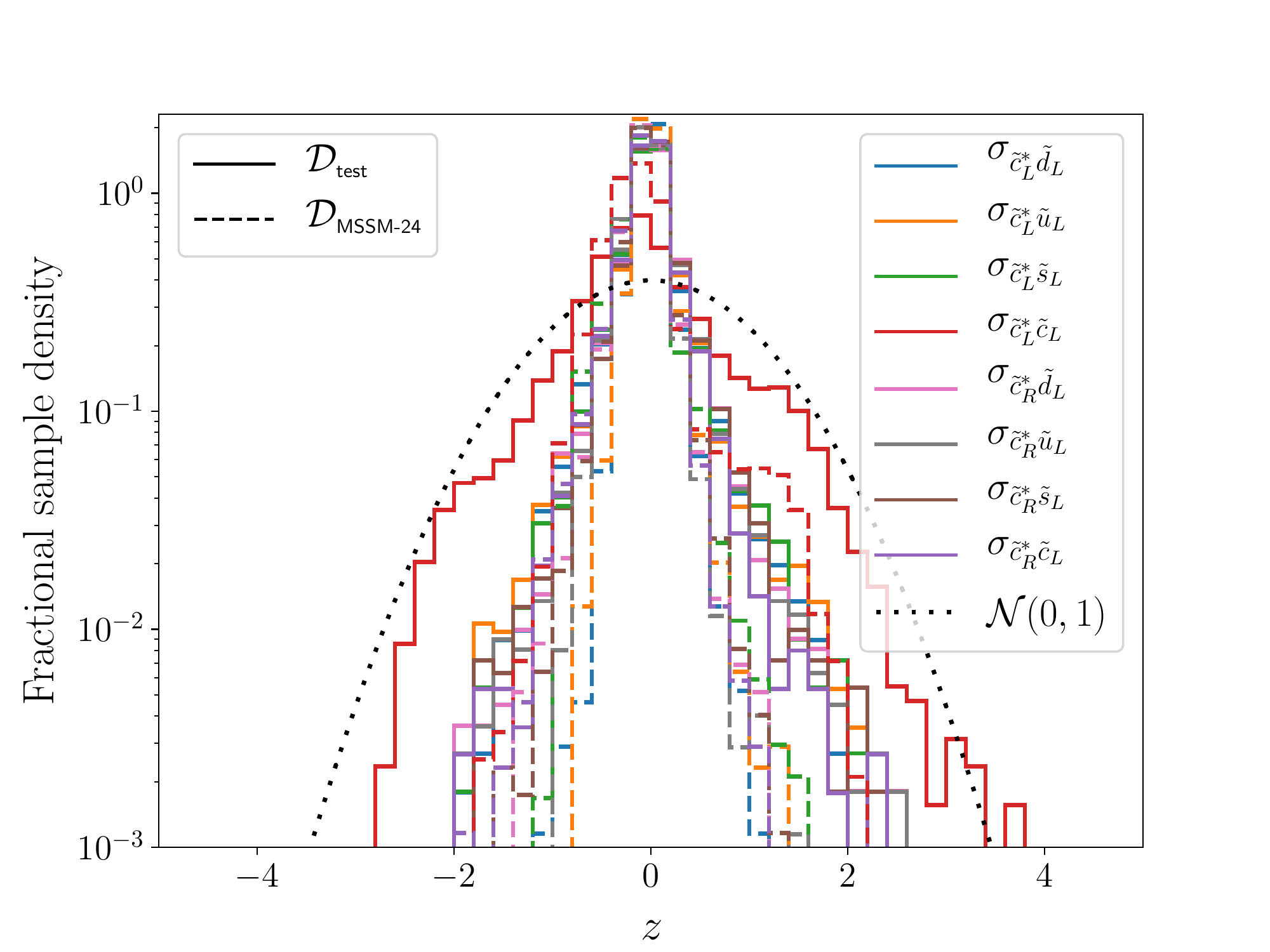}
\caption{The relative error (left) and residual (right) distributions for the first and second-generation squark--anti-squark cross-sections for the test sets $\mcDtest$ (solid) and $\mcDMSSM$ (dashed).
\label{fig:qb_rd}}
\end{figure*}

In Fig.~\ref{fig:qb_rd} we show the distributions of the relative regression error and residual (Eqs.~(\ref{eq:xsec_relative_error}) and (\ref{eq:xsec_residual})) for the points in the test sets $\mcDtest$ and $\mcDMSSM$. We have normalised all distributions to unity.
The comparison to the unit normal distribution included in Fig.~\ref{fig:qb_rd} (right column) shows that for both test sets, and all processes, the \smoking regression error is conservative with respect to the true error.

\begin{figure*}[t!]
\includegraphics[width=0.5\textwidth]{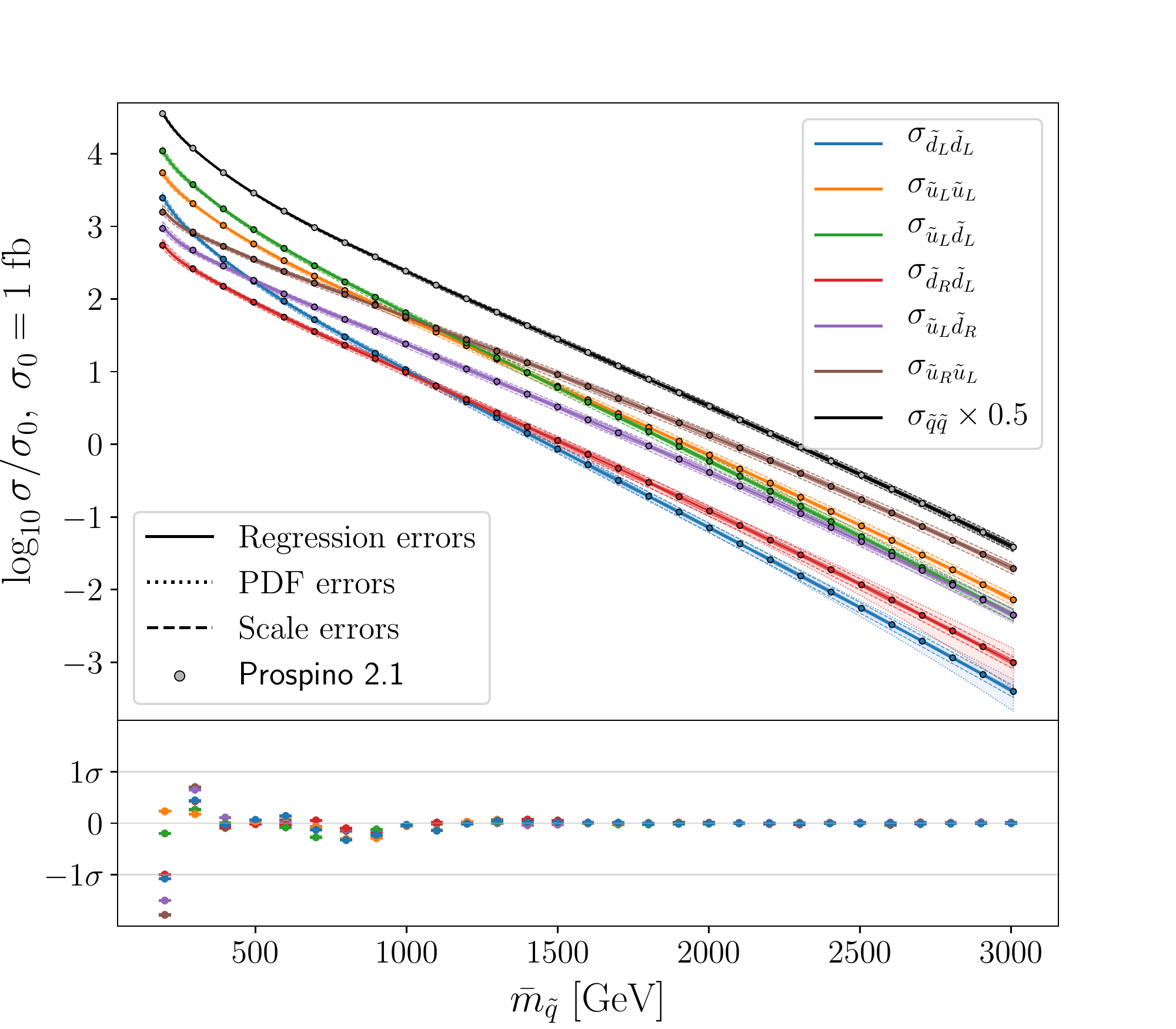}
\includegraphics[width=0.5\textwidth]{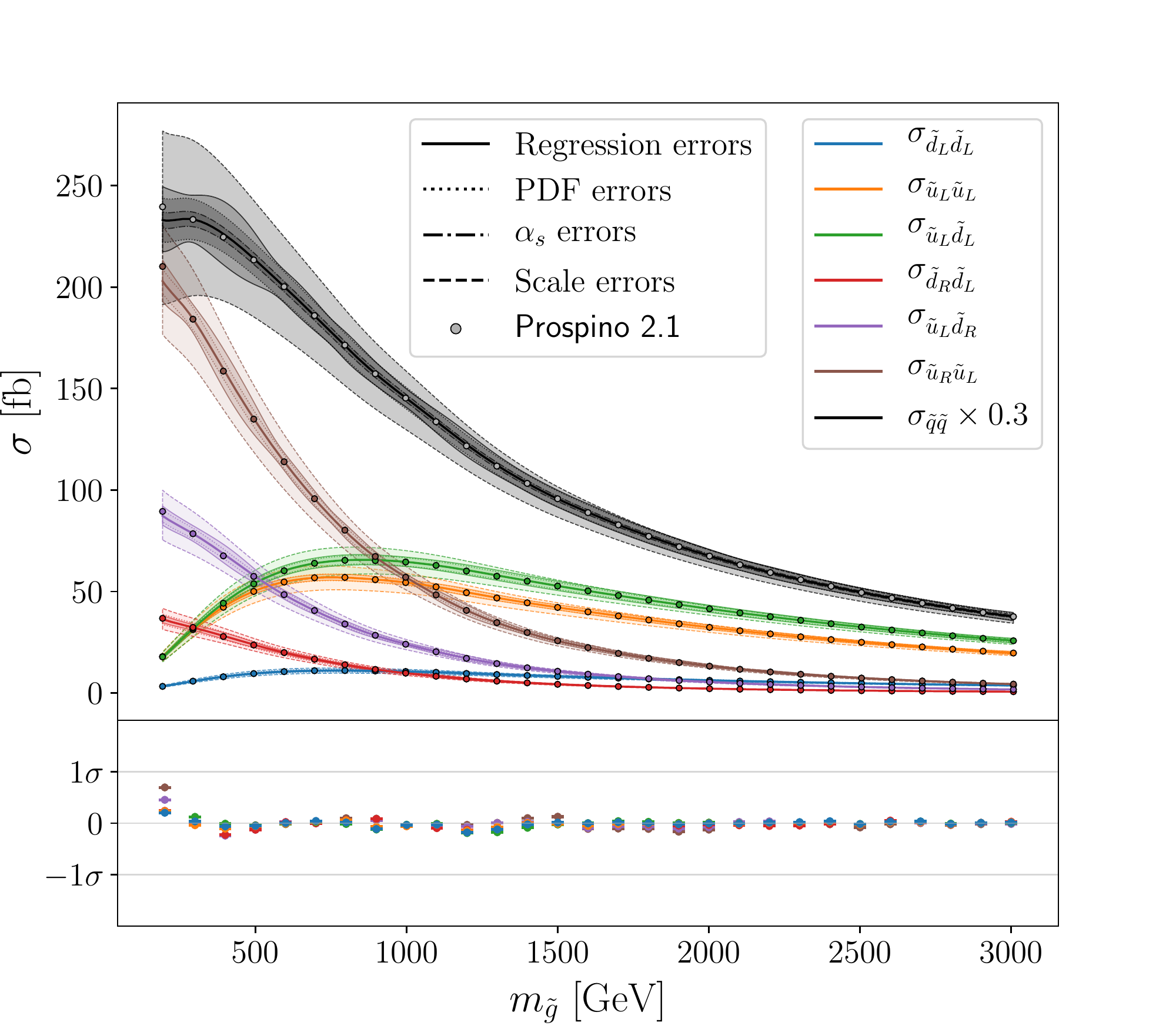}
\caption{First and second-generation squark--squark pair-production cross-section as a function of the average first and second-generation squark mass (left) and gluino mass (right), for a selection of final states (colours) and the (rescaled) total cross-section for first and second-generation squark--squark production (black). }
\label{fig:qq_masses}
\end{figure*}

The relative regression error in Fig.~\ref{fig:qb_rd} (left column) is below 10\% for all processes for the vast majority of test points.
There is again a slight tendency for \smoking to overestimate the \prospino cross-section values, in particular for the $\mcDMSSM$  test set.
As this set has individual flat priors for all the squark (soft) masses, which only a subset of $\mcDtest$ has, we expect it to be more challenging to reproduce as its points are more likely to lie on the outskirts of the validation region.

The relative error distributions are most narrow for the processes producing a squark--anti-squark pair of the same type, \ie, for the four $\tilde q_L^* \tilde q_L$ processes and the corresponding $\tilde q_R^* \tilde q_R$ processes (not shown). These cross-sections are easier to model, as the final state involves only a single mass parameter. The $\tilde s_L^* \tilde s_L$ and $\tilde c_L^* \tilde c_L$ processes have particularly small relative errors. This is likely due to the smallness of the proton PDFs for the $s$ and $c$ quarks, which effectively makes the gluino $t$-channel diagram irrelevant and thus further simplifies the parameter dependence.

\subsection{First and second-generation squark pair production}

This section looks at the validation of DGPs trained to predict cross-sections for squark--squark pair production, ${\tilde q}_{L/R} {\tilde q}^{(\prime)}_{L/R}$. As should be clear from the notation, the flavours of the pair may be identical or different, and all four combinations of squark handedness are treated as separate processes. The processes are always assumed to include the charge-conjugate state. Again, we discuss only final states with first and second-generation squarks, where at LO the final-state flavour comes from first and second-generation quarks in the proton.

As for squark--anti-squark production, within the limitations set by the design of \prospino, the LO first and second-generation squark--squark pair-production cross-sections depend on the mass(es) of the final-state squarks and the gluino mass, and the NLO corrections further depend on the mean mass of the first and second-generation squarks. We again validate our cross-sections on the sub-interval $[200,3000]$\,\GeV of the training data, for both gluino and squark masses.

If the squark masses are interchanged appropriately, the cross-sections for the process pairs $({\tilde q}_L {\tilde q}_L, {\tilde q}_R{\tilde q}_R)$ are identical in \prospino, as are those for the process pairs $({\tilde q}_R {\tilde q}_L^{\prime},{\tilde q}'_R {\tilde q}_L)$. The 64 independent processes $\tilde q_{L/R}\tilde q_{L/R}^{\prime}$ can therefore be reduced to 20 unique cross-sections. Again we use symbolic links to reuse DGPs for processes with identical cross-sections.

In Fig.~\ref{fig:qq_masses} we show the predicted first and second-generation squark--squark production cross-sections for a selection of sub-processes, as a function of $\bar m_{\tilde q}$ and $m_{\tilde g}$. All other masses are kept at 1\,\TeV. For readability, the total cross-section is rescaled. We see that \smoking reliably predicts the contribution from individual squark final-state flavours, and captures the contribution of the gluino $t$-channel diagram, which depends on the chirality of the final-state squarks, giving the qualitatively different behaviour of the cross-section as a function of the gluino mass seen in the right panel of Fig.~\ref{fig:qq_masses}.

\begin{figure*}
\includegraphics[width=0.5\textwidth]{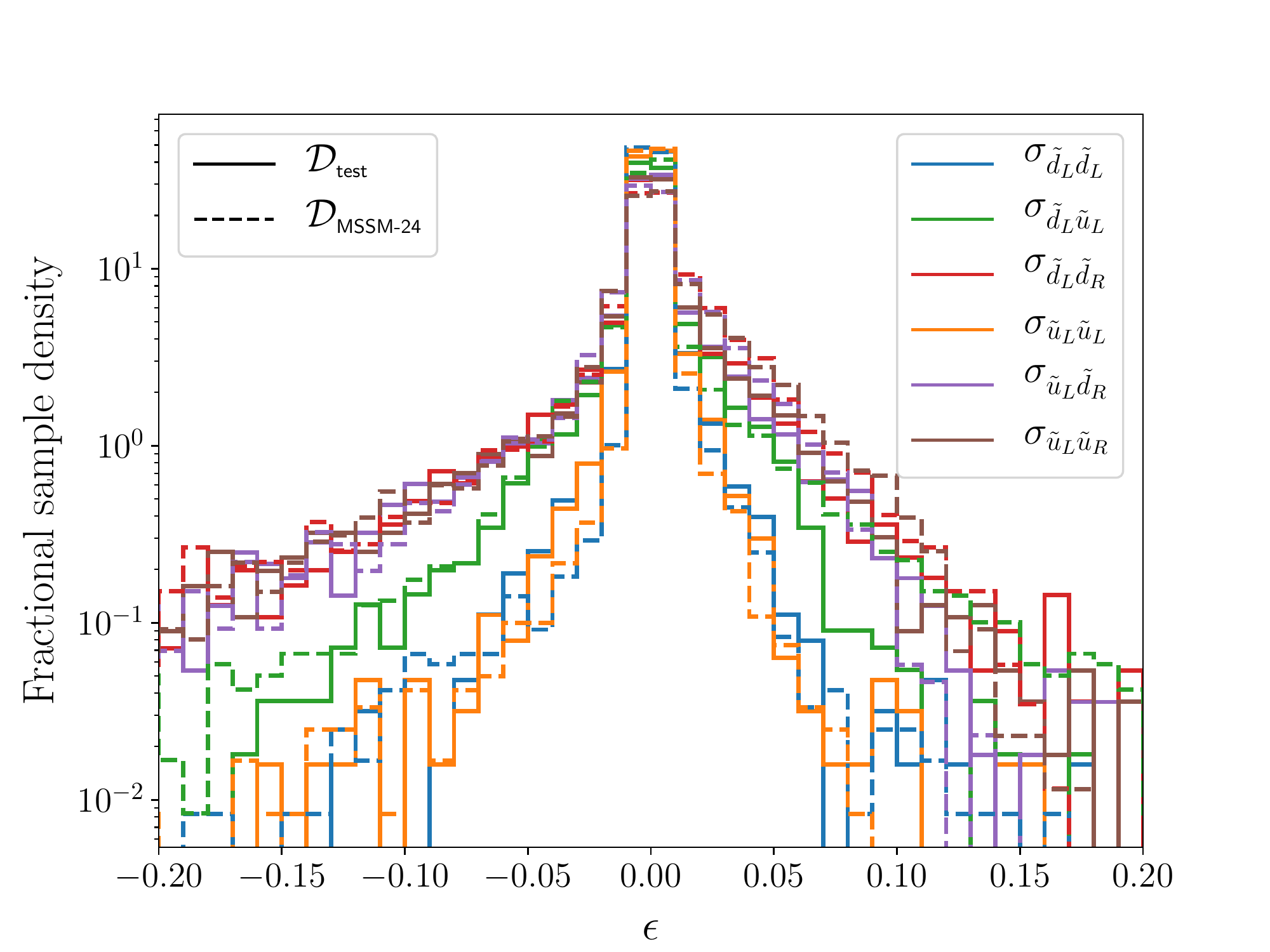}
\includegraphics[width=0.5\textwidth]{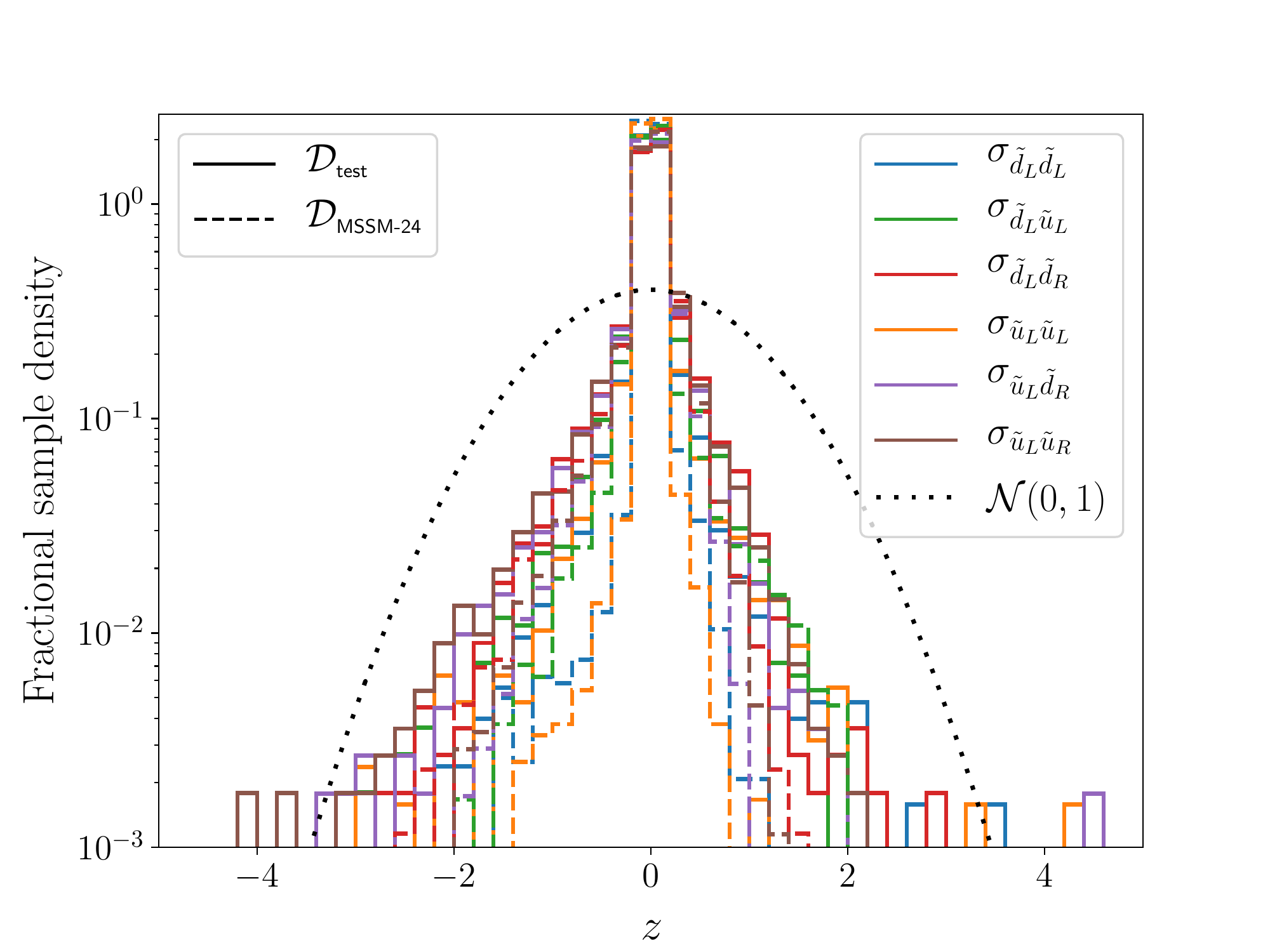}
\includegraphics[width=0.5\textwidth]{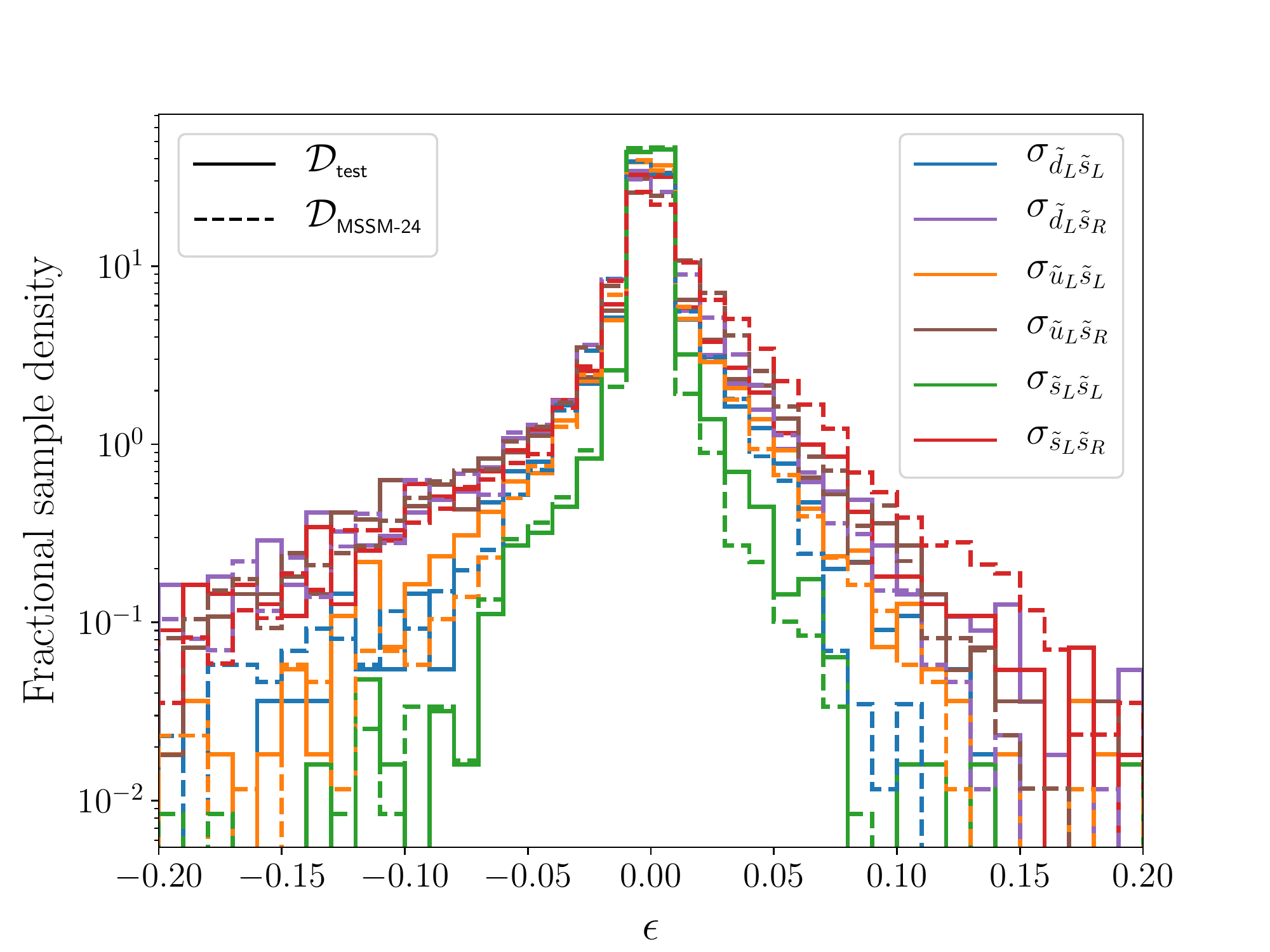}
\includegraphics[width=0.5\textwidth]{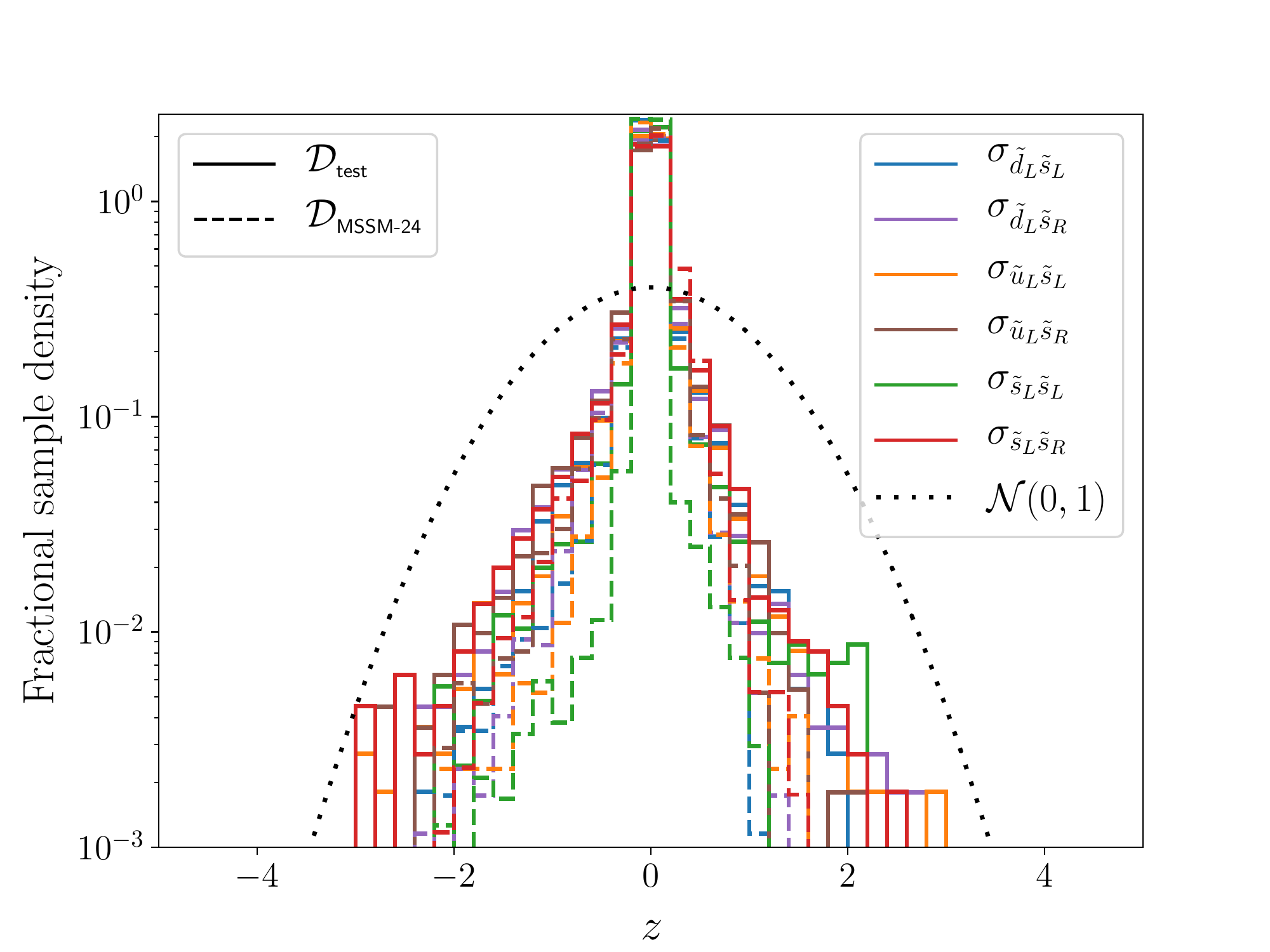}
\includegraphics[width=0.5\textwidth]{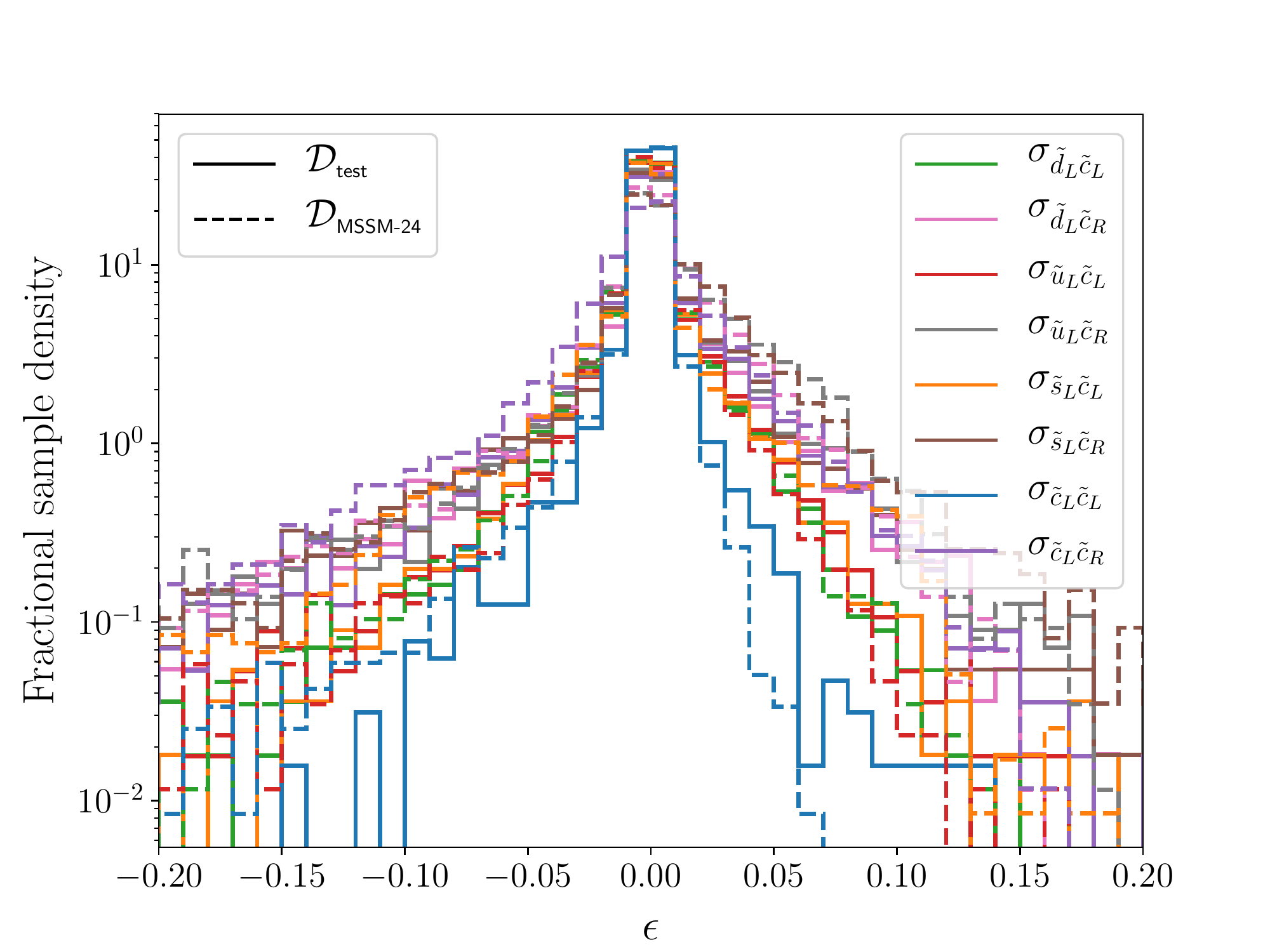}
\includegraphics[width=0.5\textwidth]{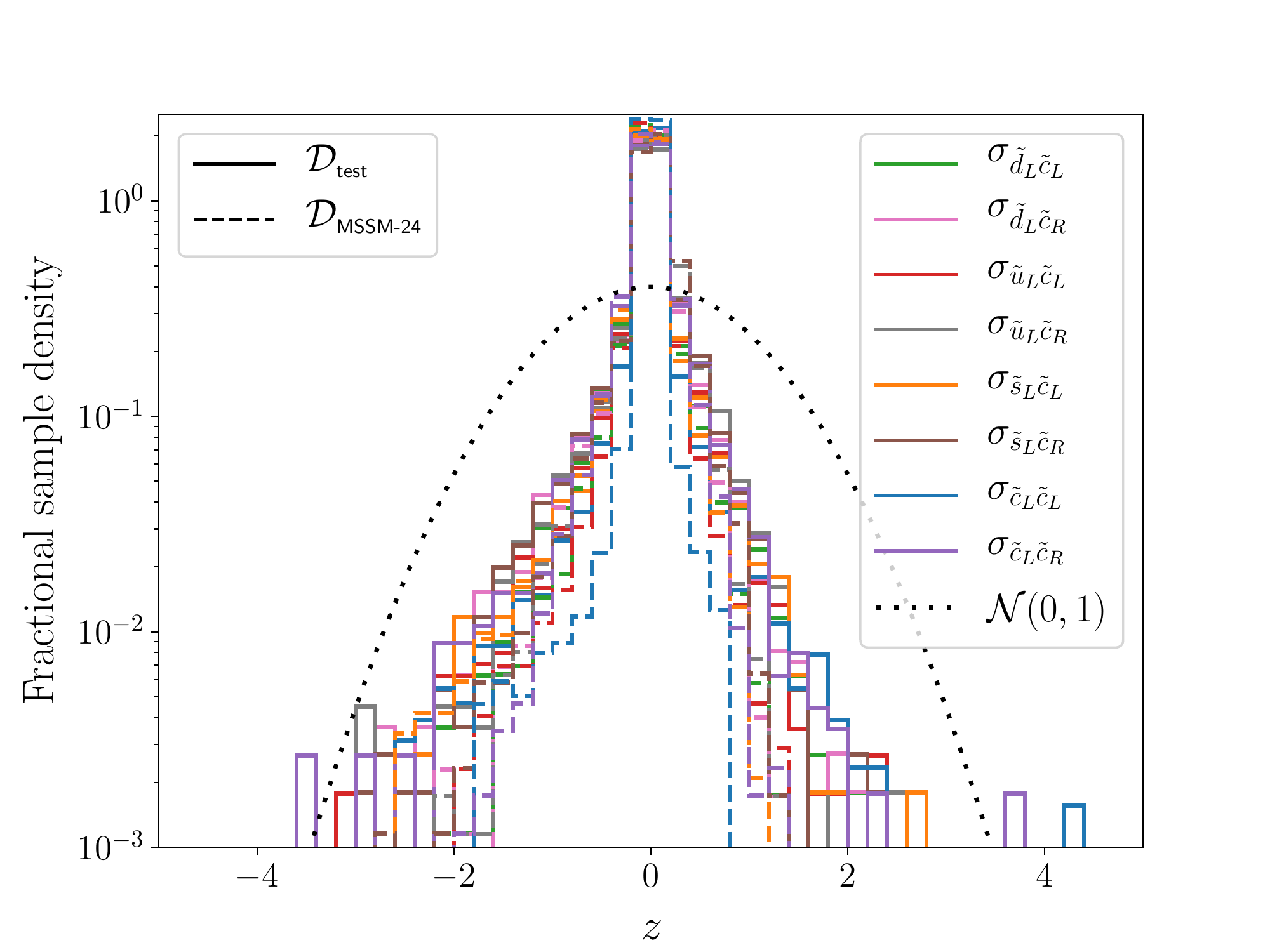}
\caption{The relative error (left) and residual (right) distributions for first and second-generation squark--squark pair-production cross-sections for the test sets $\mcDtest$ (solid) and $\mcDMSSM$ (dashed).}
\label{fig:qq_rd}
\end{figure*}

In Fig.~\ref{fig:qq_rd} we show the corresponding distributions of the relative error and residual between the \smoking prediction and the \prospino central cross-section values in $\mcDtest$ and $\mcDMSSM$, for each individual squark--squark process.  The residuals indicate that the regression errors from \smoking across all squark--squark processes are generally conservative compared to the actual difference between the true and predicted cross-sections. The relative errors are below 10\% for the large majority of test points, across all processes. There is some interesting process-dependent structure, however. For processes with two identical final-state squarks, e.g.\ $\tilde u_L\tilde u_L$ or $\tilde s_L\tilde s_L$, the relative error is quite small, due to the dependence of the cross-section upon only three instead of four masses, making it easier to predict reliably. Processes with chirally-matched squarks, i.e.\ LL or RR final states, have smaller errors than LR final states. This is due to a difference in the LO matrix element, where the LR final states depend on $(t-m_{\tilde q}^2)(u-m_{\tilde q}^2)-m_{\tilde q}^2s$, whereas LL and RR final states are proportional to $m_{\tilde g}^2s$~\cite{Ingrid}.  This more complicated kinematic dependence means that the LR final states are harder to train and show larger relative errors.

\begin{figure*}[t!]
\begin{minipage}[t]{0.485\textwidth}
\centering
\includegraphics[width=\textwidth]{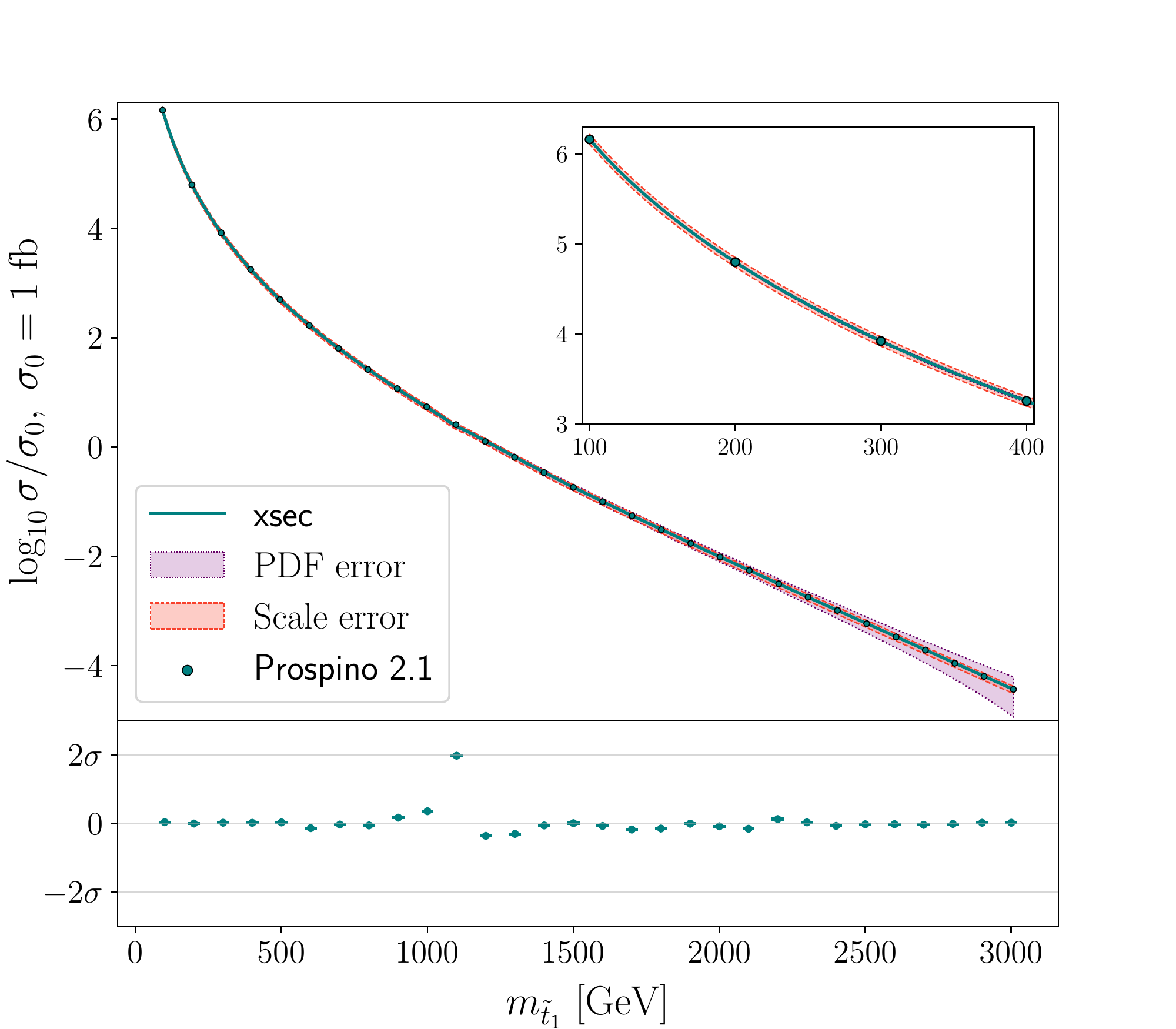}
\caption{Stop pair-production cross-section as a function of stop mass. The central-value \smoking prediction is shown in light green, the scale error in pink, and the PDF error in violet.  The $1\sigma$ $\alpha_s$ and regression error bands are too small to be visible. Superimposed on the prediction are the \prospino values (dots). Inset is a close-up of the region at low stop masses.}
\label{fig:tb_tbmass}
\end{minipage}%
\hspace{0.03\textwidth}
\begin{minipage}[t]{0.485\textwidth}
\includegraphics[width=\textwidth]{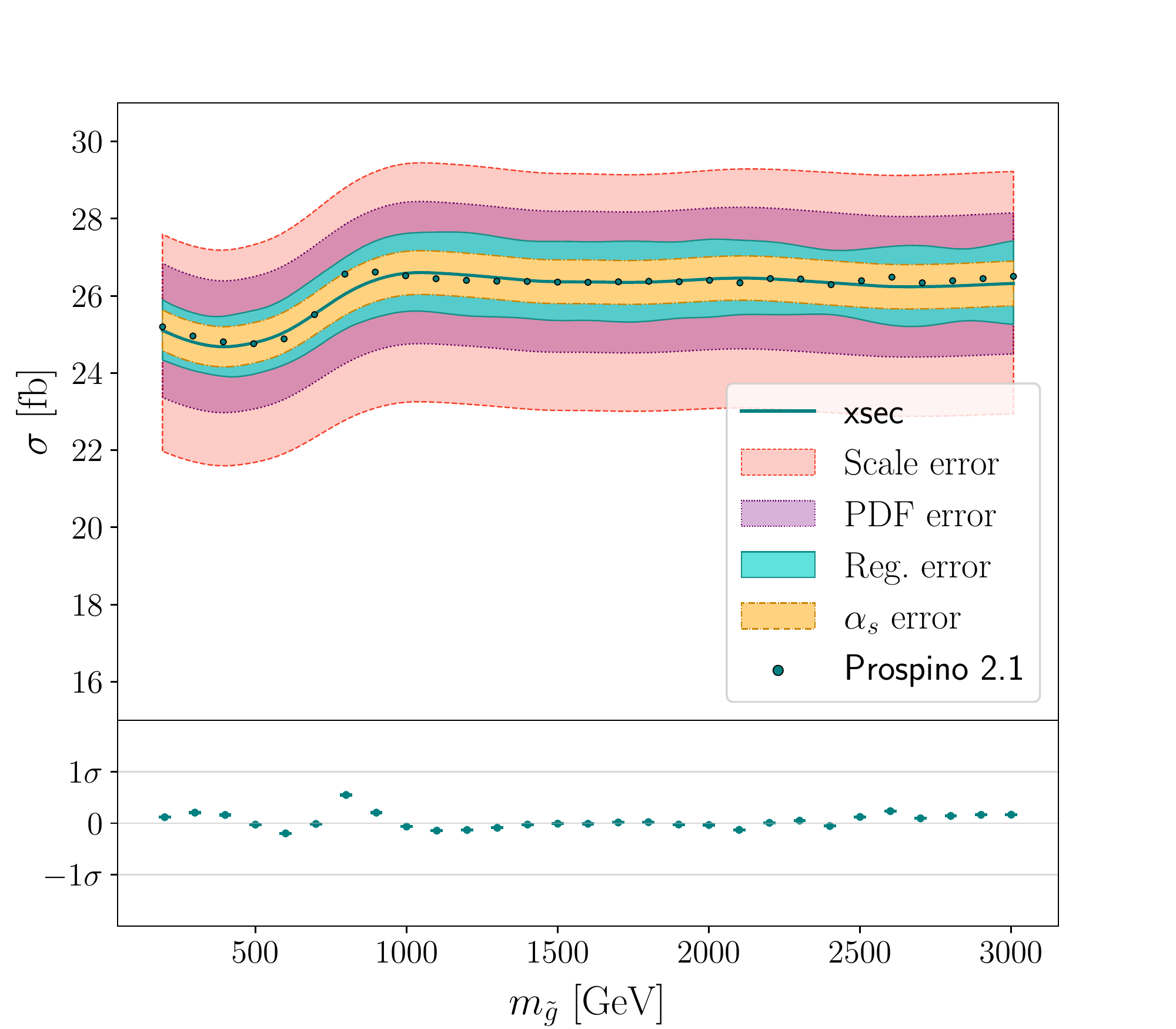}
\caption{Stop pair-production cross-section as a function of gluino mass for $m_{{\tilde t}_1}=800$\,\GeV, with all other squark masses set to 1\,\TeV.  The central-value \smoking prediction and the $1\sigma$ regression error band is shown in light green, the scale error in pink, the PDF error in violet, and the $\alpha_s$ error in yellow. Below we show the residual between the \smoking and \prospino results.}
\label{fig:tb_gmass}
\end{minipage}
\end{figure*}

\subsection{Stop and sbottom pair production}

\prospino includes PDF contributions only from light quarks and gluons.  This means that production of third-generation squarks is purely $s$-channel in our training samples, so only flavour-neutral squark--anti-squark final states are available, i.e.\ there are no processes with third-generation squarks in gluino--squark nor squark--squark production. However, the third-generation mass eigenstates used by \prospino include left-right mixing, quantified through the sbottom and stop mixing angles $\cos\theta_{\tilde b}$ and $\cos\theta_{\tilde t}$, where $\cos\theta_{\tilde t}=1$ implies that $\tilde t_1,\tilde t_2=\tilde t_L,\tilde t_R$, and similarly for sbottoms. Furthermore, \prospino does not calculate cross-sections for the production of mixed $\tilde t_1\tilde t_2^* + \textrm{c.c.}$\ (or $\tilde b_1\tilde b_2^* + \textrm{c.c.}$) states, as their cross-sections are of order $\alpha_s^4$, resulting in negligibly small rates \cite{Beenakker:1997ut}. As a result, \smoking is limited to training the third-generation processes $\tilde b_1\tilde b_1^*$,  $\tilde b_2\tilde b_2^*$, $\tilde t_1\tilde t_1^*$, and $\tilde t_2\tilde t_2^*$.

For the stop and sbottom pair-production cross-sections, only the final-state mass enters at LO as a parameter (feature). To NLO in QCD, \ie up to $\mathcal O(\alpha_s^3)$, the cross-sections depend on the stop or sbottom final-state mass, the gluino mass, the averaged first and sec\-ond-gen\-eration squark mass, and the stop or sbottom mixing angle, in roughly descending order of importance. Unlike the processes involving first and second-generation squarks, the sbottom and stop cross-sections also include the dependence on all the individual sbottom and stop masses in loops. However, for sbottom (stop) final states, the dependence on the stop (sbottom) masses was found to be very small, and we do not use them as features in the GPs.

For the NLO contributions involving heavy-quark loops, we have used top and bottom masses of $m_t=172.0$\,\GeV and $m_b=4.6$\,\GeV.  However, we have checked that the effect of changing these within current experimental uncertainties is numerically irrelevant.

Due to the symmetry of the cross-section expressions at NLO, $\sigma_{\tilde t_1\tilde t_1^*}\leftrightarrow \sigma_{\tilde t_2\tilde t_2^*}$ under the combined interchange
\begin{equation}
\begin{Bmatrix}
m_{\tilde t_1} \\
\cos\theta_{\tilde t} \\
\sin\theta_{\tilde t}
\end{Bmatrix}
\leftrightarrow
\begin{Bmatrix}
m_{\tilde t_2} \\
-\sin\theta_{\tilde t} \\
\cos\theta_{\tilde t}
\end{Bmatrix},
\end{equation}
and similarly for sbottoms. There is also very little difference between the sbottom and stop cross-sections for the same masses. Nevertheless, \smoking contains separate DGPs for all four non-zero stop and sbottom pair-production processes included in \prospino, in preparation for future extensions. Here we will focus our validation tests on the $\tilde t_1\tilde t_1^*$ cross-sections, and only discuss the other cross-sections when there are significant differences. We validate the \smoking output for the third-generation squarks over a slightly larger mass range than for the other processes ($[100,\,3000]$\,\GeV), as there is still considerable interest in light stops.

In Fig.~\ref{fig:tb_tbmass}, we show the stop pair-production cross-section predicted by \smoking as a function of the stop mass, which is by far the dominant parameter in determining the cross-section. We also show the predicted scale and PDF uncertainties, and compare to values taken directly from \prospino. In generating the plot, we have set all stop and sbottom masses degenerate, $\cos\theta_{\tilde t}=1$, and all other masses to 1\,\TeV. At low stop masses we can see that the scale error dominates, while at high masses it is the PDF error, as expected. Throughout the validation range the regression error is subdominant.

The gluino mass can be important for contributions that appear at NLO. In Fig.~\ref{fig:tb_gmass} we show the dependence of the $\tilde t_1\tilde t_1^*$ cross-section on the gluino mass, adopting a stop mass of $m_{{\tilde t}_1}=800$\,\GeV, and 1\,\TeV for all other masses. Clearly, \smoking fully captures the variation in the cross-section due to the gluino contribution at low gluino masses, although this dependence is rather small compared to the scale, PDF, and even $\alpha_s$ uncertainties, which we also show. The \prospino result seems to have some numerical jitter at high gluino masses, although the effect lies well within the regression uncertainty band.

\begin{figure*}[p!]

\begin{minipage}[t]{0.485\textwidth}
\includegraphics[width=\textwidth]{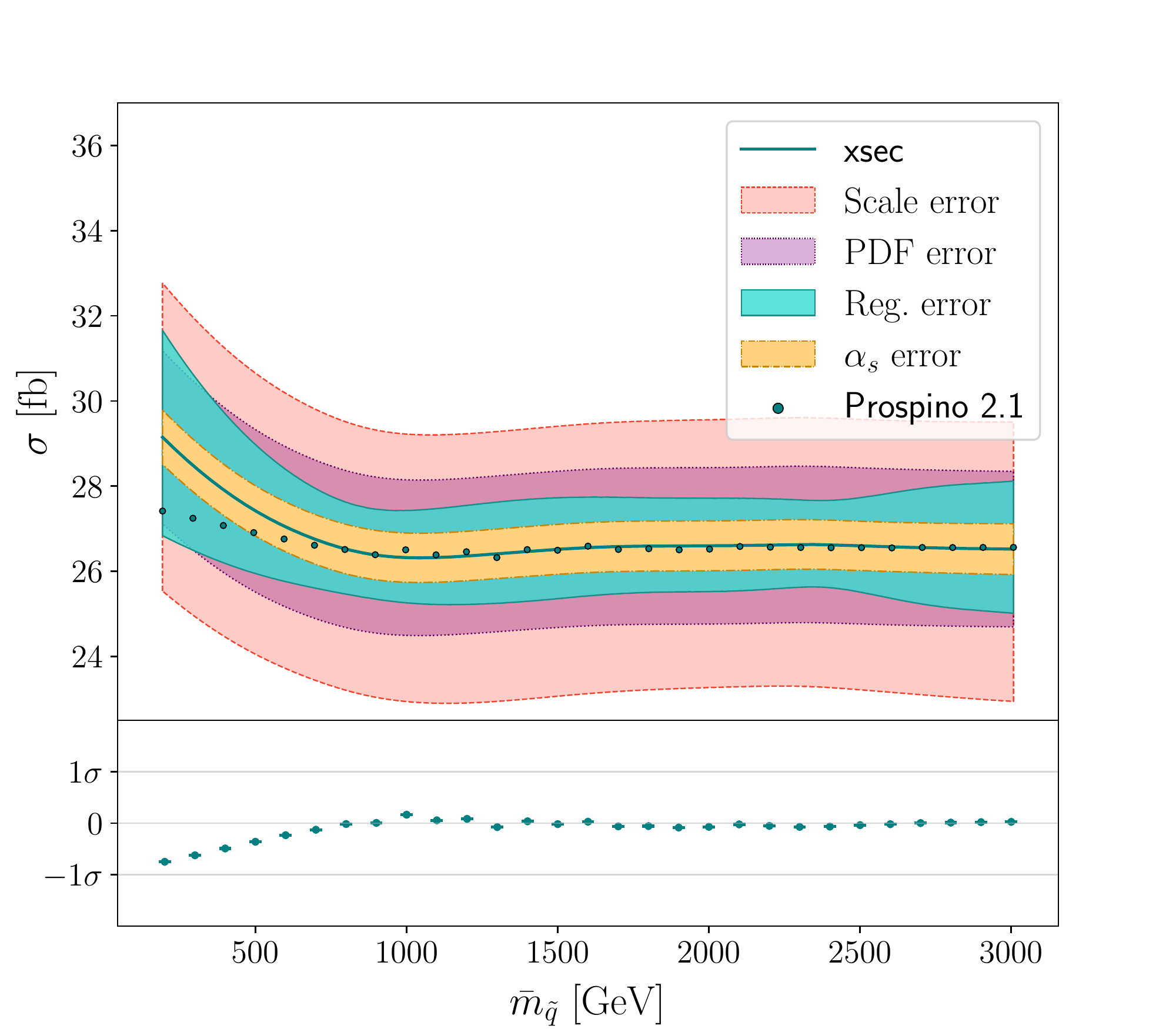}
\caption{Stop pair-production cross-sections as a function of the average of the first and second-generation squark masses, for $m_{{\tilde t}_1}=800$\,\GeV.  The \smoking prediction and $1\sigma$ regression error is in light green, the scale error in pink, the PDF error in violet, and the $\alpha_s$ error in yellow. The \prospino values are shown as dots, and below we show the residual between the \smoking regression result and \prospino.}
\label{fig:tb_msquark}
\end{minipage}%
\hspace{0.03\textwidth}
\begin{minipage}[t]{0.485\textwidth}
\includegraphics[width=\textwidth]{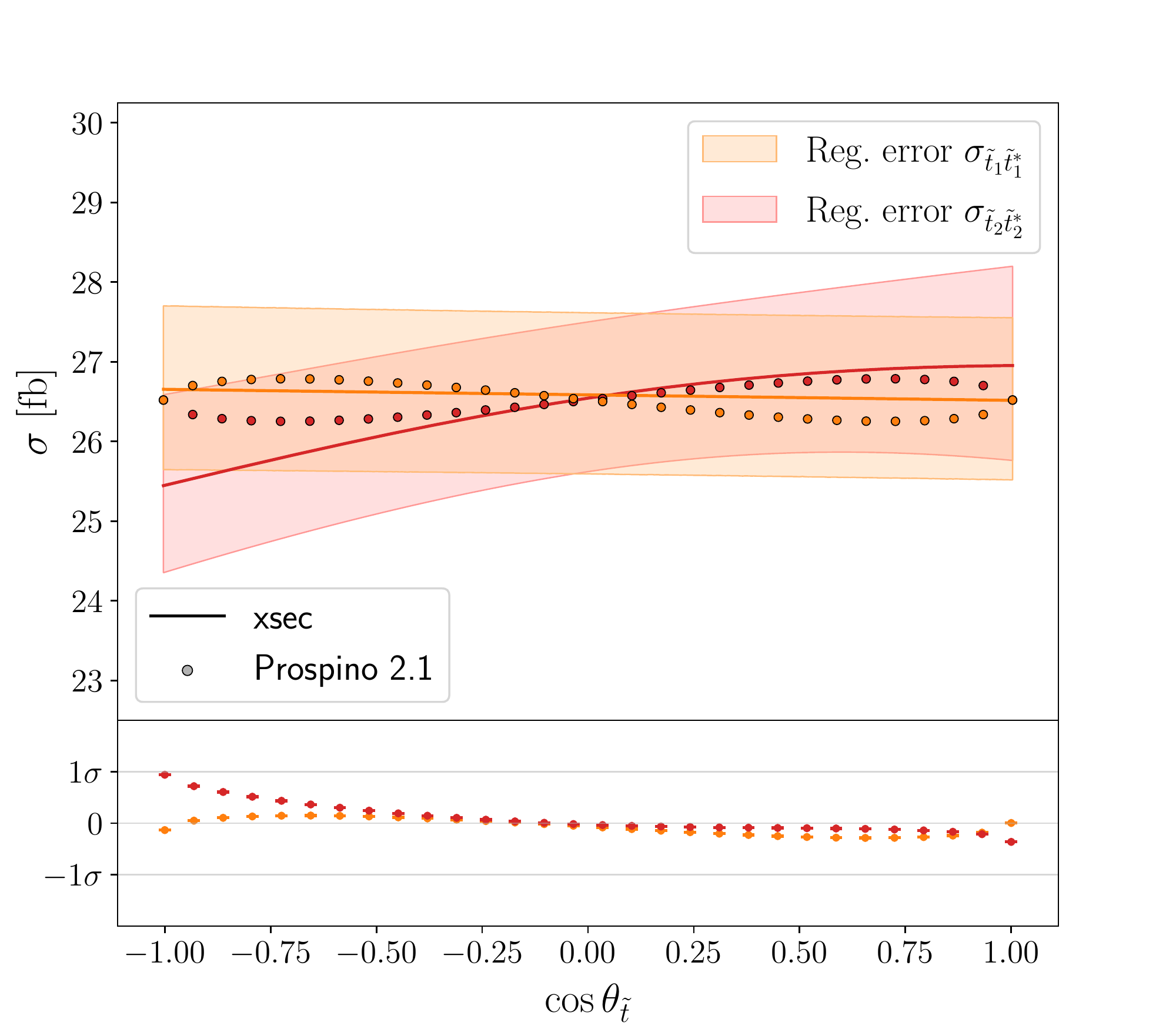}
\caption{Stop pair-production cross-sections as a function of the stop mixing angle $\cos\theta_{\tilde t}$. The central-value \smoking prediction and the regression error band is shown in orange ($\tilde t_1\tilde t_1^*$) and red ($\tilde t_2\tilde t_2^*$).  Superimposed on the prediction are the \prospino values (dots).}
\label{fig:tb_stopmix}
\end{minipage}

\includegraphics[width=0.485\textwidth]{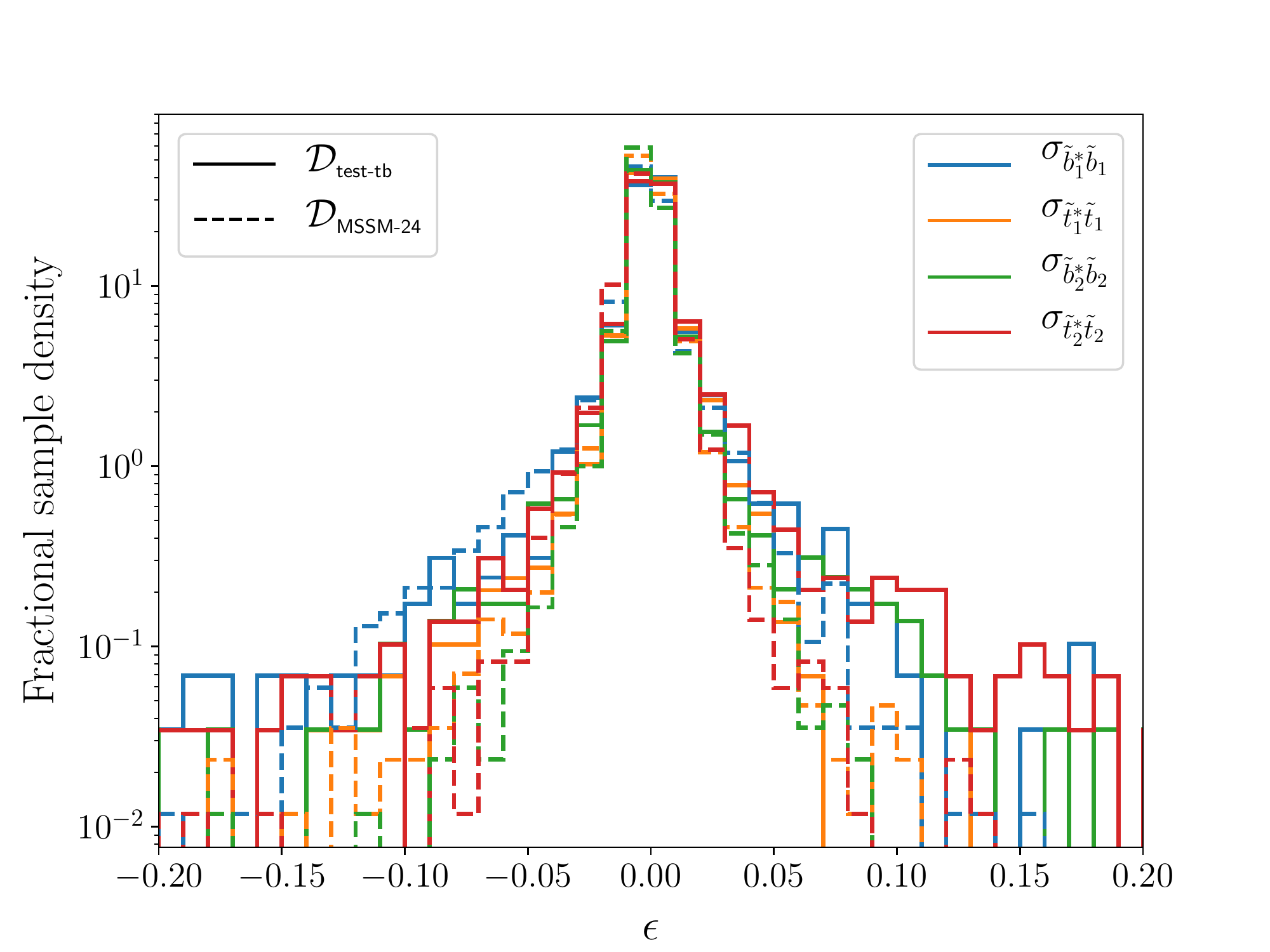}\hspace{0.03\textwidth}
\includegraphics[width=0.485\textwidth]{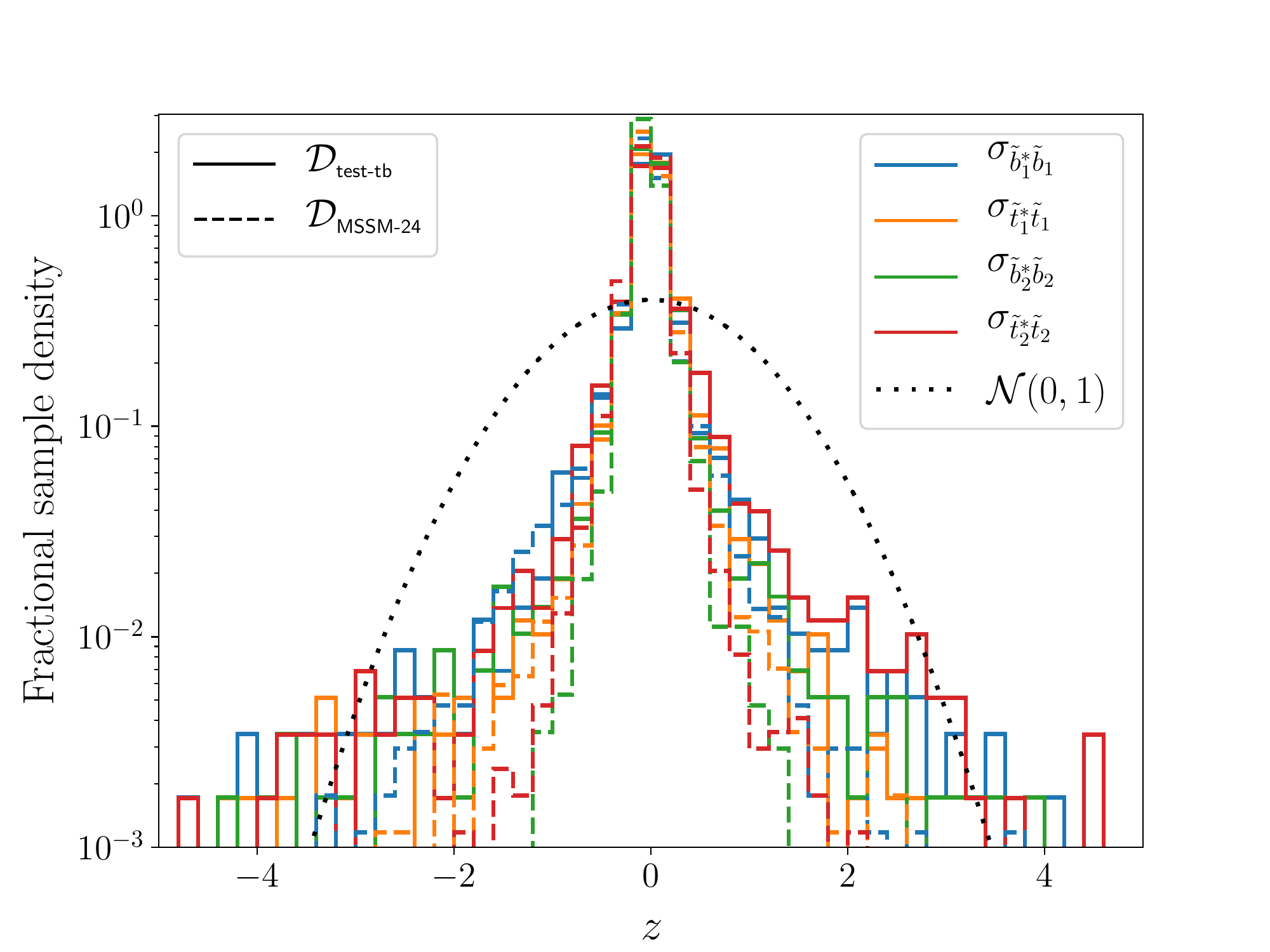}
\caption{Distributions for the relative error (left) and residual (right) for the stop and sbottom pair-production cross-sections for the test sets $\mcDtesttb$ (solid) and $\mcDMSSM$ (dashed)}
\label{fig:tb_rel_dev}
\vspace{0.02\textheight}

\includegraphics[width=0.485\textwidth]{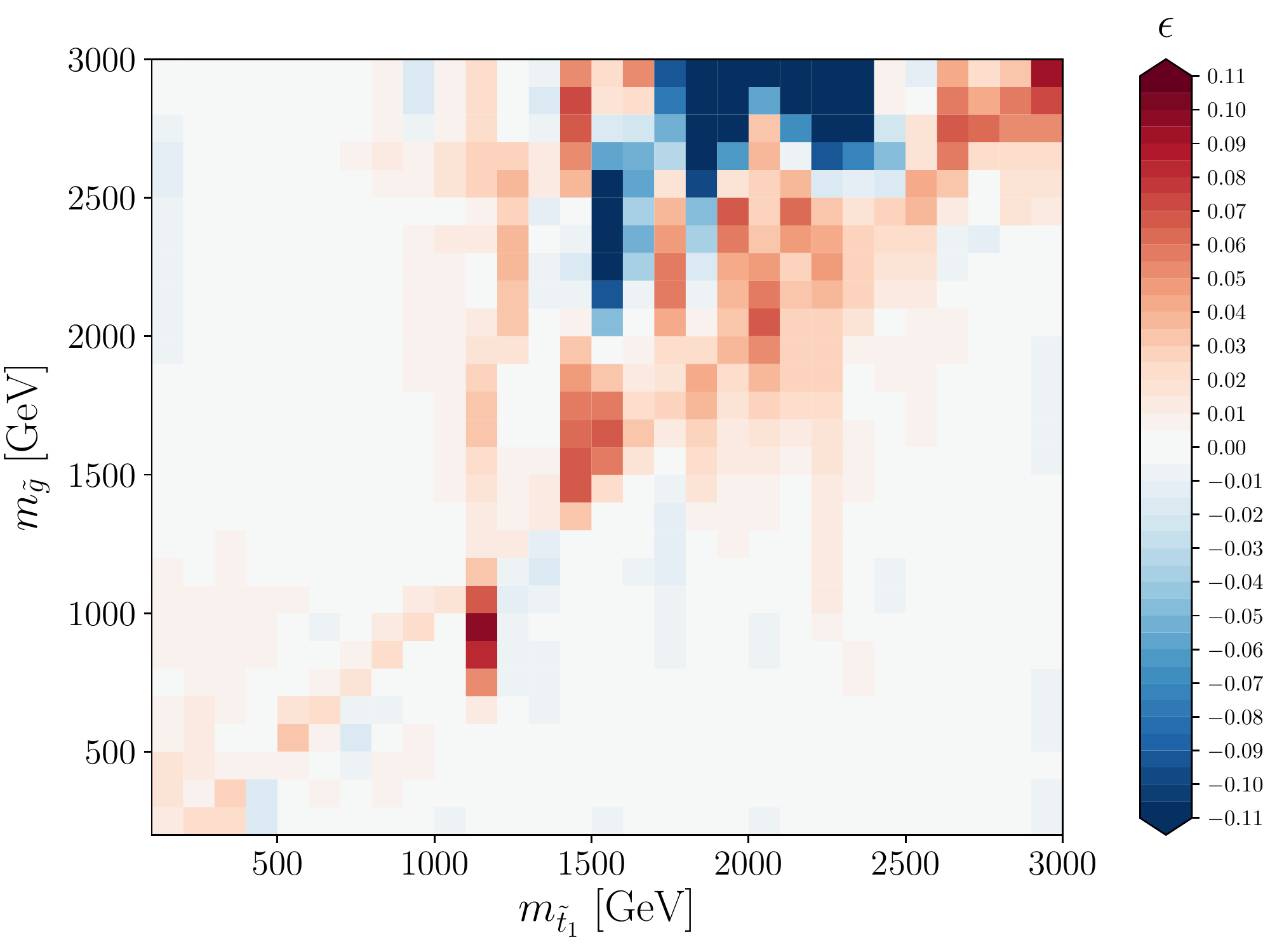}\hspace{0.03\textwidth}
\includegraphics[width=0.485\textwidth]{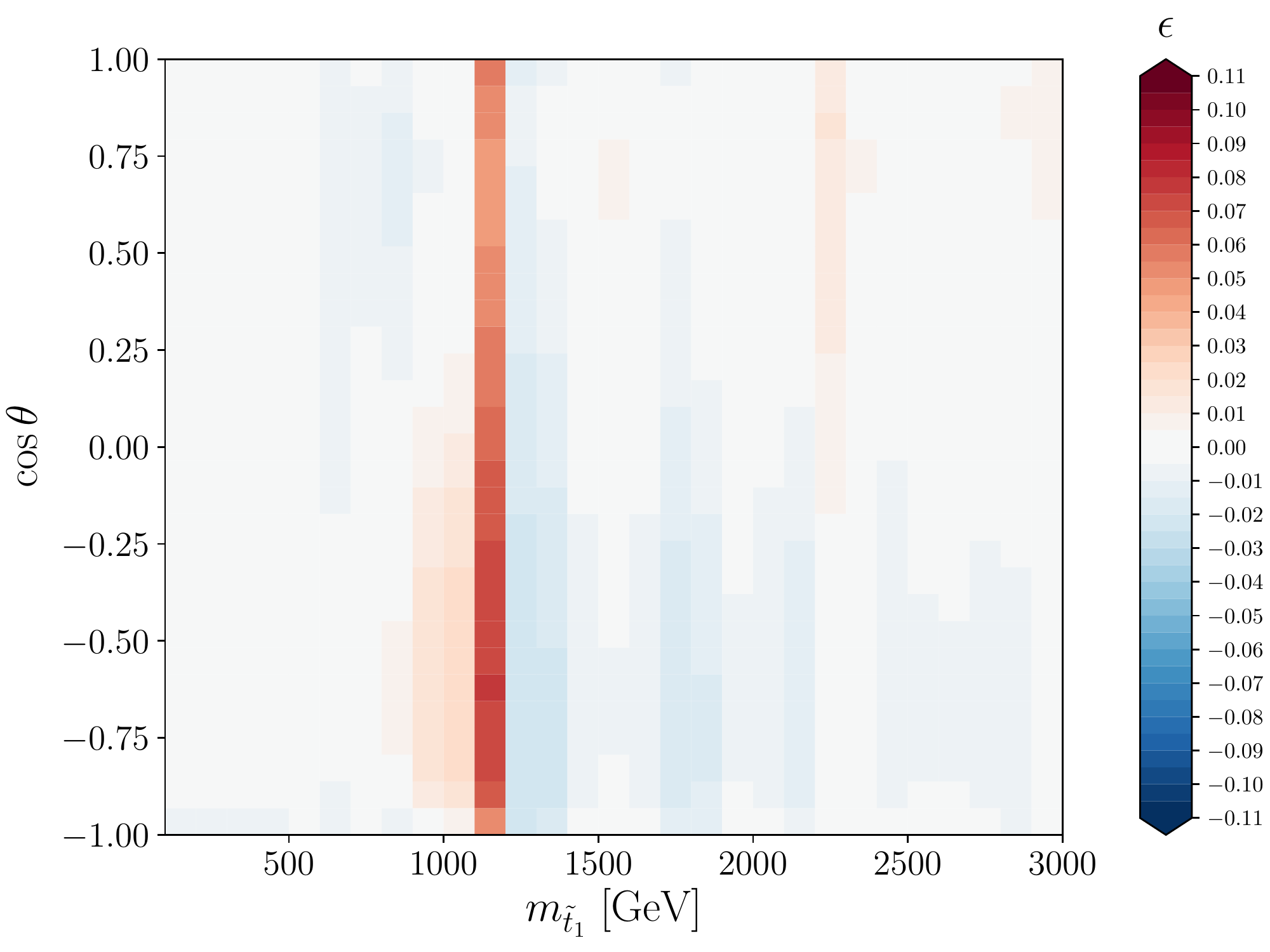}
\caption{The relative error between the predicted stop pair-production cross-section in \smoking and the \prospino value as a function of the stop mass versus gluino mass (left) and stop mass versus mixing angle (right).}
\label{fig:tb_rel_dev_2D}

\end{figure*}

In Fig.~\ref{fig:tb_msquark}, we show the stop pair-production cross-section for $m_{{\tilde t}_1}=m_{{\tilde t}_2}=800$\,\GeV as function of the average of the first and second-generation squark masses. To maximize the potential effect from squark loops, the sbottom masses are fixed to the same value as the first and second-generation squarks.
We fix the mixing angle to $\cos\theta_{\tilde t}=1$, while the gluino is decoupled at 3\,\TeV to remove the more dominant NLO contributions from gluino exchange.
We see that the dependence of the cross-section on the other squark masses is so weak that the DGP is relatively insensitive to it. However, we again observe that the \prospino results have some jitter at high squark masses. After some investigation of the sampling, we found that this jitter increases as the gluino mass is increased, and we tentatively interpret this as a numerical problem in the gluino-decoupling regime in \prospino.

We present $\tilde t_1\tilde t_1^*$ and $\tilde t_2\tilde t_2^*$ production cross-sections as a function of the mixing angle $\theta_{\tilde t}$ in Fig.~\ref{fig:tb_stopmix}, for $m_{{\tilde t}_1}=m_{{\tilde t}_2}=800$\,\GeV and all other sparticle masses set to 1\,\TeV.  We can see that the dependence on the mixing angle is so small --- of the order of 2\% --- that the limited resolution of the regression means that the DGPs cannot currently capture this behaviour.  The error is nonetheless well within that reported by \smoking.

Looking at the stop and sbottom residual distributions in $\mcDtesttb$ and $\mcDMSSM$ (right panel of Fig.\ \ref{fig:tb_rel_dev}), we see that they have somewhat longer tails than other process types.  The most notable feature is at large negative $z$, indicating a slight tendency to overestimate the cross-section reported by \prospino. This is driven by points with large ($>2$\,TeV) gluino masses, where we earlier observed the jitter in the training data.  Overall, however, the regression errors returned by \smoking for the stop and sbottom processes are conservative. The corresponding relative errors are below 10\% for the vast majority of test points, with some tails.

Finally, in Fig.~\ref{fig:tb_rel_dev_2D}, we show the relative error between the \smoking-predicted $\tilde t_1\tilde t_1^*$ production cross-section and the \prospino value, in the planes of common stop mass and gluino mass (left), and stop mass and mixing angle (right). In these plots we set all the other masses to 1 TeV, and in the left-hand plot we fix $\cos\theta_{\tilde t}=1$.

We find that \smoking predicts the values with better than 10\% accuracy in the majority of this space.  However, the left-hand panel shows again the problem with numerical jitter from \prospino at large gluino (and stop) masses. We also observe a small localised area of underestimated cross-sections (positive relative error) for $m_{\tilde t_1}\sim 1100$\,GeV and $m_{\tilde g}\sim 1000$\,GeV, seen in both plots. This region can also be found in Fig.~\ref{fig:tb_tbmass} as a single point where the residual is at the $2\sigma$ level.

\section{Code structure and usage}
\label{sec:structure}

\subsection{Speeding up cross-section evaluation}
To achieve fast cross-section predictions the \smoking code comes with fully trained GP models. These contain all relevant information about the training points, such as their inverse covariance matrix and the optimised kernel hyperparameter values. The main program, written in \python, provides a simple user interface for predicting cross-sections and their uncertainties at a given set of input parameters.

With the time-consuming processes of sampling and training carried out beforehand, two important steps remain in order to obtain predictions for a new parameter point:
\begin{enumerate}
  \item Performing the matrix-vector multiplications given in Eqs.\ (\ref{eq:GP_mean_predicted}) and (\ref{eq:GP_variance_predicted}) to compute predictions from each individual GP expert;
  \item Weighting and combining the individual predictions according to the GRBCM prescription, leading to Eq.\ (\ref{eq:GBRCM_eq14b}).
\end{enumerate}
As the aggregation step requires no matrix algebra, the first step dominates the computational cost of prediction. It scales in complexity as the square of the number of training points of the GP expert.

The code relies on \numpy~\cite{NumPy} functionality for these matrix operations. Significant speed gains are therefore possible if \numpy is properly linked to optimised numerical algebra libraries, like \blas and \lapack. Large performance differences exist between different \blas implementations. Popular ones like \mkl, \openblas and \atlas support multi-threaded calculations, enabling \numpy to take advantage of multi-core machines and \smoking to achieve the highest evaluation speeds. The \numpy package provides a function \py{show_config()} to list the numerical libraries it detected in the system it was built on.

\begin{figure}
  \centering
  \includegraphics[width=0.4\textwidth]{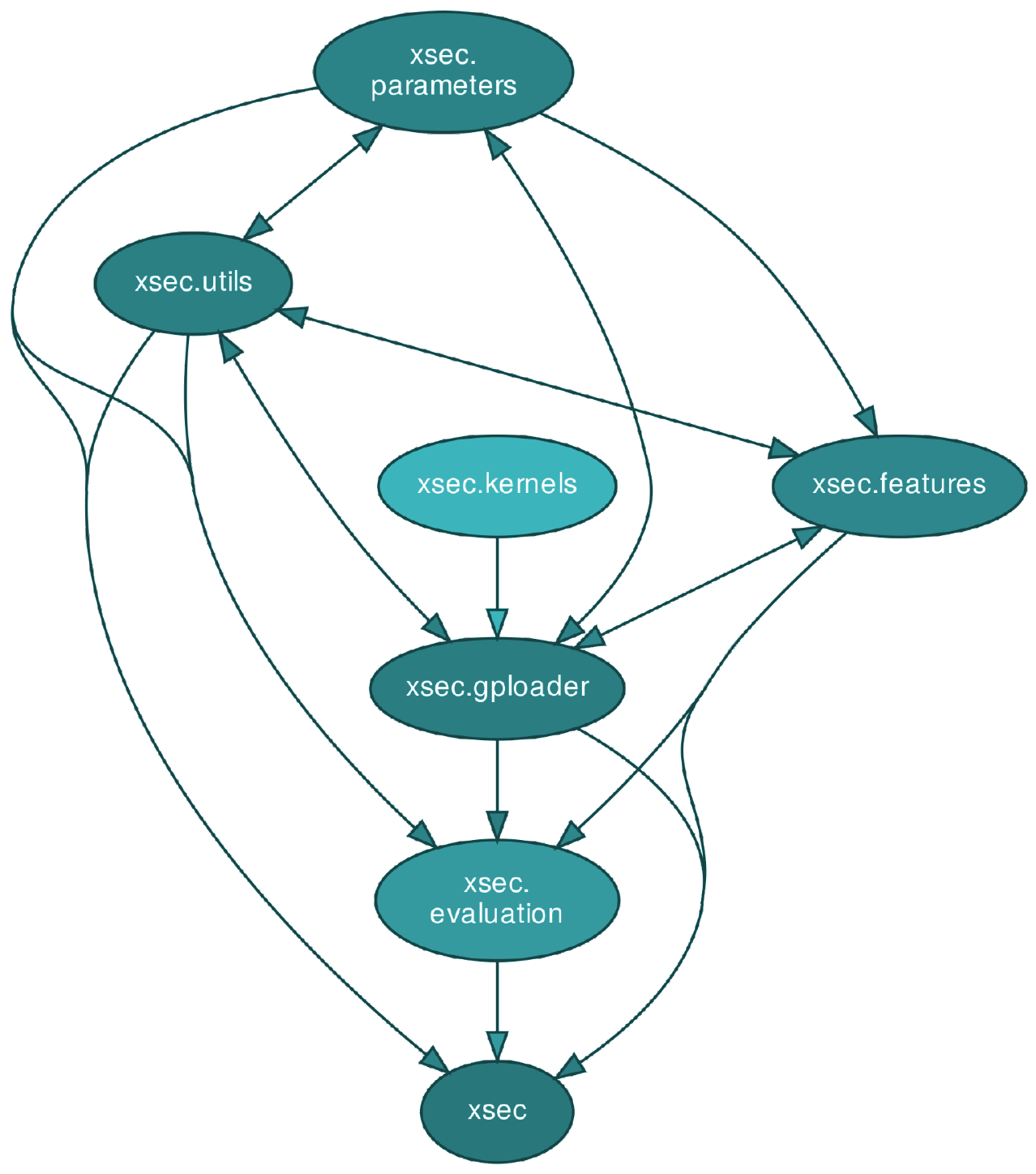}
  \caption{Module dependency graph of \smoking.}
  \label{fig:modules}
\end{figure}

The \smoking code is compatible with \python~{\textsf 2} and {\textsf 3}. While the main program consists of the six interdependent modules shown in Fig.~\ref{fig:modules}, the intended usage requires no explicit knowledge of this underlying structure. All high-level functions of relevance to the user are immediately accessible upon installing and importing the \smoking package, as detailed in the following sections.

We have not included the code used to train the GP models in the current version. Interested parties with specific use cases, or researchers trying to reproduce our results, will be given access to the private repository for this code upon request.

\subsection{Installation}
\label{subsec:installation}
The \smoking program is available from the Python Package Index (PyPI) and can be installed immediately with the \pip package manager, by executing
\begin{lstlisting}[style=terminal]
pip install xsec
\end{lstlisting}
in a terminal.
Alternatively, the user may wish to access the code directly from its GitHub repository\footnote{\url{https://github.com/jeriek/xsec}} and install a local copy:
\begin{lstlisting}[style=terminal]
git clone https://github.com/jeriek/xsec.git
pip install ./xsec
\end{lstlisting}

Due to the size of the trained GP datafiles, these are not contained in the repository, nor provided directly by \term{pip}. Instead, they are available as separate binaries in the GitHub release. We provide a script for their automatic download and extraction into a directory of choice, called by running
\begin{lstlisting}[style=terminal]
xsec-download-gprocs -g @\metavar{gp\_dir}@ -t @\metavar{process\_type}@
\end{lstlisting}
on the command line. Here, \metavar{gp\_dir} is the chosen target directory; a new directory \term{gprocs} is created in the current working directory if this argument is omitted. The other optional argument, \metavar{process\_type}, specifies the final-state type for which data will be downloaded:
\term{gg} (gluino pair production), \term{sg} (1st/2nd gen. squark--gluino pair production),
\term{ss} (1st/2nd gen. squark pair production), \term{sb} (1st/2nd gen. squark--anti-squark pair production), \term{tb} (3rd gen. squark--anti-squark pair production), or \term{all} (default).

\subsection{\python interface}
\label{sec:python-interface}

The primary envisaged use of \smoking is as a component in a parameter scan code, where the user chooses the desired particle production processes and varies the input parameters. Appendix~\ref{app:example} contains a simple example script in \python where these parameters are set by hand. Examples of SLHA file input and usage within a loop over sparticle masses are available in the GitHub repository.

\newlength{\oldparindent}
\setlength{\oldparindent}{\parindent}

To ensure that any time-consuming computations are only run when strictly necessary, the prediction process is split into four steps:
\begin{enumerate}
  \item Initialisation \\
  During the installation (see \ref{subsec:installation}), GP model files for all included processes are downloaded to a local directory \metavar{gp\_dir} of choice. At the start of a \python script, \smoking must first be pointed to that directory:
\begin{lstlisting}[style=python]
import xsec
xsec.init(data_dir="@\metavar{gp\_dir}@")
\end{lstlisting}

  \item Process selection \\
  Subsequently, the centre-of-mass energy must be set (in \GeV) through
\begin{lstlisting}[style=python]
xsec.set_energy(@\metavar{13000}@)
\end{lstlisting}
  Currently, only the 13\,\TeV dataset is available. We expect to add further energies in the near future.
  Then, one or more sparticle production processes can be specified. A single process is identified by a tuple
\begin{lstlisting}[style=python]
process = (@\metavar{pid1}@, @\metavar{pid2}@)
\end{lstlisting}
  containing the PDG codes of its final-state particles.
  A list of such tuples is required as the argument to the loading function:
\begin{lstlisting}[style=python]
xsec.load_processes([@\metavar{process1}@, @\metavar{process2}@, ...])
\end{lstlisting}
  This function call decompresses the GP model files (serialised into \python pickles to save space) for the specified processes, and loads them into memory. This step only needs to happen once per process. It may take some time, as it involves a matrix multiplication to reconstruct the inverse covariance matrix of the training points from its Cholesky decomposition.

  \hspace{\oldparindent}To show a list of all available trained processes, one can use
\begin{lstlisting}[style=python]
xsec.list_all_xsec_processes()
\end{lstlisting}

  \item Parameter input \\
  The next step is to set the values of all relevant parameters for the selected processes.  The relevant parameters are shown in Table~\ref{tab:processes}. There are three ways to enter the parameters: manually setting their values one by one, setting multiple values at once with a dictionary, or reading all parameters from an SLHA file describing the supersymmetric spectrum.
\begin{lstlisting}[style=python]
xsec.set_parameter("@\metavar{name}@", @\metavar{value}@)
xsec.set_parameters(
    {"@\metavar{name1}@": @\metavar{value1}@, "@\metavar{name2}@": @\metavar{value2}@, ...}
  )
xsec.import_slha("@\metavar{filepath}@")
\end{lstlisting}
  The example script in Appendix~\ref{app:example} lists all available parameters. The SLHA interface is based on the \pyslha package~\cite{Buckley:2013jua}, which is a dependency of \smoking. The program currently works with the SLHA1 standard, because we give the decomposition of cross-sections with squarks in terms of their flavour eigenstates (with the exception of the third generation\footnote{To avoid confusing conventions regarding the third-generation mixing angles $\theta_{\tilde t}$ and $\theta_{\tilde b}$, we use the $(1,1)$ components of the relevant mixing matrices (\texttt{STOPMIX} and \texttt{SBOTMIX} in SLHA files) as parameters.}).

  \hspace{\oldparindent}Checking and clearing the current value of one or more parameters is possible with
\begin{lstlisting}[style=python]
xsec.get_parameter("@\metavar{name}@")
xsec.get_parameters(["@\metavar{name1}@", "@\metavar{name2}@", ...])
xsec.clear_parameter("@\metavar{name}@")
xsec.clear_parameters(["@\metavar{name1}@", "@\metavar{name2}@", ...])
xsec.clear_parameters()
\end{lstlisting}
  The last line resets all parameter values.

  \item Cross-section prediction \\
  At this point, the cross-section can be evaluated by simply calling
\begin{lstlisting}[style=python]
xsec.eval_xsection()
\end{lstlisting}
  This function has two optional keywords: \py{verbose}(default \py{2}) and \py{check\_consistency} (default \py{True}). Setting \py{verbose=0} makes sure nothing is printed to the screen;
  \py{verbose=1} prints a single line per process that lists, in order: the PDG codes of the two final-state particles, the central cross-section value, and the regression, scale, PDF and $\alpha_s$ error bands (two values each);
  \py{verbose=2} prints a full description of the results. The consistency check will stop the evaluation if the features are outside the range of validity, discussed in Sec.~\ref{sec:validation}, or have not been set.
\end{enumerate}

The returned errors are always relative to the central cross-section, and a minus sign in the print-out makes it easy to distinguish the lower from the upper error bounds.
Note that currently in \smoking the PDF and $\alpha_s$ errors are symmetric by definition, following the PDF4LHC recommendations \cite{Butterworth:2015oua}, whereas the regression and scale errors are not. In order to return a symmetric $\alpha_s$ error, we average the outputs from our two GPs trained separately on the upper and lower errors. This can be changed in the code by more advanced users if desired (for example, to allow $\alpha_s$ to be treated more accurately as a nuisance parameter to be scanned over).
The regression error is asymmetric because it comes from a log-normal distribution, while the lower (upper) scale error derives from the minimum (maximum) cross-section value obtained by doubling and halving the renormalisation/factorisation scale.

The results from an evaluation for a given parameter point can be added to an existing SLHA file in the \texttt{XSECTION} block\footnote{\url{https://phystev.cnrs.fr/wiki/2013:groups:tools:slha}}:
\begin{lstlisting}[style=python]
results = xsec.eval_xsection()
xsec.write_slha("@\metavar{slha\_path}@", results)
\end{lstlisting}
The \texttt{XSECTION} structure does not allow for storing the regression error, so this is omitted. Furthermore, we do not predict individual cross-section values for all the different members of the PDF set used. Thus, in order to provide the PDF error, we follow the PDF4LHC guidelines~\cite{Butterworth:2015oua} and give the lower (upper) bound of the 68\% confidence interval by incrementing the central PDF set index by 1 (2) in the \texttt{XSECTION} block.

Finally, it is recommended to run
\begin{lstlisting}[style=python]
xsec.finalise()
\end{lstlisting}
after all evaluations have been completed. This creates a \BibTeX file in the current working directory, listing references to all original work that has been used to provide the requested results.

For advanced users, we include an option to write the decompressed GP files to disk and memory-map them for quicker access. This can reduce the memory load when many processes are requested simultaneously. The cache option can be activated by specifying
\begin{lstlisting}[style=python]
xsec.init(data_dir="@\metavar{gp\_dir}@", use_cache=True, cache_dir="@\metavar{cache\_dir}@")
\end{lstlisting}
in the initialisation step. The keyword specifying the cache directory is optional; by default a temporary directory with a random name will be created. The cache directory is removed when \py{finalise()} is called.

Another option to decrease the memory load when using \smoking for many final states is to load the data for only a few processes at a time, make predictions, and clean the memory before loading new processes. The latter can be done by
\begin{lstlisting}[style=python]
  xsec.unload_processes([@\metavar{process1}@, @\metavar{process2}@, ...])
\end{lstlisting}
If called without arguments, the data for all previously loaded processes will be cleared from memory.

\subsection{Command-line interface}
\label{sec:commandline-interface}

We also provide a command-line interface, where the user can supply a set of two final-state PDG codes and the values for the relevant features (parameters and averaged first and second-generation squark mass) involved in the corresponding cross-section. This is invoked as
\begin{lstlisting}[style=terminal]
xsec 13000 1000021 1000001 -p 1000 500 500  -g @\metavar{gp\_dir}@
\end{lstlisting}
This example calculates the cross-section for $\tilde g \tilde u_L$ production at $\sqrt{s}=13$~\TeV, using the GP data directory \metavar{gp\_dir} (see~\ref{subsec:installation}). The features in the call and their order (in this case $m_{\tilde g}$, $m_{\tilde q_i}$ and $\bar m_{\tilde q}$, all in \GeV) can be gleaned from
\begin{lstlisting}[style=terminal]
xsec 13000 1000021 1000001 --show-input
\end{lstlisting}

Alternatively, the parameters can be read from an SLHA file located at \metavar{slha\_path}:
\begin{lstlisting}[style=terminal]
xsec 13000 1000021 1000001 -r @\metavar{slha\_path}@
    -g @\metavar{gp\_dir}@
\end{lstlisting}
Executing \term{xsec --help} in a terminal provides more details on all possible inputs. We do not recommend using the command-line interface in time-sensitive applications requiring multiple cross-section evaluations for the same process. Every evaluation from the command line invokes the (inevitably slow) \py{load_processes()} step, while a lot of time can be saved by writing a \python script where this function is called only once.

\subsection{Code structure}
For the interested user we briefly describe the structure of the code and the content of the various modules. Our design objective was to keep the user interface intuitive and simple enough to be readily integrated within a larger code, even in another programming language via binding libraries.

The \py{parameters} module manages a global dictionary of parameter values, and has functions that allow for setting, getting and clearing those values. Furthermore, there is a function that checks the internal consistency of the entered values, ensuring that the mean squark mass and the centre-of-mass energy are set correctly. SLHA interface functions are also collected in this module, enabling one to read parameter values from an SLHA1 file and to write the cross-section results to an \texttt{XSECTION} block in that file.

The \py{features} module keeps track of the set of parameter values relevant to each specific production process, using a global dictionary and functions that access it.

The \py{gploader} module contains functions and global variables related to initialisation settings and to loading pre-trained GPs into memory from pickled files. When using the cache option described in Sec.~\ref{sec:python-interface}, this module is responsible for managing the cache directory on the disk.

The implementation of the GP kernels resides within the \py{kernels} module. These are mostly reproduced from the \scikit~\textsf{v0.19.2}~\cite{scikit-learn} source code under the New BSD License, apart from some new kernels and a function that reconstructs kernel function objects from the kernel parameters stored in the pickled GP files.

All functions responsible for GP predictions and combining results from multiple GPs using the GRBCM prescription are collected in the \py{evaluation} module. Each production process and cross-section type has its own data directory, containing a \py{transform} module used by the \py{evaluation} module, and relevant datafiles.  The \py{transform} module reverses the specific transformations applied to the target values during training.

Internal helper functions are defined in the \py{utils} module. These are mostly related to the internal naming system for Gaussian process model files, but also include utility functions for outputting results to screen and file, as well as the collection of \BibTeX references relevant to the requested processes.

\section{Conclusions}
\label{sec:conclusion}

In this paper, we have presented a new phenomenology code \smoking, which allows for the fast evaluation of higher-order cross-sections through regression on pre-generated training datasets. The regression in \smoking is based on Generalised Robust Bayesian Committee Machines, a method that employs distributed Gaussian processes. In addition to the central value and the related regression variance provided by the Gaussian processes, we have trained separate Gaussian processes on the scale, PDF and $\alpha_s$ errors, providing a complete ecosystem for the evaluation of cross-sections and their uncertainties.

The current version of \smoking~\textsf{1.0} includes all MSSM strong-production cross-sections at $\sqrt{s}=13$\,\TeV at NLO precision in QCD, separated into individual squark flavour final states, and allows for non-degenerate squark masses. We plan to make future updates of \smoking that will extend the code both in terms of the included energies and processes, as well as with higher-order corrections. The method used here can also be extended to other models, so long as appropriate training sets can be generated. The production of supersymmetric particles in the MSSM serves as a good starting point however, as the absence of $s$-channel resonances simplifies the training.

We would like to emphasise that the \smoking code described here is not a new calculation of the included cross-sections. It is a regression estimate based on a pre-generated sample of cross-sections taken from existing results. Thus, users of this code should also reference the original physics results on which the results of \smoking are based when using them in publications. We provide functionality within the code for easily identifying the relevant references.

\begin{acknowledgements}
This work has been supported by the Research Council of Norway (FRIPRO 230546/F20) and a NOTUR (Norway; NN9284K) grant for computation time on the Abel cluster at the University of Oslo.
This project has received funding from the European Union's Horizon 2020 research and innovation programme under the Marie Sk\l{}odow\-ska-Curie Action Innovative Training Networks MCnetITN3 and SAGEX (grant agreement Nos.~722104 and~764850), the Royal Society under grant UF160548, and the Australian Research Council under grant FT190100814.
AB wishes to thank Donatas Zaripovas and Laurynas Mince for their contributions to earlier versions of this work, and AK and PS wish to thank Ben Farmer, Nicholas Reed and Iza Veli{\v s}{\v c}ek for many useful discussions on Gaussian Processes.
\end{acknowledgements}

\appendix
\section{Code example}
\label{app:example}
Below we show a minimal \python script running the \smoking program both by specifying masses and other input parameters by hand, and by loading values from an SLHA input file. The script also demonstrates how to store the results in an SLHA file in terms of \texttt{XSECTION} blocks.

\begin{lstlisting}[style=python]
import xsec

# Set data directory
xsec.init(data_dir='gprocs')

# Set centre-of-mass energy (in GeV)
xsec.set_energy(13000)

# Load GP models for specified process(es)
xsec.load_processes([(1000021, 1000021)])

# Evaluate cross-section with given inputs
xsec.set_parameters({
    'm1000021': 1000.0,
    'm1000001': 500.0,
    'm1000002': 500.0,
    'm1000003': 500.0,
    'm1000004': 500.0,
    'm1000005': 500.0,
    'm1000006': 500.0,
    'm2000001': 500.0,
    'm2000002': 500.0,
    'm2000003': 500.0,
    'm2000004': 500.0,
    'm2000005': 500.0,
    'm2000006': 500.0,
    'sbotmix11': 0.0,
    'stopmix11': 0.0,
    'mean': 500.0,
})
xsec.eval_xsection()

# Evaluate cross-section with SLHA input
xsec.import_slha('sps1a.slha')
results = xsec.eval_xsection()

# Write results to XSECTION block in SLHA file
xsec.write_slha('sps1a.slha', results)

# Create references file
xsec.finalise()
\end{lstlisting}


\bibliographystyle{JHEP}
\providecommand{\href}[2]{#2}\begingroup\raggedright\endgroup

\end{document}